\documentclass[a4paper,11pt]{article}
\usepackage{jheppub} 

\usepackage[T1]{fontenc}
\usepackage[lofdepth,lotdepth,caption=false]{subfig}
\usepackage{amsfonts}
\usepackage{mathrsfs}
\usepackage{leftidx}
\usepackage{amssymb}
\usepackage{amsmath}
\usepackage{placeins}
\usepackage{relsize}
\usepackage{slashed}
\usepackage{multirow}
\usepackage[dvipsnames]{xcolor}
\usepackage{tikz}
\usepackage{array}
\usepackage{physics}
\usepackage{booktabs}
\usepackage{graphicx}
\usepackage{appendix}
\usepackage[caption=false]{subfig}

\usepackage{soul}
\usepackage{epsfig}
\usepackage{graphicx}               
\usepackage{url}
\usepackage{float}
\usepackage{animate}  

\usepackage{hyperref}

\hypersetup{
	bookmarksnumbered=true,
	bookmarksopen=true,
	bookmarksopenlevel=3
}

\newcommand{\be}{\begin{equation}}
\newcommand{\ee}{\end{equation}}
\newcommand{\bea}{\begin{eqnarray}}
\newcommand{\eea}{\end{eqnarray}}
\newcommand{\beq}{\begin{eqnarray}}
\newcommand{\eeq}{\end{eqnarray}}
\def\bit{\begin{itemize}}
\def\eit{\end{itemize}}
\def\ben{\begin{enumerate}}
\def\een{\end{enumerate}}

\newcommand\DN[1][\relax]{%
\ifx\relax#1\relax\else{}^{#1}\fi \!X}

\arxivnumber{2502.04108} 

\title{\boldmath Can a secluded self-interacting dark sector generate detectable gravitational waves ?}

\author[a]{Song Li,}

\author[a,e,f]{Jin Min Yang,}

\author[b,1]{Mengchao Zhang,\note{Corresponding author.}}

\author[a]{Yang Zhang,}

\author[c,d,e,f,2]{Rui Zhu\note{Corresponding author.}}

\emailAdd{chunglee@htu.edu.cn}
\emailAdd{jmyang@itp.ac.cn}
\emailAdd{mczhang@jnu.edu.cn}
\emailAdd{zhangyangphy@zzu.edu.cn}
\emailAdd{zhurui@itp.ac.cn}

\affiliation[a]{School of Physics, Henan Normal University, Xinxiang 453007, P. R.  China}
\affiliation[b]{Department of Physics and Siyuan Laboratory, Jinan University, Guangzhou 510632, P. R. China}
\affiliation[c]{State Key Laboratory of Dark Matter Physics, Tsung-Dao Lee Institute, Shanghai Jiao Tong University, Shanghai 200240, China}
\affiliation[d]{Key Laboratory for Particle Astrophysics and Cosmology (MOE) \& Shanghai Key Laboratory for Particle Physics and Cosmology, Shanghai Jiao Tong University, Shanghai 200240, China}
\affiliation[e]{Institute of Theoretical Physics, Chinese Academy of Sciences, Beijing 100190, P. R. China}
\affiliation[f]{School of Physics, University of Chinese Academy of Sciences,  Beijing 100049, P. R. China}

\abstract{In this work we study the possibility to detect the gravitational waves generated by a secluded self-interacting dark sector. ``Secluded'' means that the dark sector has almost no portal to the visible sector and thus its entropy is conserved by itself, and ``self-interacting'' means that dark matter in this model has a significant interaction to itself, making it consistent with the small-scale structure observations. A spontaneously broken $U(1)'$ is introduced for the interactions in the dark sector, and nearly massless dark radiation is also introduced to avoid the over-closure problem. Through a parameter space scan, we find that this model is highly constrained by the currently observed effective number of neutrinos ($N_{\text{eff}}$) and the large-scale structure observable Lyman-$\alpha$. Together, these two constraints exclude a large parameter space that is favored by future gravitational-wave detectors, but there is still a small portion of the model parameter space that can be detected by SKA.  }

\begin{document}
\maketitle
\flushbottom

\section{Introduction\label{sec:introduction}}

The existence of dark matter (DM) has been confirmed by its gravitational effects~\cite{Clowe:2006eq,Planck:2018vyg}. 
The abundance of DM in the universe has also been accurately determined by cosmological observations~\cite{Planck:2018vyg}.
In the $\Lambda$CDM model, DM is assumed to be cold and collisionless, which is consistent with cosmological observations at large scales~\cite{Blumenthal:1984bp,Springel:2006vs}. 
However, the assumption of collisionlessness is challenged at smaller scales ($\lesssim \mathcal{O}$(Mpc)), i.e. the so-called small-scale problems~\cite{Moore:1999gc,Moore:1994yx,Flores:1994gz,Oman:2015xda,Moore:1999nt,Klypin:1999uc,Boylan-Kolchin:2011lmk,Boylan-Kolchin:2011qkt}. 
A series of studies have shown that a velocity-dependent cross-section between DMs can solve the small-scale problems to a large extent~\cite{Kochanek:2000pi,Andrade:2020lqq,Elbert:2016dbb,Fry:2015rta,Yoshida:2000bx,Moore:2000fp,Zavala:2012us,Elbert:2014bma,Rocha:2012jg,Peter:2012jh,Kaplinghat:2015aga,Tulin:2017ara,Vogelsberger:2014pda,Dooley:2016ajo,Dave:2000ar,Robertson:2018anx,Spergel:1999mh,Colin:2002nk,Vogelsberger:2012ku,Harvey:2015hha,Burkert:2000di,Sagunski:2020spe,Yoshida:2000uw,Yang:2022mxl}.
Specifically, if the scattering cross-section between DMs inside halo is large enough, the evolution history and density distribution of DMs in the halo will be different from those of collisionless DM, and thus be in better agreement with astronomical observations.
This requirement on DM properties has inspired many model-building works~\cite{Aboubrahim:2020lnr,Bellazzini:2013foa,Boddy:2014qxa,Bringmann:2013vra,Buckley:2009in,Chen:2023rrl,Duerr:2018mbd,Feng:2009hw,Feng:2009mn,Foot:2014osa,Foot:2014uba,Foot:2016wvj,Kainulainen:2015sva,Kamada:2018kmi,Kamada:2018zxi,Kamada:2019gpp,Kamada:2019jch,Kamada:2020buc,Kang:2015aqa,Kitahara:2016zyb,Ko:2014bka,Ko:2014nha,Loeb:2010gj,Ma:2017ucp,Schutz:2014nka,Tulin:2012wi,Tulin:2013teo,vandenAarssen:2012vpm,Wang:2016lvj}, and many of them introduce a light mediator particle and use it to induce the velocity-dependent cross-section between DMs. 
This light mediator is generally not massless, and thus, due to the limited force range, the success of cold DM on large scales will not be ruined. 

It has been shown by previous studies that, in order to be consistent with small-scale data, the masses of DM and light mediator should be located in $\mathcal{O}(10)$ GeV -- $\mathcal{O}(100)$ GeV and  $\mathcal{O}(1)$ MeV -- $\mathcal{O}(10)$ MeV respectively. 
The existence of a MeV-scale dark mediator causes many problems. 
The dark mediator is generally chosen to be spin-0 scalar or spin-1 vector.
Therefore, there are renormalizable portals that link dark sector to visible sector, i.e. Higgs portal or kinetic mixing. 
If the MeV-scale dark mediators were populated in the early universe, then, because the MeV scale is close to the Big Bang nucleosynthesis (BBN) temperature, the decay of these dark mediators will definitely affect BBN observables~\cite{Depta:2020zbh,Hufnagel:2018bjp,Ibe:2021fed,Ghosh:2023ilw,Ghosh:2024cxi}. 
If dark mediators are very long-lived, they will also affect cosmic microwave background (CMB). 
Combined with other limits like supernova~\cite{Chang:2016ntp,Caputo:2025aac}, direct detection~\cite{DelNobile:2015uua}, and rare hadron decay measurements~\cite{Gori:2020xvq}, these MeV-scale dark mediators are highly constrained. 

A popular method used in the literature to evade the above constraints is turning off those portals or assuming them to be negligibly small, and thus there is no energy injection into the visible sector from the dark mediator during the BBN or CMB period. 
On the other hand, to avoid the overclosure problem, in the dark sector we also need to introduce some nearly massless particles to which dark mediators can decay~\cite{Blennow:2012de}. 
Those nearly massless dark sector particles are generally called dark radiation (DR). 
Such a ``DM+dark mediator+DR'' scenario with self-interaction can easily escape traditional DM search limits (i.e., direct searches, indirect searches, and collider searches) and meet the DM properties observed on different scales. 
However, due to the lack of portal between dark sector and visible sector, the detection of this scenario is also difficult. 

In recent years, the detection of stochastic gravitational waves (GWs) from new physics models has been intensively studied, which can serve as a good method to detect the dark sector~\cite{Addazi:2017gpt,Addazi:2020zcj,Azatov:2021ifm,Azatov:2022tii,Bai:2018dxf,Breitbach:2018ddu,Costa:2022lpy,Costa:2022oaa,Dent:2022bcd,Fairbairn:2019xog,Ghosh:2020ipy,Chen:2023rrl,Han:2023olf,Hashino:2018zsi,Huang:2017rzf,Jaeckel:2016jlh,Ratzinger:2020koh,Schwaller:2015tja,Soni:2016yes,Tsumura:2017knk,Wang:2022akn,Wang:2022lxn,Witten:1984rs,Kang:2025nhe,Costa:2025csj,Goncalves:2025uwh,Li:2025nja,Banik:2024zwj,Croon:2018erz,Athron:2023xlk}.
If a first-order phase transition (FOPT) ever happened in the dark sector, then a certain amount of latent heat will be transferred to GWs. 
After being produced, GWs propagate and get red-shifted in the universe, and thus their relic might be detected by current or future GW detectors. 

When investigating the application of stochastic GWs to detect this secluded ``DM+dark mediator+DR'' self-interacting scenario, even if the portal to the visible sector is negligibly small, there are two constraints that must be taken into account:
\begin{itemize}
\item In order to prevent the dark sector GWs from being too weak, the temperature of the dark sector (or, equivalently, the temperature of DR) cannot be too low. 
In addition, to enhance the GW signals, the dark sector phase transition usually needs to be strongly super-cooling. 
However, in the strongly super-cooling case, the entropy released from the energy difference between false vacuum and true vacuum finally goes to the DR.
DR, as a relativistic degree of freedom in addition to the SM neutrinos, will cause a deviation from the observed effective number of neutrinos ($N_{\text{eff}}$). 
Therefore, the GW detection of the ``DM+dark mediator+DR'' scenario inevitably encounters the constraint from $N_{\text{eff}}$ measurement~\cite{Bai:2021ibt,Breitbach:2018ddu}. 
\item When the temperature of the DR is not very low, the DMs in the DR bath undergo frequent scattering with the DR.
This DM-DR scattering will cause the amplitude of perturbations to be suppressed at small scales. 
Since this phenomenon is similar to baryon acoustic oscillations (BAO) caused by baryon-photon scattering, it is also called dark acoustic oscillations (DAO) in the literature~\cite{Cyr-Racine:2013fsa,Cyr-Racine:2012tfp,Buckley:2014hja}. DAO will leave an observable impact on the matter power spectrum and cosmic microwave background (CMB), which in turn limits dark-sector models with DR and self-interaction. 
\end{itemize}

In this paper we propose a concrete but economical model to realize this ``DM+dark mediator+DR'' scenario. 
A dark $U(1)'$ gauge symmetry is introduced to provide self-interaction in the dark sector, with both DM and DR charged under it~\footnote{For a secluded dark sector with non-Abelian strong interaction see e.g. ~\cite{Blinnikov:1982eh,Beylin:2020bsz}}. 
Furthermore, this dark $U(1)'$ is spontaneously broken by a dark Higgs mechanism, and the corresponding FOPT process generates stochastic GWs. 
By analyzing DM self-scattering cross sections inside halos, the stochastic GWs signal, the modification of $N_{\text{eff}}$, and the  DAO, we conduct a comprehensive phenomenological study of this secluded dark sector model which is difficult to detect by traditional methods.

In Sec.~\ref{sec:Model} we introduce the dark sector model that we will study. 
The evolution of the dark sector as the universe cools will also be described in this section. 
In Sec.~\ref{sec:detection} we explain in detail how various observables or limits are calculated. 
A comprehensive scan of the model parameter space and the presentation of results will be given in Sec.~\ref{sec:results}. 
We conclude this work in Sec.~\ref{sec:conclu}.

\section{Dark sector model and its evolution\label{sec:Model}}

In this section we introduce a concrete dark sector model that can embrace all the phenomena introduced in Sec.~\ref{sec:introduction},
and discuss how the temperature in this dark sector changes as the universe evolves.

\subsection{Lagrangian of a secluded dark sector\label{sec:Model1}}

In this work we consider a secluded dark sector gauged by a $U(1)'$ that induces interactions inside the dark sector. 
Such a setting is common in the literature~\cite{Falkowski:2011xh,Dutta:2022knf,An:2009vq,Chen:2023rrl}, but we additionally introduce a complex scalar charged under $U(1)'$ to serve as DR. 
The complete Lagrangian of the dark sector is as follows ($(-,+,+,+)$ metric): 
\begin{eqnarray}
\mathcal{L}_\text{Dark} = \bar{\chi}( i \slashed{D} - m_{\chi} ) \chi - (D_{\mu} S)^{\dagger} D^{\mu} S - (D_{\mu} \psi )^{\dagger} D^{\mu} \psi - \frac{1}{4} F'_{\mu\nu}  F'^{\mu\nu} - V(S,\psi)  
\end{eqnarray} 
Here, $\chi$, as vector-like fermion, is the DM particle. 
$S$ is the complex dark Higgs field that obtains a vacuum expectation value (VEV) below the critical temperature and break $U(1)'$. 
Complex scalar $\psi$ is the lightest particle in the dark sector and serves as DR. 
$D_{\mu} \equiv \partial_{\mu} + i g' Q_i A'_{\mu} $ ($i=\chi,S,\psi$) is the covariant derivative, and $g'$ is the gauge 
coupling for $U(1)'$. 
$F'_{\mu\nu} \equiv \partial_{\mu}A'_{\nu} - \partial_{\nu}A'_{\mu} $ is the field strength of $U(1)'$ gauge boson $A'$. 
$A'$ is also called dark photon. Different with the SM photon, dark photon $A'$ becomes massive after spontaneous symmetry breaking (SSB) of $U(1)'$. 

The $U(1)'$ charge assignment can be quite arbitrary~\footnote{If one consider other issues like asymmetry generation in the dark sector, or the $\chi-\bar{\chi}$ oscillation, $U(1)'$ charge assignment will be a subtle problem. See, e.g. ~\cite{Tulin:2012re,Buckley:2011ye,Han:2023olf,Chen:2023rrl}, for more discussion. }. 
In this work we consider following two kinds of $U(1)'$ charge assignments: 
\begin{eqnarray}
\text{Charge assignment A: } \ \ \{Q_{\chi}, Q_{S}, Q_{\psi}\} &=& \{+1,+2/3, +1  \} \\
\text{Charge assignment B: } \ \ \{Q_{\chi}, Q_{S}, Q_{\psi}\} &=& \{+1,+3, +1/2  \} 
\end{eqnarray}
The complete dark scalar potential is given by 
\begin{eqnarray}
V(S,\psi)  &=& \tilde{m}^2_\psi \psi^\dagger \psi  - \mu^2 S^{\dagger} S +  \frac{\lambda_S}{4} \left( S^{\dagger}S \right)^2 + \frac{\lambda_\psi}{4} \left( \psi^{\dagger} \psi \right)^2 + \lambda_{S\psi} \left( S^{\dagger}S \right) \left( \psi^{\dagger} \psi \right)
\end{eqnarray} 
Our charge assignments prevent terms like $S^\dagger \psi^2$ or $S^\dagger \psi^3$, and thus the form of dark scalar potential is relatively simple. 
Yukawa terms like $\psi^\dagger \bar{\chi}^{\text{c}}\chi $ or $S^\dagger \bar{\chi}^{\text{c}}\chi $ are also forbidden due to our charge assignments. 

After $U(1)'$ SSB, $S$ obtains a VEV, i.e. $\langle S \rangle = v/\sqrt{2}$ with $v \equiv 2 \mu / \sqrt{\lambda_S} $. 
Then the mass square of $\psi$ becomes $m^2_{\psi} \equiv \tilde{m}^2_\psi + \frac{1}{2}\lambda_{S\psi}v^2 $. 
To make $\psi$ a DR, we simply require $m^2_{\psi} \ll (0.1\text{ eV})^2 $. 
In addition, dark Higgs mass square $m_s^2 = \frac{1}{2} \lambda_S v^2$ and dark photon mass square $m^2_{A'} = (Q_Sg')^2 v^2$ are also fixed after SSB. 
The values for coupling $\lambda_\psi$ and $\lambda_{S\psi}$ are not directly relevant in this work, and thus the input parameters of this model (for a certain charge assignment) are: 
\begin{eqnarray}
m_\chi \ , \ m_{A'} \ , \ m_s \ , \ g' \ (\text{or $\alpha'$}),
\end{eqnarray}
where $\alpha' \equiv \frac{g'^2}{4\pi}$ is the dark fine structure constant.

The parameter range we will study is  as follows
\begin{eqnarray}
& 10\text{ GeV} < m_\chi < 600\text{ GeV} \ , \ 0.1\text{ MeV}  < m_{A'} < 10\text{ MeV} \ , \ 0.005 < \alpha' < 0.4 \ , \ \\\nonumber
&     \sqrt{ \frac{ 3 }{4\pi} Q_S^2 \alpha' m^2_{A'} } < m_s < \frac{1}{2} m_{A'} \ . \ \ 
\end{eqnarray}
The parameter ranges in the first line are favored by the small-scale structure, as we briefly explained in Sec.~\ref{sec:introduction}. 
More details will be given in the following sections.  
We set $\alpha'$ greater than 0.005 to make the DM relic density easy to satisfy, see next subsection for explanation. 
$\alpha' < 0.4$ is simply used to avoid Landau pole at a high energy scale~\cite{Li:2024wqj}. 
In the calculation of effective potential we further require $Q_S^2 \alpha' < 0.2$ to make perturbative calculation valid. 
The lower bound of the dark Higgs mass $m_s$ is the Linde-Weinberg bound which says $m_s$ cannot be too light; otherwise, the SSB cannot happen when loop corrections to the effective potential are considered. 
The lower bound present here is an approximation, while the complete formula can be found in our previous paper~\cite{Li:2024wqj}. 
The upper bound of $m_s$ is required by a FOPT in the dark sector. 
Lattice simulations show that FOPT in Abelian Higgs model can only happen when Higgs mass is smaller or much smaller than gauge boson mass~\cite{Karjalainen:1996wx,Dimopoulos:1997cz}. 
Thus we dictate this upper bound to give our model the potential to generate detectable stochastic gravitational waves from a FOPT.

It should be noticed that our dark sector model has three renormalizable portals to the visible sector, i.e. $(S^\dagger S)(H^\dagger H)$, $(\psi^\dagger \psi)(H^\dagger H)$, and $F'_{\mu\nu}F^{\mu\nu}$, where $H$ is the SM Higgs doublet and $F^{\mu\nu}$ the electromagnetic field strength.  
In this work we set the couplings of these three terms to zero and thus make the dark sector fully secluded. 

\subsection{Evolution of the secluded dark sector\label{sec:Model2}}

In this work we assume the dark sector to be thermalized at a very high temperature. 
This thermalization process in the dark sector can also be called dark reheating, which can be realized via mechanisms like asymmetric reheating~\cite{Sandick:2021gew,Adshead:2016xxj}. 
This mechanism allow the reheating temperatures of the visible sector and the dark sector to be roughly on the same order of magnitude. 
In this work we use $T'$ to label the temperature of dark sector. 
The evolution of the dark sector after dark reheating is roughly as follows
\begin{itemize}
\item Dark reheating temperature $T'_\text{reh}$ is much larger than the mass of dark matter, i.e. $T'_\text{reh} \gg m_\chi$. 
Thus all the dark sector particles we listed in the previous subsection are highly relativistic after dark reheating, for a period of time. The effective degrees of freedom (d.o.f.) for energy density in the dark sector (labeled as ${g'_{\star,e}}$) right after dark reheating are $g'_{\star,e}(T'_\text{reh}) = \frac{7}{8}\times 4 + 2 + 2 + 2 =  9.5$.  
\item As the temperature drops, DM $\chi$ and anti-DM $\bar{\chi}$ gradually become non-relativistic and their abundances decrease through the annihilation process $\chi\bar{\chi}\to A'A'$. 
After freeze-out, the sum of DM abundance and anti-DM abundance $Y_\chi + Y_{\bar{\chi}}$ is frozen and it constitutes the observed relic density.

In the symmetric DM case where $Y_\chi = Y_{\bar{\chi}}$, the relic density is inversely proportional to annihilation cross-section $\sigma_{\chi\bar{\chi}\to A'A' } $, and thus the available parameter space is greatly compressed. 
To enlarge the parameter space, we assume there is an asymmetry between DM and anti-DM particles, i.e. $Y_{\Delta\chi} \equiv Y_\chi - Y_{\bar{\chi}} > 0$. 
It has been shown in previous work that when $\alpha' > 0.005$, the relic density is mainly determined by $Y_{\Delta\chi}$~\cite{Chen:2023rrl,Baldes:2017gzw,Graesser:2011wi}. 
Our work~\cite{Chen:2023rrl} also shows that $Y_{\Delta\chi}$ can vary from 0 to $10^{-5}$ in a complete model, this is mainly because CP violation phase in the dark sector can be quite large.
 The following formula can be used to quickly estimate the current abundance of DM:
\begin{eqnarray}
\Omega_{\chi} h^2 \approx m_{\chi} Y_{\Delta\chi} s_0 / \rho_{\text{cr}} \approx Y_{\Delta\chi} \left( \frac{m_{\chi}}{\text{GeV}} \right) \times 2.72 \times 10^8
\end{eqnarray}
Thus for DM heavier than 10 GeV, the currently observed relic density $\Omega_{\chi} h^2 = 0.12$ can always be satisfied by a suitable value of $Y_{\Delta\chi}$. 
In the rest of this paper, we can simply assume $\Omega_{\chi} h^2 = 0.12$ is true.

\item When $T'$ is around the MeV scale, the FOPT happens in the dark sector and dark photon $A'$ becomes massive. 
Bubble collisions and subsequent processes generate stochastic gravitational waves that might be detectable. 
Released latent heat also heat up the dark sector. 
When the FOPT lasts much less than a Hubble time (this is actually always true in our dark sector model), we can expect the energy density to be approximately conserved before and after FOPT: 
\begin{equation}
u'_{\rm before}=u'_{\rm after}\,,\label{eq:EnergyConservation}
\end{equation}
where $u'$ represents the internal energy density of the dark sector. $u'$ can be calculated from the effective potential in thermal equilibrium:
\begin{equation}
u'=V(v,T)+sT=V(v,T)-T\pdv{V(v,T)}{T}\,.
\label{eq:energydensity}
\end{equation}
We can use the percolation temperature $T'_{\rm perc}$ (the temperature at which FOPT is nearly complete) as $T'_{\rm before}$ to calculate $T'_{\rm after}$, with the corresponding equation being:
\begin{equation}
V(0,T'_{\rm perc})-T\left.\pdv{V(0,T)}{T}\right|_{T=T'_{\rm perc}}=V(v(T'_{\rm after}),T'_{\rm after})-T\left.\pdv{V(v,T)}{T}\right|_{\substack{T=T'_{\rm after}\\ v=v(T'_{\rm after})}}\label{eq:exactReheating}
\end{equation}
Another way to estimate the internal energy density is through effective degrees of freedom and the zero-temperature effective potential:
\begin{equation}
u'\approx V(v,T=0)+\rho'= V(v,T=0) +g'_{\star,e}\frac{\pi^2}{30}T^4
\end{equation}
Therefore, Eq.~\eqref{eq:EnergyConservation} can be approximated as:
\begin{equation}
\rho'_{\rm after}=\rho'_{\rm before}+ \Delta V(v,T=0) 
\end{equation}
where $\Delta V(v,T=0)$ is the vacuum-energy difference between two phases and $\rho'$ the radiation energy density in the dark sector. 
The above equation is used in previous studies such as~\cite{Bringmann:2023opz}.
It is clear that the reheating after the phase transition mainly comes from the release of vacuum energy, which increases the temperature of the dark sector by a factor of approximately $\left(1+\frac{\Delta V}{\rho'_{\rm before}}\right)^{1/4}$, provided $g'_{\star,e}$ doesn't change much during the FOPT.  This result helps us quickly estimate the reheating effect, but in this work we use the more precise formula Eq.~\eqref{eq:exactReheating} to calculate $T'_{\rm after}$.

\item After FOPT, as the dark sector temperature decreases, the remaining massive $A'$ and $s$ will all eventually decay to DR. 
Then the dark sector will be composed of DM $\chi$ and DR $\psi$/$\psi^\dagger$.
The dark-sector d.o.f. $g'_\star$ during that time is 2.  
\end{itemize}

It is convenient to define a temperature ratio between dark sector and visible sector:
\begin{eqnarray}
\xi \equiv \frac{T'}{T_{\gamma}}
\end{eqnarray}
Here $T_{\gamma}$ is the photon temperature in the visible sector. 
We can consider temperature ratio $\xi$ as a function of $T_{\gamma}$, i.e. $\xi(T_{\gamma})$.
The value of $\xi$ right after dark reheating (to be labeled as $\xi_\text{reh}$) is actually an input parameter that depends on the details of the very early universe. 
The value of $\xi(T_{\gamma})$ can be determined by the entropy conservation (note the jump during the FOPT): 
\begin{eqnarray}
\xi(T_{\gamma}) = \begin{cases}\xi_\text{reh} \left( \frac{9.5 \times g_{\star,s}(T_\gamma) }{106.75\times g'_{\star,s}(T_\gamma \xi) } \right)^{\frac{1}{3}}\,,\quad \text{before FOPT;}\\
\xi_\text{after} \left( \frac{g'_{*,s}(T'_{\rm after}) \times g_{\star,s}(T_\gamma) }{g_{*,s}(T'_{\rm after}/\xi_\text{after})\times g'_{\star,s}(T_\gamma \xi) } \right)^{\frac{1}{3}}\,,\quad \text{after FOPT,}
\end{cases}
\end{eqnarray}
with $g_{\star,s}$ and $g'_{\star,s}$ the effective d.o.f. for entropy density in the visible sector and dark sector, respectively. 
In dark sector, approximation $g'_{\star,s} \approx g'_{\star,e}$ always holds true. 
While in the visible sector, $g_{\star,s} \approx g_{\star,e}$ is valid only before neutrino decoupling. 
Effective d.o.f. in both sectors can be determined numerically, see~\cite{Husdal:2016haj} for detailed formula. 

\begin{figure}[ht]
\centering
\includegraphics[width=4.0in]{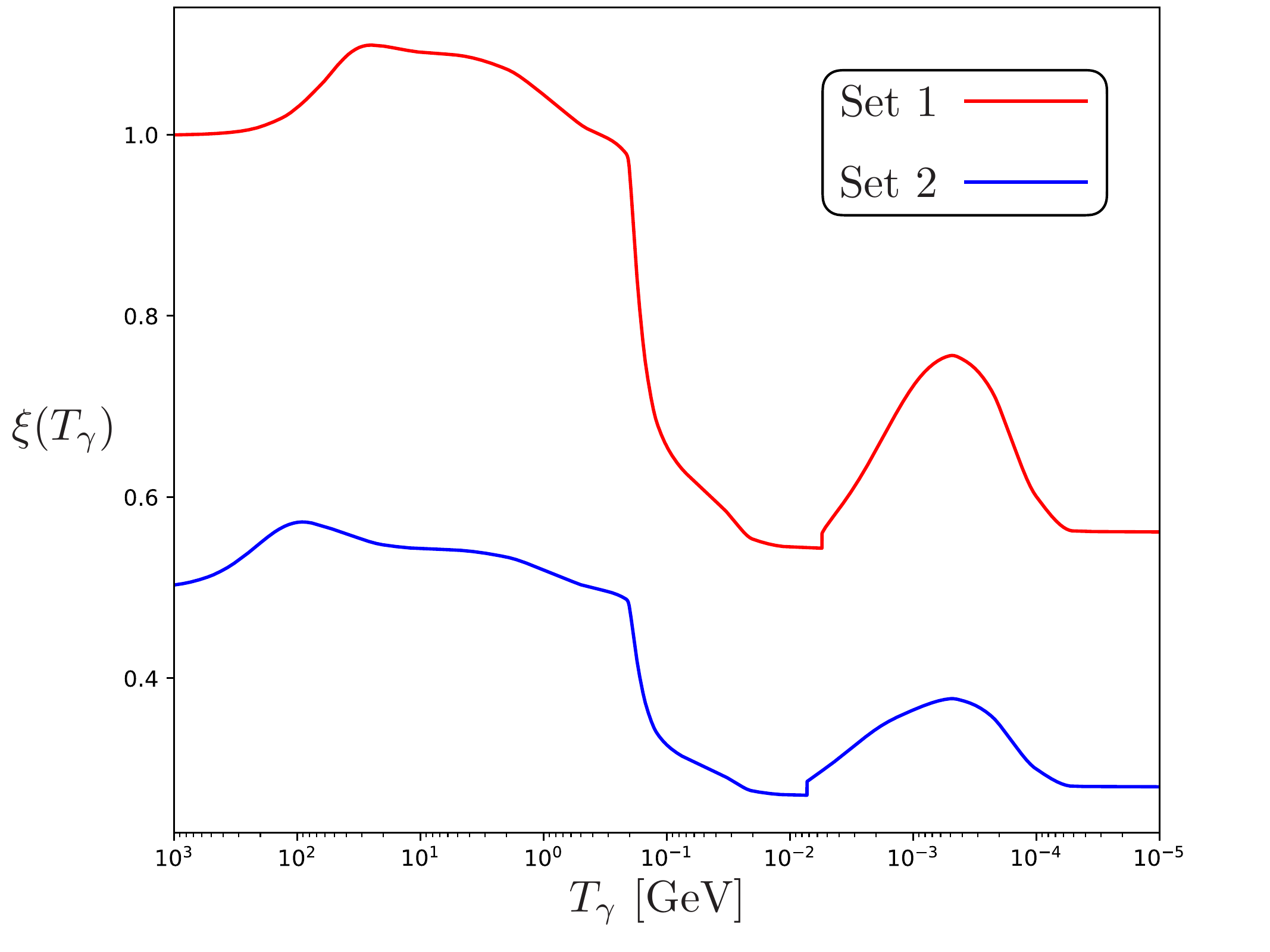}
\vspace{-20pt}
\caption{Temperature ratio $\xi$ as functions of $T_\gamma$ for two benchmark sets. }
\label{fig:Tratio}
\end{figure}

To illustrate the evolution of $\xi$, we consider the following two benchmark sets (both with charge assignment A):
\begin{eqnarray}
&\nonumber \text{Set 1: } m_\chi = 200 \text{ GeV}  , \ m_{A'} = 5 \text{ MeV}  , \ m_s = 2 \text{ MeV}  , \ \alpha'=0.2 , \ \xi_\text{reh} = 1.0   \\  
&\nonumber \text{Set 2: } m_\chi = 400 \text{ GeV}  , \ m_{A'} = 4 \text{ MeV}  , \ m_s = 1 \text{ MeV}  , \ \alpha'=0.1  , \ \xi_\text{reh} = 0.5  
\end{eqnarray}
In Fig.~\ref{fig:Tratio} we present the temperature ratio $\xi$ as a function of $T_\gamma$ for the above two benchmark sets. 
As we discussed before, FOPT happens at $T'_{\text{perc}}$ (around $\mathcal{O}(1)$ MeV) instantly increases the value of $\xi$. 
On the other hand, when $T_\gamma \ll $ MeV, $\xi$ will approach its asymptotic value.
The asymptotic value of $\xi$ will be labeled as $\xi_{\infty}$ and plays an important role in the calculation of the next section. 

\section{Calculation of observables and constraints of the self-interacting secluded dark sector\label{sec:detection}}

In this section we present in detail the calculations of the observables and constraints of the dark sector, including DM scattering cross sections inside halos, DAO caused by the DM-DR scattering, the additional contribution to $N_{\text{eff}}$, and FOPT-induced GW spectra. 

\subsection{DM self-interacting inside halos\label{sec:detection1}}

The behavior of DM inside halos, such as the density distribution of DM and the rotation curves of galaxies, depends on the average scattering cross section between DMs over DM mass, $\overline{\sigma}/m_\chi$.
Due to the Virial theorem, different astronomical systems (from dwarf galaxies to galaxy clusters) with different masses have different average DM relative velocities $\langle v_{\text{rel}} \rangle$. 
In order for DMs in different astronomical systems to behave as observed, $\overline{\sigma}/m_\chi$ needs to depend on the average velocity $\langle v_{\text{rel}} \rangle$. 
Below we summarize $\langle v_{\text{rel}} \rangle$ of different astronomical systems and the requirements for $\overline{\sigma}/m_\chi$ from different observables:
\begin{itemize}
\item Diversity of galactic rotation curves for isolated spiral galaxies over a wide mass range~\cite{Kamada:2016euw,Ren:2018jpt,Kaplinghat:2019dhn}: 
\begin{eqnarray}
  \langle v_{\text{rel}} \rangle \sim 43.2 - 432 \text{km/s} \ , \  \overline{\sigma}/m_\chi > \text{2 cm$^{2}$g$^{-1}$}
\end{eqnarray}
\item To avoid subhalo evaporation of Milky Way (MW) size halo~\cite{Vogelsberger:2012ku,Rocha:2012jg,Zavala:2012us}: 
\begin{eqnarray}
  \langle v_{\text{rel}} \rangle \gtrsim 200 \text{km/s} \ , \  \overline{\sigma}/m_\chi < \text{10 cm$^{2}$g$^{-1}$}
\end{eqnarray}
\item The anti-correlation between the central DM densities and orbital pericenter distances of Milky Way’s dwarf spheroidal galaxies (MW dSphs)~\cite{Correa:2020qam}:
\begin{eqnarray}
  \langle v_{\text{rel}} \rangle \sim 20 - 60 \text{km/s} \ , \  \overline{\sigma}/m_\chi \sim \mathcal{O}(10) - \mathcal{O}(100)  \text{ cm$^{2}$g$^{-1}$}
\end{eqnarray}
Detailed values are given in Tab.~\ref{tab:data}. 
\item Dense DM substructure which perturbs the strong lens galaxy SDSSJ0946+1006~\cite{Minor:2020hic,Nadler:2023nrd}: 
\begin{eqnarray}
  \langle v_{\text{rel}} \rangle \sim 144 \text{km/s} \ , \  \overline{\sigma}/m_\chi \sim 14 - 25  \text{ cm$^{2}$g$^{-1}$}
\end{eqnarray}
This observation is labeled as `perturber' in Tab.~\ref{tab:data}.
\item Rotation curves of isolated and gas-rich ultradiffuse galaxies (UDGs)~\cite{Nadler:2023nrd}~\footnote{The lower bound comes from private discussion with the author of Ref.~\cite{Nadler:2023nrd}.}:
\begin{eqnarray}
  \langle v_{\text{rel}} \rangle \sim 57.6 \text{km/s} \ , \  \overline{\sigma}/m_\chi \sim 47-86  \text{ cm$^{2}$g$^{-1}$}
\end{eqnarray}
\item For groups and clusters with $\langle v_{\text{rel}} \rangle$ about 1152 km/s and 1440 km/s, $\bar{\sigma}/m_\chi$ need to be smaller than 1 cm$^{2}$g$^{-1}$ and 0.1 cm$^{2}$g$^{-1}$ to be consistent with observations~\cite{Sagunski:2020spe,Kaplinghat:2015aga,Andrade:2020lqq}. 
\end{itemize}

\begin{table}[htp]
\begin{center}
\begin{tabular}{ c c c }
\hline
\hline
Observables &\ \ \ $ \langle v_{\text{rel}} \rangle $ (km/s)  \ \ \ &  \ \ \  $\overline{\sigma}/ m_\chi $ (cm$^2$/g)  \ \ \  \\ 
\hline
\hline
MW dSph: UM & \ \ \  30.87 \ \ \  & \ \ \ 40 -- 50 \ \ \  \\
\hline
MW dSph: Draco & \ \ \  56.34 \ \ \  & \ \ \ 20 -- 30 \ \ \  \\ 
\hline
MW dSph: LeoII & \ \ \  20.98 \ \ \  & \ \ \ 90 -- 150 \ \ \  \\ 
\hline
MW dSph: Sextans & \ \ \  32.34 \ \ \  & \ \ \ 70 -- 120 \ \ \  \\ 
\hline
MW dSph: Carina & \ \ \  48.42 \ \ \  & \ \ \ 40 -- 50 \ \ \  \\ 
\hline
MW dSph: CVnI & \ \ \  38.29 \ \ \  & \ \ \ 50 -- 80 \ \ \  \\  
\hline
MW dSph: Sculptor & \ \ \  62.83 \ \ \  & \ \ \ 30 -- 40 \ \ \  \\ 
\hline
MW dSph: Fornax & \ \ \  57.95 \ \ \  & \ \ \ 30 -- 50 \ \ \  \\ 
\hline
MW dSph: LeoI & \ \ \  58.27 \ \ \  & \ \ \ 50 -- 70 \ \ \  \\ 
\hline
Perturber & \ \ \  144 \ \ \  & \ \ \ 14 -- 25 \ \ \  \\ 
\hline
UDGs & \ \ \  57.6  \ \ \  & \ \ \ 47 -- 86  \ \ \  \\ 
\hline
\hline
\end{tabular}
\caption{Small-scale data used for $\chi^2$ fit. }
\label{tab:data}
\end{center}
\end{table}

In Tab.~\ref{tab:data} we present all the data with an error bar of $\overline{\sigma}/ m_\chi $.
These data will be used to perform a $\chi^2$ fit to obtain the corresponding 2$\sigma$ region in parameter space which is consistent with small-scale observations. 
Constraints other than those listed in Tab.~\ref{tab:data} are used to limit the parameter space directly. 

The methods for calculating the DM scattering cross-section depend on model parameters $\{ m_{\chi},\ m_{A'},\ \alpha'  \}$ and $v_{\text{rel}}$. 
In the Born regime with  $ \frac{ Q_\chi^2 \alpha'  m_{\chi} }{m_{A'}} \ll 1 $, analytic formula can be obtained via perturbative calculation~\cite{Tulin:2013teo,Feng:2009hw,Buckley:2009in,Kahlhoefer:2017umn}. 
In the classical regime with $ \frac{ Q_\chi^2\alpha'  m_{\chi} }{m_{A'}} \gtrsim 1 $ and $ \frac{ m_{\chi} v_{\text{rel}} }{ m_{A'} } \gg 1 $, formulae are obtained via fitting with numerical results~\cite{Feng:2009hw,Buckley:2009in,Khrapak:2003kjw,Cyr-Racine:2015ihg}. 
In the quantum regime with $ \frac{Q_\chi^2 \alpha'  m_{\chi} }{m_{A'}} \gtrsim 1 $ and $ \frac{ m_{\chi} v_{\text{rel}} }{ m_{A'} } \lesssim 1$, the cross-section can be calculated via Hulth$\acute{\text{e}}$n approximation~\cite{Tulin:2013teo}. 
In the semi-classical regime $ \frac{Q_\chi^2 \alpha'  m_{\chi} }{m_{A'}} \gtrsim 1 $ and $ \frac{ m_{\chi} v_{\text{rel}} }{ m_{A'} } \gtrsim 1 $, an analytic formula is also given in~\cite{Colquhoun:2020adl}. 
Recent study on DM scattering cross-section see also~\cite{Kamada:2023iol}.
We use viscosity cross section $\sigma_\text{V} \equiv \int d \Omega \sin^2\theta \frac{d\sigma}{d \Omega} $ as the proxy of DM elastic scattering. As suggested by~\cite{Colquhoun:2020adl},  $ \overline{ \sigma } \equiv \langle \sigma_\text{V} v^3_{\text{rel}}  \rangle / 24 \sqrt{\pi} v^3_0 $, with $v_0$ the velocity dispersion, is used to do the thermal averaging. 
Above results have been implemented in public package CLASSICS~\cite{Colquhoun:2020adl}, and we will use it to calculate $\overline{\sigma}$ in this work.

\subsection{Dark acoustic oscillations  (DAO) caused by the DM-DR scattering\label{sec:detection2}}

The relics in the dark sector contain not only stable DM $\chi$, but also stable DR $\psi$. 
The scattering between DM and DR induces pressure in the dark plasma, and thus erases matter perturbations at small scales. 
That is to say, compared with the traditional cold DM, the matter power spectrum in our dark sector model is suppressed in the high-$k$ region, which affects structure formation and leaves imprints on the CMB~\cite{Cyr-Racine:2013fsa,Cyr-Racine:2012tfp,Buckley:2014hja}. 

To handle the impact of DM-DR scattering, here we use the ETHOS (effective theory of structure formation) parameterization to simplify our analysis~\cite{Cyr-Racine:2015ihg}. 
In the ETHOS framework, we only need to know the interaction strength between DR and DM (labeled by $a_{\mathrm{dark}}$), the dependence of the comoving interaction rate of DR and DM (labeled by $\Gamma_{\mathrm{DR}-\mathrm{DM}}$) on dark sector temperature $T'$~\footnote{$\Gamma_{\mathrm{DR}-\mathrm{DM}}\propto {T'}^n$, and thus this dependence is described by the value of $n$.}, and the energy density of DR (described by temperature ratio $\xi_\infty$). Because perturbations grow slowly before the matter-dominated era, by that time $\xi$ has already reached its asymptotic value $\xi_\infty$). 
The comoving interaction rate $\Gamma_{\mathrm{DR}-\mathrm{DM}}$ is given by 
\begin{equation}
\Gamma_{\mathrm{DR}-\mathrm{DM}}=\left(\frac{3 \bar{n}_\chi m_\chi}{4 \rho_{\mathrm{DR}}}\right) \frac{a(T')}{16 \pi m_\chi^3} \frac{\eta_{\mathrm{DR}}}{3} \int \frac{p^2 \dd p}{2 \pi^2} p^2 \frac{\partial \bar{f}_{\mathrm{DR}}(p)}{\partial p}\left[A_0(p)-A_1(p)\right]\,,\label{eq:DM-DR-ScatteringRate}
\end{equation}
with $\bar{n}_\chi$, $\rho_{\text{DR}}$, $a(T')$, and $\bar{f}_{\mathrm{DR}}(p)$ the averaged DM number density, DR energy density, scale factor, and averaged DR phase-space density with temperature $T'$, respectively. $\eta_i$ is the inner d.o.f. of particle $i$. The functions $A_l(p)$ are given by 
\begin{align}
A_l(p) &= \left.\frac{1}{2} \int_{-1}^1 \dd{\tilde{\mu}} P_l(\tilde{\mu})\left(\frac{1}{\eta_\chi \eta_{\mathrm{DR}}} \sum_{\text {states }}|\mathcal{M}|^2\right)\right|_{\substack{t=2 p^2(\tilde{\mu}-1) \\ s=m_\chi^2+2 p m_\chi}}
\end{align}
with $|\mathcal{M}|^2$ the squared amplitude of the DM-DR scattering process $\chi \psi \to \chi \psi$.  
$P_l(\tilde{\mu})$ is the Legendre polynomial and $\tilde{\mu}$ is the cosine of the angle between the incident and scattered particles.

In our model, the scattering of DM and DR occurs by exchanging a t-channel $A'$, and the corresponding summed amplitude squared is 
\begin{equation}
\sum_{\text {states }}|\mathcal{M}|^2 =  32 
Q_\chi^2 Q_\psi^2 g'^4\frac{pm_\chi(pm_\chi+p^2(\tilde{\mu}-1))}{(2p^2(\tilde{\mu}-1)-m_{A'}^2)^2} \approx 32 Q_\chi^2 Q_\psi^2 g'^4\frac{p^2 m^2_\chi }{m_{A'}^4} \ , \  \text{for} \ \ p \ll m_{A'} < m_\chi \ .
\label{eq:DAO1}
\end{equation}
Thus,
\begin{equation}
A_0(p) = \frac{32 Q_\chi^2 Q_\psi^2 g'^4}{\eta_\chi \eta_{\mathrm{DR}}} \frac{p^2 m^2_\chi }{m_{A'}^4} \ , \ A_1(p)=0 \ .
\end{equation}

Integrating with respect to $p$, we get
\begin{equation}
\Gamma_{\mathrm{DR}-\mathrm{DM}}= - a(T')\frac{4\pi^3}{21}\frac{\bar{n}_\chi}{\rho_{\mathrm{DR}}}\frac{ Q_\chi^2 Q_\psi^2 g'^4 T'^6 }{ \eta_\chi m^4_{A'} } = -\frac{40\pi}{7} \frac{\rho_{\chi,0}}{\eta_\mathrm{DR} \eta_\chi} \frac{ Q_\chi^2 Q_\psi^2 g'^4}{(\xi_\infty T_{\text{CMB},0})^2} \frac{T'^4}{m_\chi m^4_{A'}}
\end{equation}
Here $\rho_{\chi,0}$ is current DM energy density and $T_{\text{CMB},0}$ is current CMB temperature. By this expression we know that $\Gamma_{\mathrm{DR}-\mathrm{DM}}\propto {T'}^4$. 

\begin{figure}[ht]
\centering
\includegraphics[width=4.0in]{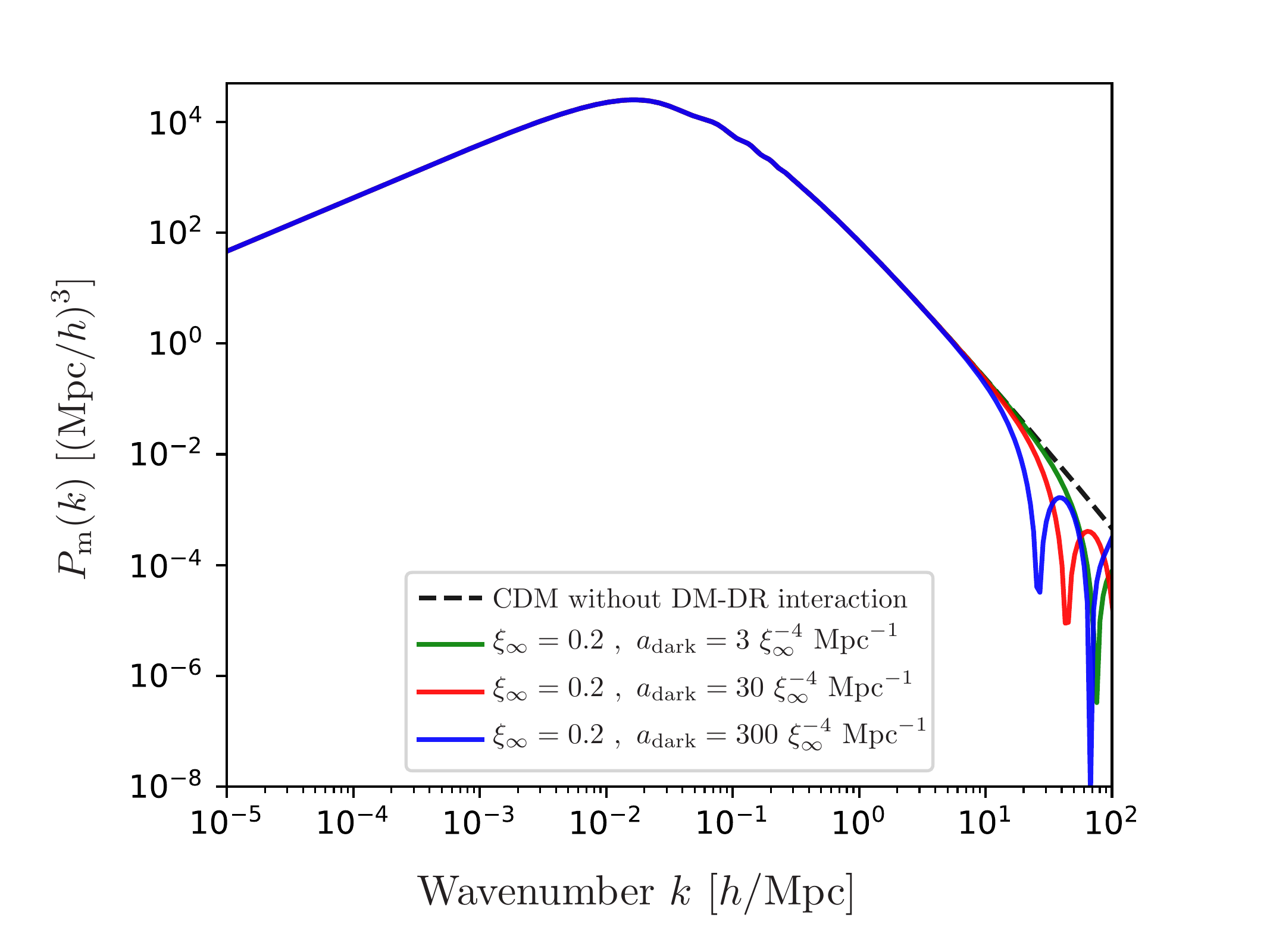}
\vspace{-20pt}
\caption{Current ($z=0$) linear matter power spectra. Black dashed line corresponds to CDM without DR. Other solid lines correspond to DM-DR interacting model with different interaction strength. }
\label{fig:LPS}
\end{figure}

On the other hand, the value of interaction strength $a_{\mathrm{dark}}$ can be inferred from $\Gamma_{\mathrm{DR}-\mathrm{DM}}$ via
\begin{equation}
\Gamma_{\mathrm{DR}-\mathrm{DM}} = - \Omega_\chi h^2 a_{\mathrm{dark}} \left( \frac{1+z}{1+z_d} \right)^4
\end{equation}
with $z_d = 10^7$ a normalization factor. 
Noting $1+z = T'/(\xi_\infty T_{\text{CMB},0})$, we can get an expression of $a_{\mathrm{dark}}$ that is convenient for our model
\begin{eqnarray}
\label{eq:DAO2}
a_{\mathrm{dark}}& =& \frac{40\pi\times 10^{28}\times Q_\chi^2 Q_\psi^2 g'^4 }{7 \eta_\mathrm{DR} \eta_\chi m_\chi  m^4_{A'} } \left( \rho_{\text{crit},0} h^{-2} \right) \left( \xi_\infty T_{\text{CMB},0} \right)^2 \\\nonumber
&=& 3.07\times 10^7 \times \xi_\infty^2\times Q_\chi^2 Q_\psi^2 g'^4 \left( \frac{\text{GeV}}{m_\chi} \right) \left( \frac{\text{MeV}}{m_{A'}} \right)^4 \text{Mpc}^{-1}
\end{eqnarray}

To illustrate the effect of DAO, we use the ETHOS framework implemented in \texttt{CLASS}~\cite{Becker:2020hzj,Blas:2011rf} to calculate the linear matter power spectrum for different values of $a_{\mathrm{dark}}$. 
Linear matter power spectra are shown in Fig.~\ref{fig:LPS}.
We also show the linear matter power spectrum of CDM for comparison. 
It clearly demonstrates that DAO suppresses the power spectrum at small scales. Additionally, stronger DM-DR interactions lead to a decline in the power spectrum at larger scales.

Here we need to emphasize that the bound on $a_{\mathrm{dark}}$ and $\xi_\infty$ cannot be easily induced from the modified linear matter power spectrum at $z=0$ as shown in Fig.~\ref{fig:LPS}. A complete and reliable analysis can only be obtained after utilizing a large set of pre-computed hydrodynamical simulations~\cite{Murgia:2018now,Archidiacono:2019wdp}. This is mainly because the Lyman-$\alpha$ forest depends on detailed knowledge of the intervening hydrogen clouds. Here we adopt the bound from previous studies~\cite{Archidiacono:2019wdp}.  
The bound on $a_{\mathrm{dark}}$ strongly relate to the  dependence of $\Gamma_{\mathrm{DR}-\mathrm{DM}}$ on $T'$. 
For  $\Gamma_{\mathrm{DR}-\mathrm{DM}}\propto {T'}^4$, the most stringent restriction on $a_{\mathrm{dark}}$ currently comes from Lyman-$\alpha$~\cite{Archidiacono:2019wdp}: 
\begin{equation}
a_{\mathrm{dark}} < 30 \ \xi_\infty^{-4} \ \text{Mpc}^{-1}
\end{equation}
Thus this constraint becomes weaker when temperature ratio $\xi_\infty$ is smaller than 1.

\subsection{Contribution  of DR to $N_\text{eff}$\label{sec:detection3}}

$\Delta N_{\mathrm{eff}}$ is the effective number of neutrino species in addition to the SM neutrinos,
and in our model it come from DR $\psi$/$\psi^\dagger$. 
The calculation of $\Delta N_{\mathrm{eff}}$ is straightforward: 
\begin{equation}
\Delta N_{\mathrm{eff}}=\frac{8}{7}\left(\frac{T_\nu}{T_\gamma}\right)^{-4} \frac{\rho_{\mathrm{DR}}}{\rho_\gamma}=\frac{8}{7}\left(\frac{4}{11}\right)^{-4 / 3}\left(\frac{2}{2}\right)\left(\frac{T^{\prime}}{T_\gamma}\right)^4 = \frac{8}{7}\left(\frac{4}{11}\right)^{-4 / 3} \xi_\infty^{4}
\end{equation}
Compared with our previous work~\cite{Chen:2023rrl}, the above equation also takes into account the dark sector heating caused by FOPT, as we discussed in Sec.~\ref{sec:Model2}.  

Current limit on $\Delta N_{\mathrm{eff}}$ from joint Planck + BAO data analysis~\cite{Planck:2018vyg} and precise SM calculation~\cite{deSalas:2016ztq} is: 
\begin{equation}
\Delta N_\text{eff} < 0.29
\end{equation}
Thus, $\Delta N_{\mathrm{eff}}$ gives a very strong bound on dark sector temperature and the strength of FOPT in the dark sector. 

\subsection{GW signals from dark sector FOPT\label{sec:detection4}}

In this subsection we briefly introduce the calculation procedure of GW signals caused by the dark sector FOPT, and the formulas used will be shown.
The FOPT happens around MeV scale, and thus the DM in our model, which is much heavier than MeV, is irrelevant in phase transition. 
Furthermore, DR $\psi$/$\psi^\dagger$ is much lighter than the MeV scale and has a very tiny coupling with dark Higgs, and so DR can be ignored in the thermal effective potential. 
The relevant Lagrangian is actually an Abelian Higgs model~\cite{Wainwright:2011qy,Chiang:2017zbz}:
\begin{equation}
\mathcal{L}_{U(1)'}  = -  \left(\partial_{\mu} S + i Q_S g' A'_{\mu} S  \right)^{\dagger} \left(\partial^{\mu} S + i Q_S g' A'^{\mu} S  \right) -\frac{1}{4} F'_{\mu\nu} F'^{\mu\nu} 
 +\mu^2 S^{\dagger} S - \frac{\lambda_S}{4} \left( S^{\dagger} S \right)^2 
  \label{lagrangian}
\end{equation}

Under Landau gauge, ghost is decoupled, and the field value dependent mass squares for dark Higgs $s$, Goldstone $a$, and dark photon $A'$, are:
\begin{eqnarray}
& & m^2_{s}(\phi) = \frac{3}{4}\lambda_S \phi^2 - \mu^2      \ , \ 
 m^2_a(\phi) =  \frac{1}{4}\lambda_S \phi^2 - \mu^2  \ , \ 
 m^2_{A'}(\phi) = (Q_Sg')^2 \phi^2 
 \label{mass2}
\end{eqnarray}

The thermal effective potential at temperature $T$ (in this subsection the temperature of dark sector is labeled by $T$ for convenience) is  schematically expressed as~\footnote{It should be emphasized that the $V(v,T)$ used in Eq.~(\ref{eq:energydensity}) also include contribution from DR.}: 
\begin{eqnarray}
V(\phi,T) = V^0(\phi) + V^{\text{1-loop}}(\phi) + V^\text{T}(\phi,T) + V^\text{daisy}(\phi,T)
\end{eqnarray}
Here $V^0$ is the tree-level potential, $V^{\text{1-loop}}$ is the sum of one-loop Coleman-Weinberg potential with on-shell counter terms,
$V^\text{T}$ is thermal correction, and $V^\text{daisy}$ is the thermal daisy resummation.
Tree-level potential is directly given by:
\begin{eqnarray}
V^0(\phi) = - \frac{1}{2}\mu^2 \phi^2 + \frac{1}{16} \lambda_S \phi^4 
\end{eqnarray}
$V^{\text{1-loop}} = V^\text{CW} + \delta_{\textbf{c.t.}}$ with $V^\text{CW}$ the one-loop Coleman-Weinberg potential~\cite{Coleman:1973jx}: 
\begin{eqnarray}
V^\text{CW}(\phi) &=& \frac{1}{64\pi^2} \Bigg\{  m_{s}^4(\phi) \left[ \ln\left( \dfrac{m_{s}^2(\phi)}{Q^2} \right)-\dfrac{3}{2}\right] 
    +  m_a^4(\phi) \left[ \ln\left( \dfrac{m_a^2(\phi)}{Q^2} \right)-\dfrac{3}{2}\right]   \\ \nonumber
  & & \ \ \ \ \ \  + 3 m_{A'}^4(\phi) \left[ \ln\left( \dfrac{m_{A'}^2(\phi)}{Q^2} \right)-\dfrac{5}{6}\right]     \Bigg\}
\end{eqnarray}
The counter terms $\delta_{\textbf{c.t.}}$ is added to make the one-loop potential satisfy following on-shell conditions: 
\begin{eqnarray}
\left. \frac{d }{ d \phi} V^{\text{1-loop}}(\phi) \right|_{\phi=v} = 0 \ , \ \left. \frac{d^2 }{ d \phi^2} V^{\text{1-loop}}(\phi) \right|_{\phi=v} = -\Delta \Sigma
  \label{onshell}
\end{eqnarray}
where $\Delta \Sigma \equiv \Sigma(m^2_{s}) - \Sigma(0) $ is the difference of scalar self-energy. 
$\Delta \Sigma$ is used to deal with the infrared (IR) divergence caused by massless Goldstone under Landau gauge~\cite{Delaunay:2007wb}. 
Final expression of $V^{\text{1-loop}}$ is~\cite{Anderson:1991zb}: 
\begin{eqnarray}
V^{\text{1-loop}}(\phi) &=& \frac{1}{64\pi^2} \left[ m^4_{s}(\phi) \left( \ln \frac{m^2_{s}(\phi)}{m^2_{s}} -\frac{3}{2} \right) + 2 m^2_{s} m^2_{s}(\phi)  \right] \\\nonumber
 &+& \frac{3}{64\pi^2} \left[ m^4_{A'}(\phi) \left( \ln \frac{m^2_{A'}(\phi)}{m^2_{A'}} -\frac{3}{2} \right) + 2 m^2_{A'} m^2_{A'}(\phi)  \right] \\\nonumber
 &+& \frac{1}{64\pi^2} \left[ m^4_a(\phi) \left( \ln \frac{m^2_a(\phi)}{m^2_{s}} -\frac{3}{2} \right)  \right]
  \label{CW}
\end{eqnarray}
with $m_{A'}$, $m_a$, and $m_s$ the zero-temperature masses.
Thermal correction is given by~\cite{Dolan:1973qd,Quiros:1999jp}:
\begin{eqnarray}
V^\text{T}(\phi,T) = \frac{T^4}{2\pi^2} \Bigg[  J_B\left( \dfrac{m_{s}^2(\phi)}{T^2} \right)  
    + J_B\left( \dfrac{m_a^2(\phi)}{T^2} \right) 
    + 3 J_B\left( \dfrac{m_{A'}^2(\phi)}{T^2} \right)  \Bigg]
\end{eqnarray}
with bosonic thermal function: 
\begin{eqnarray}
J_B(x) = \int^{\infty}_0 k^2 \ln\left[1- \exp\left( -\sqrt{k^2 + x} \right)  \right]  dk
\end{eqnarray}
Finally, daisy resummation, which is used to solve the IR problem caused by the bosonic zero-mode, is given by~\cite{Arnold:1992rz}: 
\begin{eqnarray}
V^\text{daisy}(\phi,T) &=& - \frac{T}{12\pi} \Bigg\{  \left[ \left(  {m}^2_{s}(\phi) + \Pi_{s}(T) \right)^{\frac{3}{2}} - \left(  {m}^2_{s}(\phi)  \right)^{\frac{3}{2}} \right]  \\\nonumber
 & & \ \   + \left[ \left( {m}^2_a(\phi)   + \Pi_{a}(T)  \right)^{\frac{3}{2}} - \left(  {m}^2_a(\phi) \right)^{\frac{3}{2}} \right]  + \left[ \left( {m}^2_{A'}(\phi)  + \Pi_{A'}(T)  \right)^{\frac{3}{2}} - \left( {m}^2_{A'}(\phi)  \right)^{\frac{3}{2}} \right]      \Bigg\}
\end{eqnarray}
with $\Pi_{s}(T) = \left( \lambda_S/6 + (Q_Sg')^2/4 \right)T^2 $, 
$\Pi_{a}(T) = \left( \lambda_S/6 + (Q_Sg')^2/4 \right)T^2$, and 
$\Pi_{A'}(T) = \left(  (Q_Sg')^2/3 \right)T^2$ the thermal Debye mass squares. 

\subsubsection{Phase transition parameters\label{sec:nc}}

The most important parameter for the FOPT GWs signal calculation is the characteristic temperature for abundant GWs production, $T_*$. 
In the literature, $T_*$ is generally taken to be the percolation temperature $T_{\rm perc}$.
Percolation temperature $T_{\rm perc}$ is determined by the temperature where the probability for a point in the false vacuum is 70\%~\cite{Guth:1981uk,Guth:1979bh,Megevand:2016lpr,Turner:1992tz,Kobakhidze:2017mru,Ellis:2018mja,Wang:2020jrd,Cai:2017tmh}:
\begin{eqnarray}
P(T_{\rm perc}) = e^{-I(T_{\rm perc})} = 70 \% \ , \  
I(T)=\frac{4 \pi v_w^3}{3} \int_T^{T_c} \frac{d T^{\prime} \Gamma(T^{\prime})}{H(T^{\prime}) T^{\prime 4}}\left(\int_T^{T^{\prime}} \frac{d \tilde{T}}{H(\tilde{T})}\right)^3
\end{eqnarray}
with $v_w$ the bubble wall velocity, $\Gamma(T)$ the phase transition rate per unit volume at temperature $T$, $H(T)$ the Hubble expansion rate, and $T_c$ the critical temperature where two phases have the same free energy density. 
The determination of $v_w$ will be discussed later on. 
The thermal transition rate $\Gamma(T)$ is given by~\cite{Coleman:1977py,Callan:1977pt,Linde:1981zj}: 
\begin{eqnarray}
\Gamma(T) = T^4 \left( \frac{S_3(T)}{2\pi T} \right)^{3/2} e^{-S_3(T)/T}
\end{eqnarray}
with  3-D Euclidean action $S_3(T)$ given by: 
\begin{eqnarray}
S_3(T) = \int d^3 x \left[ \frac{1}{2} (\nabla \phi)^2 + V(\phi,T) \right] = 4\pi \int_0^{\infty} r^2 dr \left[ \frac{1}{2} \left( \frac{d\phi}{dr} \right)^2 + V(\phi,T) \right] 
\end{eqnarray}
$\phi(r)$ is determined by equation of motion:
\begin{eqnarray}
\frac{d^2\phi}{d r^2} + \frac{2}{r} \frac{d\phi}{d r} = \frac{\partial V(\phi,T)}{\partial \phi}
\label{Eom}
\end{eqnarray}
and boundary conditions $\lim\limits_{r \to \infty} \phi(r) = 0$ and $\left. \frac{d\phi}{dr} \right|_{r=0} = 0$. 
The equation of motion for $\phi(r)$ can be solved numerically by shooting method~\cite{Apreda:2001us,Wainwright:2011kj,Athron:2024xrh}. 

After determining the phase transition temperature $T_*$, which is actually percolation temperature $T_{\rm perc}$, other phase transition parameters that need to be calculated are the FOPT strength parameter $\mathbb{A}$, the duration parameter $\beta/H_\ast$,bubble wall velocity $v_w$, and the efficiency parameters $\kappa_{b,s,t}$. 

$\mathbb{A}$ is the difference of the trace of energy-momentum tensor between two phases divided by radiation energy density:
\begin{eqnarray}
\mathbb{A} = \frac{1}{\rho_\ast} \left[ \Delta V - \frac{T}{4}\frac{d \Delta V}{d T} \right]_{T=T_\ast}
\end{eqnarray}
with $\rho_\ast$ is the radiation energy density at $T_\ast$.  $\beta/H_\ast$ is given by: 
\begin{eqnarray}
\frac{\beta}{H_{\ast}} = T \left. \frac{d }{d T} \left( \frac{S_3(T)}{T}  \right) \right|_{T=T_\ast}
\end{eqnarray}

To determine $v_w$ and $\kappa_{b,s,t}$, first we need to confirm whether the FOPT is ``runaway'' or not. 
It is convenient to define a dark sector strength parameter $\mathbb{A}'$ which is given by $\mathbb{A}' = \frac{\rho_\ast}{\rho'_\ast} \mathbb{A}$ where $\rho'_\ast$ is the radiation energy density in the dark sector. 

If $\mathbb{A}'$ is larger than the so-called threshold value $\mathbb{A}'_{\infty} = \frac{1}{\rho'_\ast } \frac{T^2_\ast}{24} \left( \sum_i c_i n_i \Delta m_i^2  \right)$~\cite{Espinosa:2010hh} ($c_i = 1 \ (1/2)$ for bosons (fermions), $n_i$ is the number of d.o.f., and $\Delta m_i^2$ the mass square difference between two phases), then FOPT is ``runaway'' and bubble wall velocity $v_w$ can be approximated by 1~\footnote{More discussion on the bubble wall velocity in ``runaway'' case see e.g.~\cite{Bodeker:2017cim}.}. 
$\kappa_{b,s,t}$ are given by~\cite{Hindmarsh:2015qta,Schmitz:2020syl}: 
\begin{eqnarray}
\kappa_b = 1 - \frac{\mathbb{A}'_{\infty}}{\mathbb{A}'} \ , \ \kappa_s = \frac{\mathbb{A}'_{\infty}}{\mathbb{A}'} \frac{\mathbb{A}'_{\infty}}{0.73+0.083\sqrt{\mathbb{A}'_{\infty}} + \mathbb{A}'_{\infty} }  \ , \ \kappa_t = 0.1\kappa_s
\end{eqnarray}

If $\mathbb{A}' < \mathbb{A}'_{\infty}$, then the FOPT is "non-runaway" and the bubble wall will eventually reach a subluminal velocity. 
Due to the difficulty in the calculation of $v_w$ in this case~\cite{Moore:1995si,Megevand:2009gh,Huber:2013kj,Konstandin:2014zta,Dorsch:2018pat,Laurent:2022jrs,Wang:2020zlf}, in this work we simply take $v_w=0.9$ for a quick and optimistic GW signal estimate. 
In this case, $\kappa_{b,s,t}$ are approximately given by~\cite{Espinosa:2010hh}: 
\begin{eqnarray}
\kappa_b \simeq 0 \ , \ \kappa_s =  \frac{\mathbb{A}'}{0.73+0.083\sqrt{\mathbb{A}'} + \mathbb{A}' }  \ , \ \kappa_t = 0.1\kappa_s
\end{eqnarray}

\subsubsection{ Gravitational waves spectrum\label{sec:gw}}

There are three main GW generation mechanisms during the FOPT: bubble walls collision~\cite{Kamionkowski:1993fg,Kosowsky:1991ua,Kosowsky:1992vn,Kosowsky:1992rz,Caprini:2007xq,Huber:2008hg}, sound waves~\cite{Hindmarsh:2013xza,Giblin:2014qia,Giblin:2013kea,Hindmarsh:2015qta}, and magnetohydrodynamic turbulence~\cite{Caprini:2006jb,Kahniashvili:2008pe,Kahniashvili:2008pf,Kahniashvili:2009mf,Caprini:2009yp,Kisslinger:2015hua}.
The spectrum of GWs from these three processes can be found in the literature~\cite{Caprini:2015zlo,Caprini:2019egz,Huber:2008hg,Hindmarsh:2015qta,Caprini:2009yp}. 
The total GWs signal is the linear superposition of these three: 
\begin{eqnarray}
h^2\Omega_{\text{GW}} (f)  = h^2\Omega_{{b}} (f)  + h^2\Omega_{{s}} (f) + h^2 \Omega_{{t}} (f).
\end{eqnarray}
Each term on the right hand side of above equation can be decomposed into peak amplitudes ($\Omega^{\text{peak}}_{{b,s,t}}$) and  spectral shape functions ($\mathcal{S}_{{b,s,t}}$): 
\begin{eqnarray}
& & h^2\Omega_{{b}} (f) \simeq h^2 \Omega^{\text{peak}}_{{b}} \mathcal{S}_{{b}}(f,f^{\text{peak}}_{{b}}), \\\nonumber
& & h^2\Omega_{{s}} (f) \simeq h^2 \Omega^{\text{peak}}_{{s}} \mathcal{S}_{{s}}(f,f^{\text{peak}}_{{s}}),  \\\nonumber
& & h^2\Omega_{{t}} (f) \simeq h^2 \Omega^{\text{peak}}_{{t}} \mathcal{S}_{{t}}(f,f^{\text{peak}}_{{t}}),
\end{eqnarray}
where the peak amplitudes are determined by 
\begin{equation}
  h^2 \Omega^{\text{peak}}_{{b}} = 1.24\times 10^{-5} \left( \frac{  {h_{\text{eff}}}_0   }{ {h_{\text{eff}}}_\ast  } \right)^{4/3} ({g_{\text{eff}}}_\ast) (\xi_\infty)^4  \left( \frac{0.11 v_w}{0.42+v_w^2} \right) \left( \frac{\kappa_b \mathbb{A}}{ 1+\mathbb{A} } \right)^2 \left( \frac{v_w}{\beta/H_\ast} \right)^2 \ , 
  \end{equation}

  \begin{equation}
  h^2 \Omega^{\text{peak}}_{{s}} = 1.97\times 10^{-6} \left( \frac{  {h_{\text{eff}}}_0   }{ {h_{\text{eff}}}_\ast  } \right)^{4/3} ({g_{\text{eff}}}_\ast) (\xi_\infty)^4    \left( \frac{\kappa_s \mathbb{A}}{ 1+\mathbb{A} } \right)^2 \left( \frac{v_w}{\beta/H_\ast} \right) \Upsilon  \ , 
   \end{equation}
   
   \begin{equation}
  h^2 \Omega^{\text{peak}}_{{t}} = 2.49\times 10^{-4} \left( \frac{  {h_{\text{eff}}}_0   }{ {h_{\text{eff}}}_\ast  } \right)^{4/3} ({g_{\text{eff}}}_\ast) (\xi_\infty)^4    \left( \frac{\kappa_t \mathbb{A}}{ 1+\mathbb{A} } \right)^{3/2} \left( \frac{v_w}{\beta/H_\ast} \right) \ .
   \end{equation}
Here $h_{\text{eff}}$ and $g_{\text{eff}}$ indicate the total effective d.o.f. for entropy and energy respectively~\footnote{It should be noticed that the definition of $h_{\text{eff}}$ and $g_{\text{eff}}$ is the same as our previous work~\cite{Chen:2023rrl}, i.e. we let the total radiation energy density and entropy density to be  proportional to the fourth and third powers of the dark sector temperature, respectively. This definition makes the expression of $h^2 \Omega^\text{peak} $ looks different with general formula in the literature, e.g. \cite{Breitbach:2018ddu}. But transfer to the normal $T_\gamma$ definition and count the $(\xi_\infty)^4$ factor, our formula will go back to the usual form.    }, and ``$0$'' corresponds to current time. Suppression factor $\Upsilon$ in $h^2 \Omega^{\text{peak}}_{{s}} $ is~\cite{Guo:2020grp}
  \begin{equation} \nonumber 
\Upsilon = 1- \frac{1}{\sqrt{1 + \tau_{sw} H_\ast  }}   \approx  1- \frac{1}{\sqrt{1 + R_\ast H_\ast/ \bar{U}_f  }} 
   \end{equation}
with mean bubble separation $R_\ast$ and  root mean square fluid velocity $\bar{U}_f$ given by 
  \begin{equation} \nonumber 
R_\ast = (8\pi)^{1/3} \frac{v_w}{\beta}  \ , \ \bar{U}_f = \sqrt{\frac{3}{4}  \frac{\kappa_s \mathbb{A}}{ 1+\mathbb{A} }  }
   \end{equation}

The spectral shape functions are given by: 
\begin{eqnarray}
& &  \mathcal{S}_b(f, f^{\text{peak}}_b )  =  \left( \frac{f}{f^{\text{peak}}_{{b}}} \right)^{2.8} \left[ \frac{3.8}{1 + 2.8 (f/f^{\text{peak}}_{{b}})^{3.8}} \right],  \\\nonumber
& &  \mathcal{S}_s(f, f^{\text{peak}}_s )  =  \left( \frac{f}{f^{\text{peak}}_{{s}}} \right)^{3} \left[ \frac{7}{4 + 3 (f/f^{\text{peak}}_{{s}})^{2}} \right]^{7/2},   \\\nonumber
& &  \mathcal{S}_t(f, f^{\text{peak}}_t )  = \left( \frac{f}{f^{\text{peak}}_{{t}}} \right)^{3} \left[ \frac{1}{1 +  (f/f^{\text{peak}}_{{t}})} \right]^{11/3} \frac{1}{1 + 8\pi f/h_\ast },
\end{eqnarray}
with
\begin{eqnarray}
h_\ast = \frac{a_\ast}{a_0} H_\ast = 7.44\times 10^{-11} \text{Hz} \left( \frac{   {g_{\text{eff}}}^{1/2}_\ast  }{ {h_{\text{eff}}}^{1/3}_\ast  } \right)  \left( \frac{  {h_{\text{eff}}}_0  }{ 3.91  } \right)^{1/3} \xi_\infty \left( \frac{ T_\ast }{1 \text{ MeV} } \right).
\end{eqnarray}
The peak frequencies in shape functions are given by:
\begin{equation}
 f^{\text{peak}}_b = 7.44\times 10^{-11} \text{Hz} \left( \frac{   {g_{\text{eff}}}^{1/2}_\ast  }{ {h_{\text{eff}}}^{1/3}_\ast  } \right)  \left( \frac{  {h_{\text{eff}}}_0  }{ 3.91  } \right)^{1/3} \xi_\infty \left( \frac{ T_\ast }{1 \text{ MeV} } \right) \left( \frac{ \beta/H_\ast }{v_w} \right) \left( \frac{0.62 v_w}{ 1.8 - 0.1 v_w + v_w^2 } \right)  ,
    \end{equation}
\begin{eqnarray}
f^{\text{peak}}_s &=& 8.59\times 10^{-11} \text{Hz} \left( \frac{   {g_{\text{eff}}}^{1/2}_\ast  }{ {h_{\text{eff}}}^{1/3}_\ast  } \right)  \left( \frac{  {h_{\text{eff}}}_0  }{ 3.91  } \right)^{1/3} \xi_\infty \left( \frac{ T_\ast }{1 \text{ MeV} } \right) \left( \frac{ \beta/H_\ast }{v_w} \right)   ,
\end{eqnarray}
\begin{eqnarray}
f^{\text{peak}}_t &=& 13.0\times 10^{-11} \text{Hz} \left( \frac{   {g_{\text{eff}}}^{1/2}_\ast  }{ {h_{\text{eff}}}^{1/3}_\ast  } \right)  \left( \frac{  {h_{\text{eff}}}_0  }{ 3.91  } \right)^{1/3} \xi_\infty \left( \frac{ T_\ast }{1 \text{ MeV} } \right) \left( \frac{ \beta/H_\ast }{v_w} \right) .
\end{eqnarray}

\section{Parameter space scan results\label{sec:results}}

In this section we present a parameter space scan. 
There are 5 input parameters in this work, i.e. \{ $m_\chi,\ m_{A'},\ \alpha',\ m_s,\ \xi_\text{reh}$ \}. 
Their ranges have been given in Sec.~\ref{sec:Model1}. 
In addition, we consider two charge assignments: 
\begin{eqnarray}
\nonumber \text{Charge assignment A: } \ \ \{Q_{\chi}, Q_{S}, Q_{\psi}\} &=& \{+1,+2/3, +1  \} \\
\nonumber \text{Charge assignment B: } \ \ \{Q_{\chi}, Q_{S}, Q_{\psi}\} &=& \{+1,+3, +1/2  \} 
\end{eqnarray}
 For $\xi_\text{reh}$ we use uniform linear sampling, while for other parameters we use uniform logarithmic sampling. 
To fully cover the parameter space, we choose the numbers of sampling points for parameters $m_\chi$, $m_{A'}$, $\alpha'$, $m_s$, and $\xi_\text{reh}$ to be  
100, 60, 100, 100, and 10 respectively. Thus the total number of parameter points is 600 million.

\subsection{Scan results\label{sec:results1}}

\begin{figure}[ht]
\centering
\includegraphics[width=6.5in]{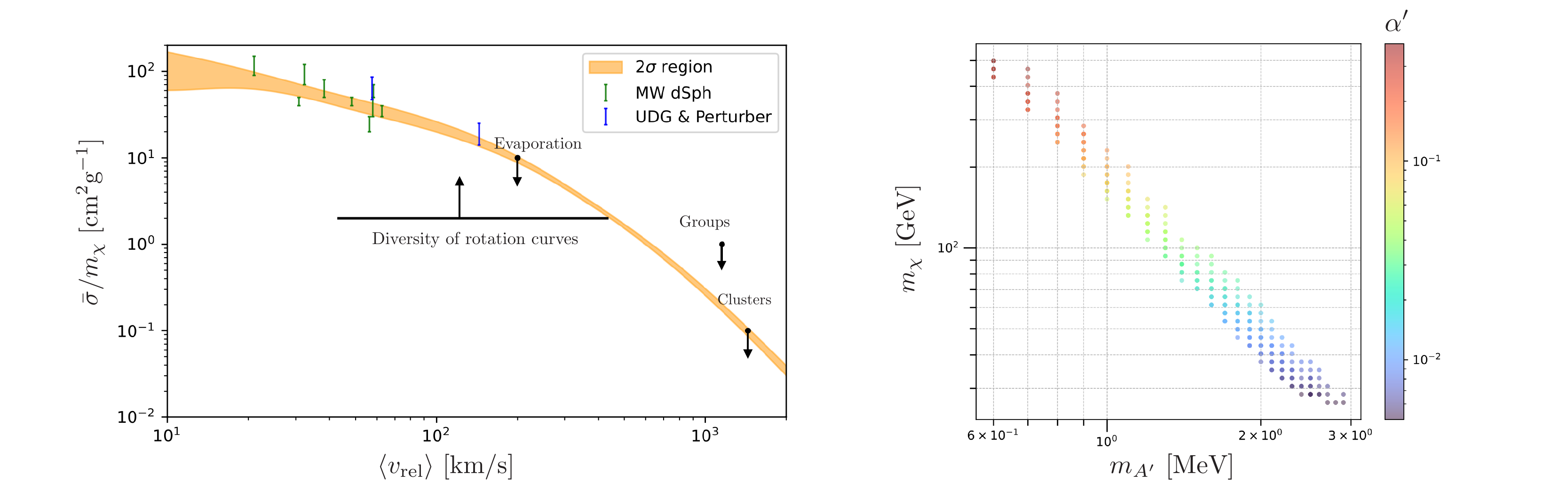}
\vspace{-30pt}
\caption{Left: $\overline{\sigma}/m_\chi$ as function of average DM relative velocity $\langle v_\text{rel} \rangle$. Small-scale data and limits are also shown. Orange region corresponds to 2$\sigma$ region. 
Right: the parameter space composed by \{ $m_\chi,\ m_{A'},\ \alpha'$ \} which is within the 2$\sigma$ region. }
\label{fig:fit}
\end{figure}

The first restriction we impose on the parameter space is that the  velocity-dependent DM scattering cross section $\overline{\sigma}$ given by a parameter point is within the 2$\sigma$ confidence interval obtained from small-scale data, as we explained in Sec.~\ref{sec:detection1}.
In Fig.~\ref{fig:fit} we present the scan results, which show that the small-scale data and limits compress the parameter sub-space composed by \{ $m_\chi,\ m_{A'},\ \alpha'$ \} into a quite small region. 
Since $Q_\chi$ is the same in charge assignments A and B, this scan result is applicable to both charge assignments. 

\begin{figure}[htb]
\centering
\hspace*{-0.8cm}
\includegraphics[width=6.5in]{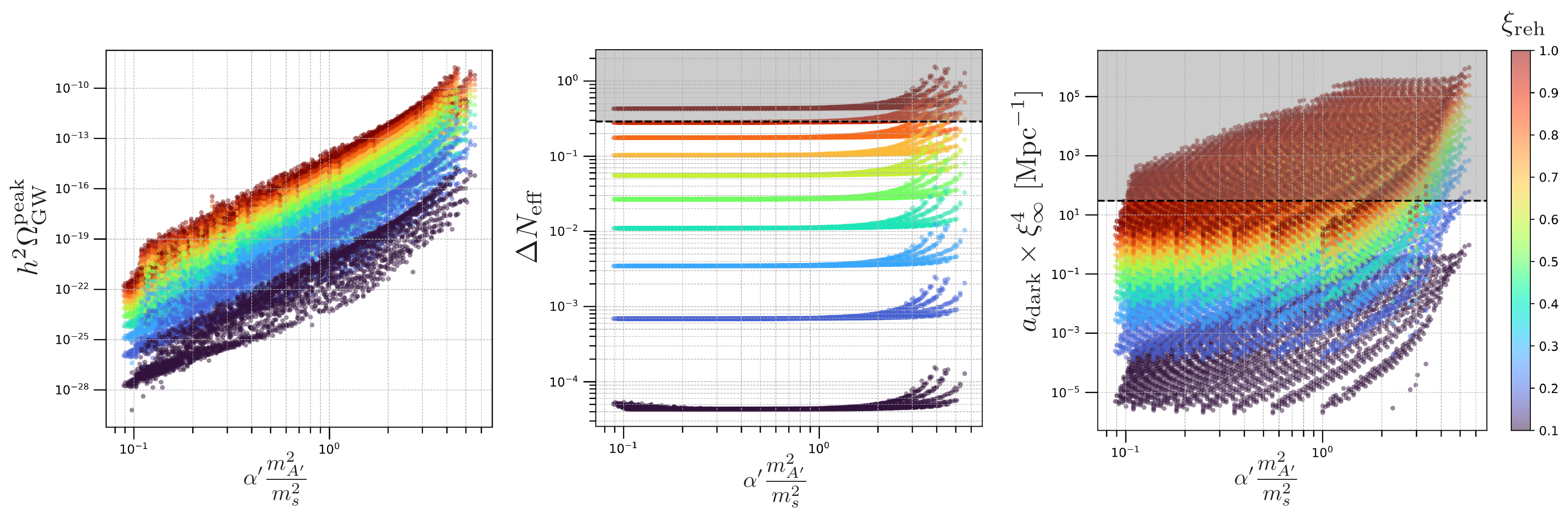}
\vspace{-30pt}
\caption{In this figure we consider charge assignment A. All the samples are consistent with the small-scale data. 
Left: peak amplitudes of GWs as functions of parameter combination $\alpha'\frac{m^2_{A'}}{m^2_s}$, color-mapped by initial temperature ratio $\xi_\text{reh}$.
Middle: $\Delta N_{\mathrm{eff}}$ as functions of parameter combination $\alpha'\frac{m^2_{A'}}{m^2_s}$, color-mapped by initial temperature ratio $\xi_\text{reh}$. Gray region has been excluded by current data. 
Right: $a_\text{dark}\times \xi^4_{\infty}$ as functions of parameter combination $\alpha'\frac{m^2_{A'}}{m^2_s}$, color-mapped by initial temperature ratio $\xi_\text{reh}$. Gray region has been excluded by current data.}
\label{fig:scanA}
\end{figure}

\begin{figure}[htb]
\centering
\hspace*{-0.8cm}
\includegraphics[width=6.5in]{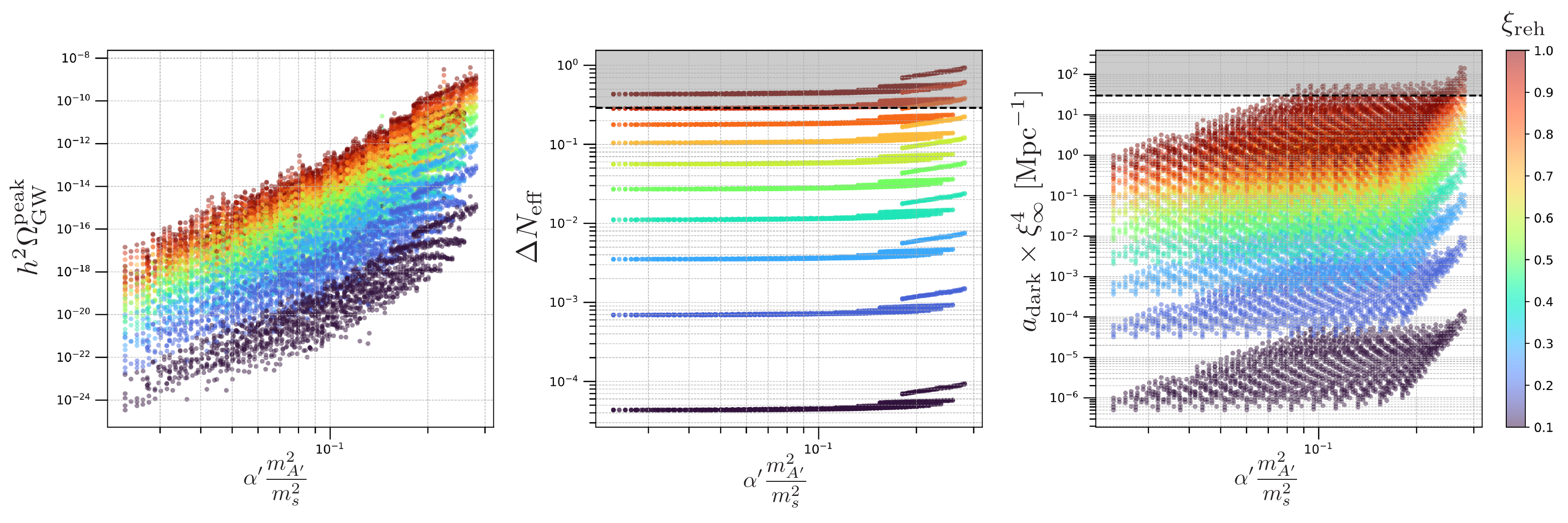}
\vspace{-30pt}
\caption{Same as Fig.~\ref{fig:scanA}, but for charge assignment B.}
\label{fig:scanB}
\end{figure}

The parameter points that are consistent with small-scale data are used for further analysis. 
In Fig.~\ref{fig:scanA} and Fig.~\ref{fig:scanB} we present the peak amplitudes of GWs (labeled as $h^2\Omega^\text{peak}_\text{GW}$),    $\Delta N_{\mathrm{eff}}$, and DR-DM interaction strength parameter $a_\text{dark}$ calculated from those parameter points. 
We find that $h^2\Omega^\text{peak}_\text{GW}$ is mainly determined by parameter combination $\alpha'\frac{m^2_{A'}}{m^2_s}$ and initial temperature ratio $\xi_\text{reh}$.
Generally speaking, $\alpha'\frac{m^2_{A'}}{m^2_s}$ controls the strength of FOPT and $\xi_\text{reh}$ roughly determines the energy density in the dark sector. 
Without considering other limits, as presented in Fig.~\ref{fig:scanA} and~\ref{fig:scanB}  (Left) $h^2\Omega^\text{peak}_\text{GW}$ can be even larger than $10^{-10}$ for both charge assignments, which is promising to be detected by future GW detectors like SKA telescope~\cite{Janssen:2014dka} or space-based LISA interferometer~\cite{LISA:2017pwj}, provided the peak frequency is not far from the sensitive regions of SKA or LISA. 
However, as presented in Fig.~\ref{fig:scanA} and~\ref{fig:scanB} (Middle) $\Delta N_{\mathrm{eff}}$ uniformly excludes dark-sector models which are too hot after dark FOPT. 
Lyman-$\alpha$ also puts a bound on the parameter space, especially for charge assignment A, as shown in Fig.~\ref{fig:scanA} (Right). 
The Lyman-$\alpha$ bound on charge assignment B is much weaker, as shown in Fig.~\ref{fig:scanB} (Right). 
This result is mainly because charge assignment B can obtain strong GWs with small $\alpha'$ (as shown in Fig.~\ref{fig:scanB} (Left)). 
In addition, as shown in Eq.~\ref{eq:DAO1} and~\ref{eq:DAO2}, the DM-DR interaction strength $a_\text{dark}$ is proportional to $Q^2_\chi Q^2_\psi$, and thus assignment B with a much smaller $Q_\psi$ can more easily evade the Lyman-$\alpha$ constraint. 

\begin{table}[t]
\centering
\caption{Benchmark points used for the CLASS-ETHOS validation of the linear matter power spectrum. All benchmark points use charge assignment B.}
\label{tab:ethos_benchmark_points}
\begin{tabular}{c|cccccc}
\hline
 
& $m_\chi~[{\rm GeV}]$ & $m_s~[{\rm MeV}]$ & $m_{A'}~[{\rm MeV}]$ & $\alpha'$ & $\xi_{\rm ini}$ & $a_{\rm dark}\xi_\infty^4~[{\rm Mpc}^{-1}]$ \\
\hline
Benchmark point 1 & $37.6$ & $0.381$ & $2.1$ & $0.00779$ & $1.0$ & $560.0$ \\
Benchmark point 2 & $70.5$ & $0.472$ & $1.7$ & $0.0217$ & $1.0$ & $95.3$ \\
Benchmark point 3 & $70.5$ & $0.629$ & $1.7$ & $0.0207$ & $1.0$ & $29.9$ \\
Benchmark point 4 & $37.6$ & $0.433$ & $2.0$ & $0.00714$ & $0.8$ & $1.0$ \\
\hline
\end{tabular}
\end{table}

\begin{figure}[ht]
\centering
\hspace*{-0.8cm}
\includegraphics[width=4.in]{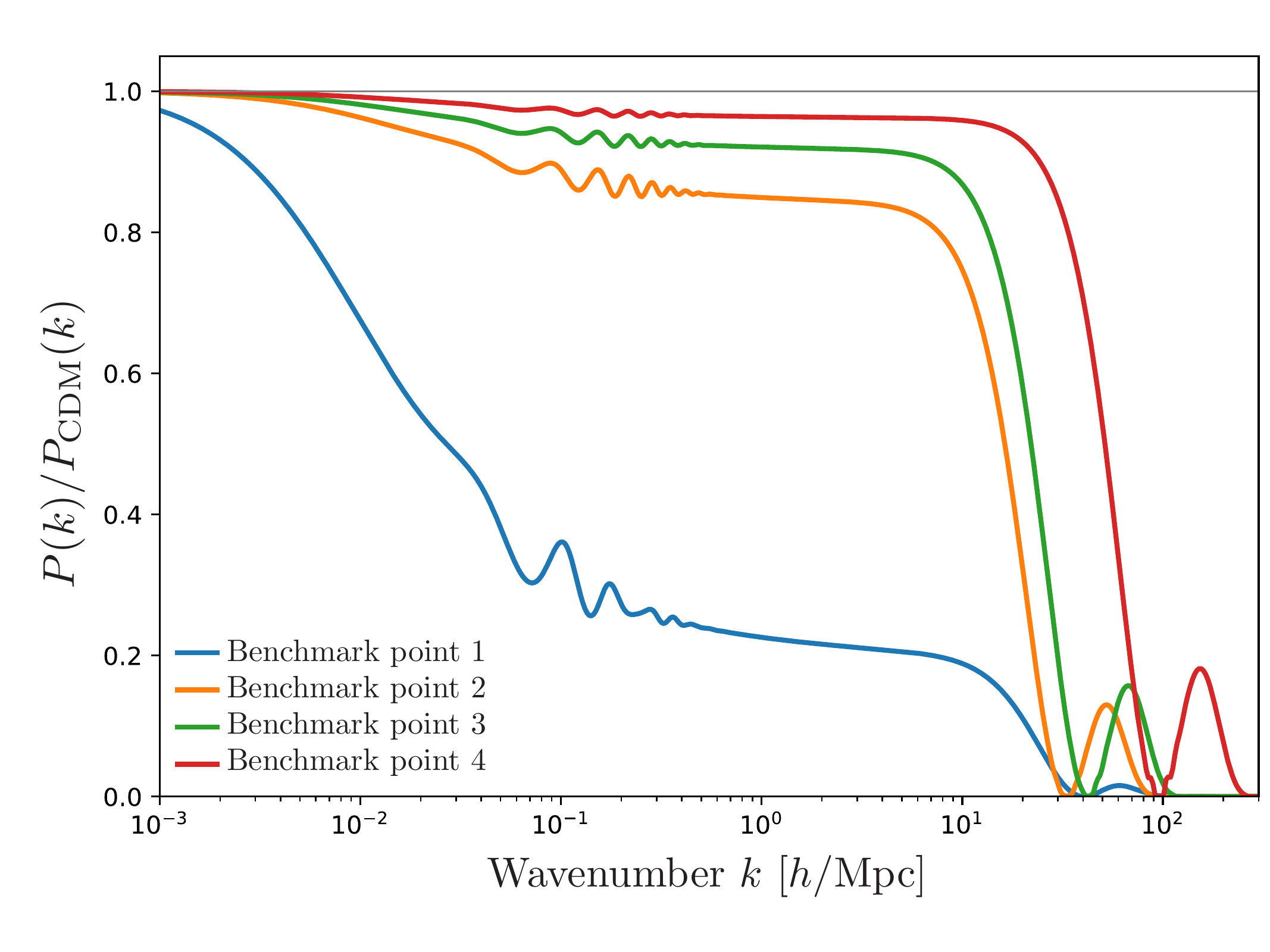}
\vspace{-10pt}
\caption{Validation of the ETHOS mapping for four representative benchmark points in charge assignment B. We show the ratio of the linear matter power spectrum to the corresponding CDM spectrum. The four curves correspond to the benchmark points listed in Tab.~\ref{tab:ethos_benchmark_points}. The spectra exhibit the expected monotonic weakening of the small-scale suppression as $a_{\rm dark}\xi_\infty^4$ decreases.}
\label{fig:ethos_power_ratio}
\end{figure}

To validate the use of the single-parameter Lyman-$\alpha$ criterion in our model, we perform an explicit CLASS-ETHOS calculation for several representative points in charge assignment B. 
The benchmark points are listed in Tab.~\ref{tab:ethos_benchmark_points}. 
They are chosen from the same scan sample used in the main analysis and cover four qualitatively different regimes of the quantity $a_{\rm dark}\xi_\infty^4$: far above the adopted Lyman-$\alpha$ bound, moderately above the bound, close to the boundary, and safely below the bound.  
The resulting ratios $P(k)/P_{{\rm CDM}}(k)$ are shown in Fig.~\ref{fig:ethos_power_ratio}. The behavior is as expected. The point with the largest value of $a_{\rm dark}\xi_\infty^4$ gives the strongest suppression of the matter power spectrum, together with visible dark acoustic oscillation features. As $a_{\rm dark}\xi_\infty^4$ is decreased from benchmark point 1 to benchmark point 4, the suppression scale moves to larger $k$ and the spectrum becomes progressively closer to the standard CDM result. 
This explicit CLASS-ETHOS check supports the application of the adopted Lyman-$\alpha$ bound to the present dark-sector model and confirms that the exclusion is controlled, to a good approximation, by the combination $a_{\rm dark}\xi_\infty^4$.

\begin{figure}[ht]
\centering
\hspace*{-0.8cm}
\includegraphics[width=6.5in]{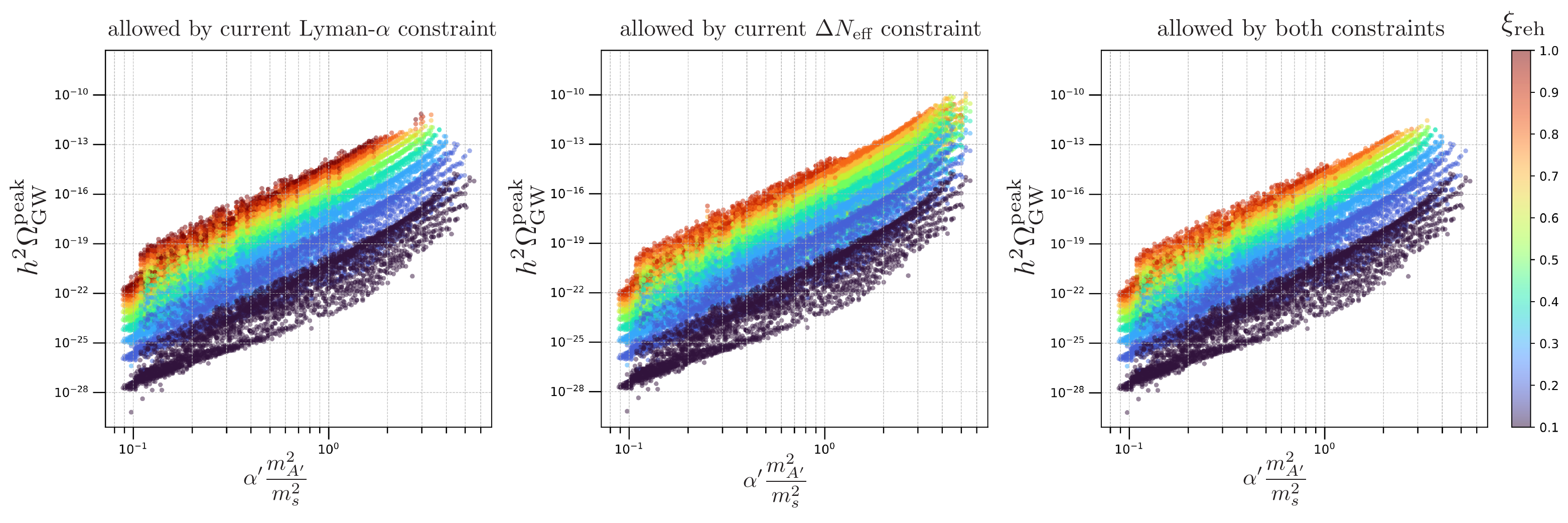}
\vspace{-30pt}
\caption{In this figure we consider charge assignment A. All the samples are consistent with the small-scale data. 
Left: parameter points allowed by current Lyman-$\alpha$ constraint.
Middle: parameter points allowed by current $\Delta N_\text{eff}$ constraint.
Right: parameter points allowed by both constraints.}
\label{fig:limitA}
\end{figure}

\begin{figure}[ht]
\centering
\hspace*{-0.8cm}
\includegraphics[width=6.5in]{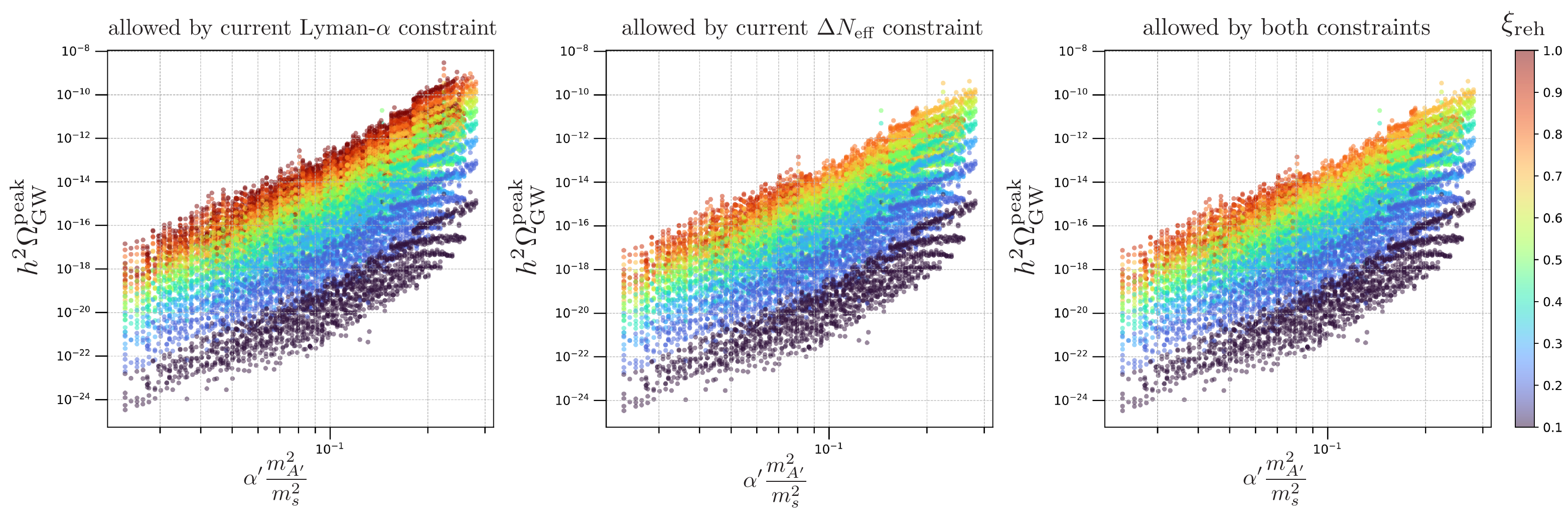}
\vspace{-30pt}
\caption{Same as Fig.~\ref{fig:limitA}, but for charge assignment B.}
\label{fig:limitB}
\end{figure}

To further illuminate the impact from $\Delta N_{\mathrm{eff}}$ and Lyman-$\alpha$, we present Fig.~\ref{fig:limitA} and Fig.~\ref{fig:limitB} which clearly show how $h^2\Omega^\text{peak}_\text{GW}$ is constrained by bounds. 
The maximum of $h^2\Omega^\text{peak}_\text{GW}$ is suppressed by about three orders of magnitude for charge assignment A and by less than two orders of magnitude for charge assignment B. 

\begin{figure}[ht]
\centering
\hspace*{-0.5cm}
\includegraphics[width=6.2in]{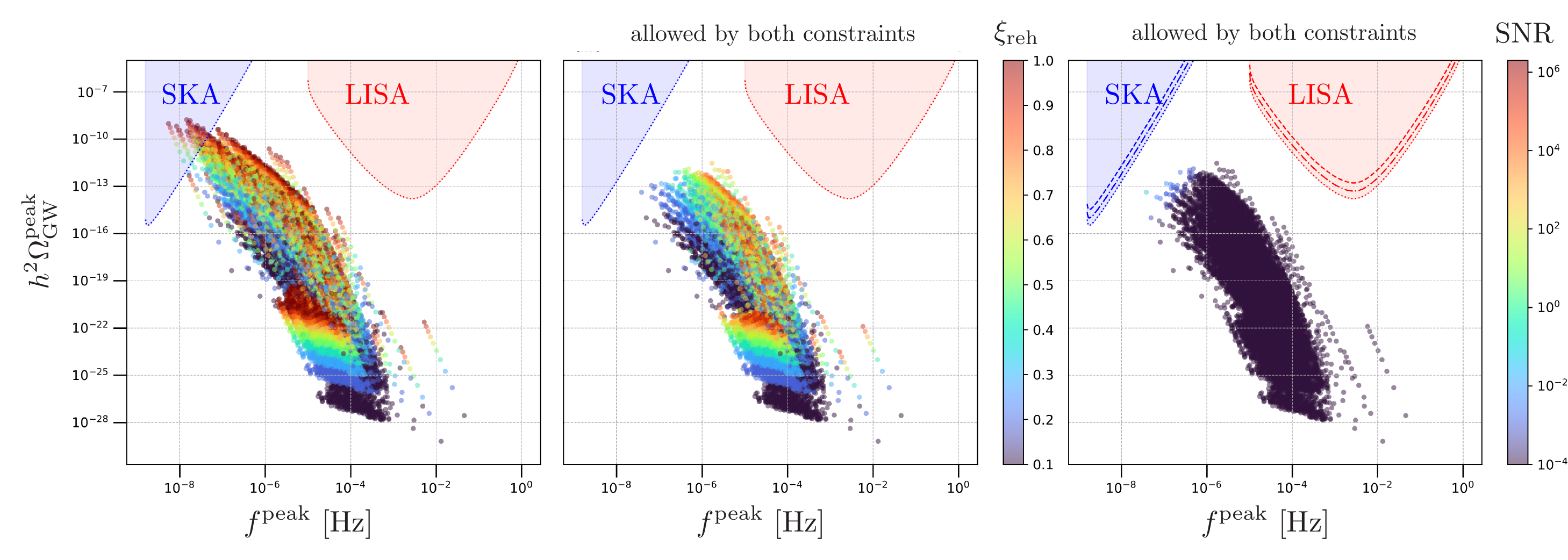}
\vspace{-30pt}
\caption{In this figure we consider charge assignment A. All the samples are consistent with the small-scale data.  The blue dotted line, blue dotdash line, and blue dashed line are PLISCs of SKA with ``observing time = 20 years, SNR = 1'', ``observing time = 20 years, SNR = 3'', and ``observing time = 20 years, SNR = 10'', respectively. 
The red dotted line, red dotdash line, and red dashed line are PLISCs of LISA with ``observing time = 1 year, SNR = 1'', ``observing time = 1 year, SNR = 3'', and ``observing time = 1 year, SNR = 10'', respectively. 
Left: the current $\Delta N_\text{eff}$ and Lyman-$\alpha$ constraints are not considered. Color-mapped by $\xi_\text{reh}$.
Middle: both $\Delta N_\text{eff}$ and Lyman-$\alpha$ constraints are considered. Color-mapped by $\xi_\text{reh}$.
Right: both $\Delta N_\text{eff}$ and Lyman-$\alpha$ constraints are considered. Color-mapped by SNR. }
\label{fig:gwsA}
\end{figure}

\begin{figure}[ht]
\centering
\hspace*{-0.5cm}
\includegraphics[width=6.2in]{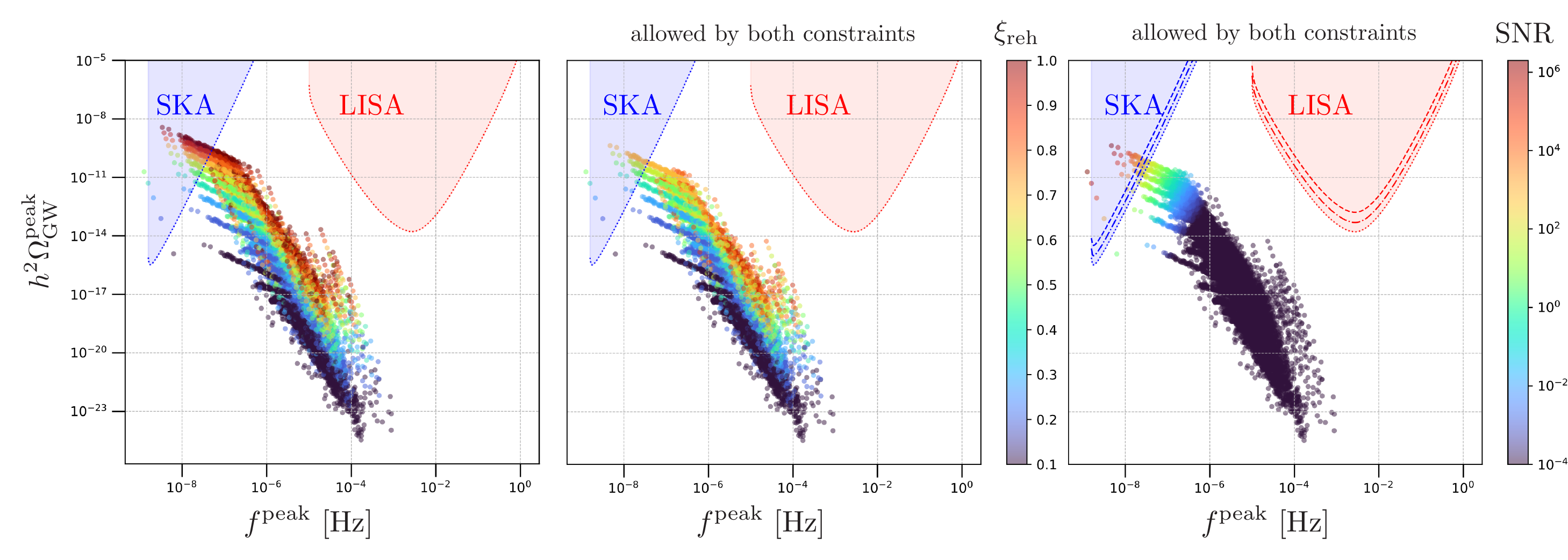}
\vspace{-30pt}
\caption{Same as Fig.~\ref{fig:gwsA}, but for charge assignment B.}
\label{fig:gwsB}
\end{figure}

In Fig.~\ref{fig:gwsA} and Fig.~\ref{fig:gwsB} we further project parameter points on the standard ``frequency-amplitude'' plane, with the ``power-law-integrated sensitivity curves'' (PLISCs) of SKA and LISA also presented~\cite{Schmitz:2020syl,Thrane:2013oya}. 
The PLISC of SKA is normalized to ``observing time = 20 years'' and ``SNR (signal-to-noise ratio) = 1'', while the PLISC of LISA is normalized to ``observing time = 1 year'' and ``SNR = 1''. 
If the GW spectrum from a parameter point intersects a PLISC, then this parameter point can be treated as detectable at the corresponding experiment. 
To avoid clutter in the figure, we only depict the peaks of the GW spectra in Fig.~\ref{fig:gwsA} and Fig.~\ref{fig:gwsB} instead of showing the complete spectrum.

In Fig.~\ref{fig:gwsA} (Left) we see that without considering current $\Delta N_\text{eff}$ and Lyman-$\alpha$ constraints, a considerable number of parameter points in charge assignment A can be detected by SKA. 
However, Fig.~\ref{fig:gwsA} (Middle) shows that, taking both constraints into account, all the peaks given by charge assignment A are below the PLISC of SKA.  
We have checked numerically that not only the peaks of these parameter points are not inside the PLISC, but their entire spectra also do not intersect with the PLISC. 
To make this point clearer, in Fig.~\ref{fig:gwsA} (Right) we present the same plot as Fig.~\ref{fig:gwsA} (Middle) but color-mapped by SNR. 
It shows that all the parameter points have a SNR smaller than 1. 
Therefore, for charge assignment A, we conclude that, considering current $\Delta N_\text{eff}$ and Lyman-$\alpha$ constraints, we cannot detect the secluded self-interacting dark sector discussed in this paper. 

For charge assignment B, however, due to a much weaker Lyman-$\alpha$ bound, there are a small number of parameter points that can be detected by SKA after considering both constraints, as shown in Fig.~\ref{fig:gwsB} (Middle) and Fig.~\ref{fig:gwsB} (Right). 
Thus we can say that, even though $\Delta N_\text{eff}$ puts a strong and uniform limit on all dark FOPTs, the secluded self-interacting dark sector scenarios that can escape the Lyman-$\alpha$ bound would still be promising targets for SKA.

\subsection{Bubble wall velocity dependence\label{sec:results2}}

\begin{figure}[htb]
\centering
\hspace*{-0.8cm}
\includegraphics[width=6.5in]{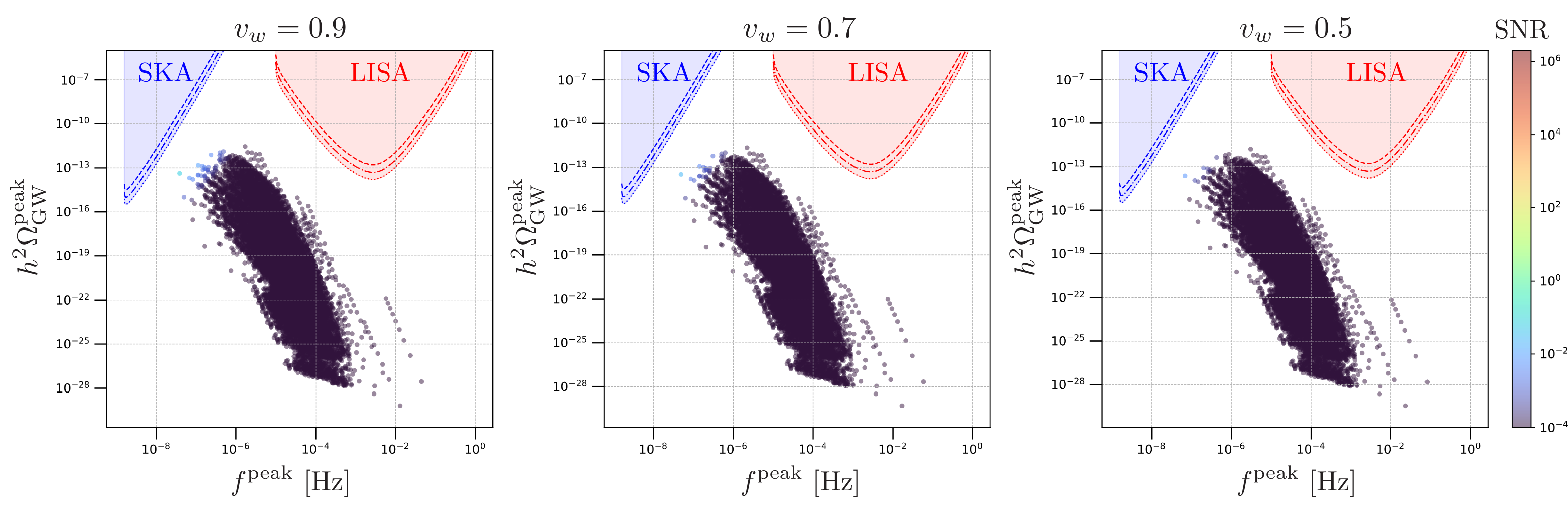}
\vspace{-30pt}
\caption{In this figure we consider charge assignment A. All the samples are consistent with the small-scale data. Both $\Delta N_\text{eff}$ and Lyman-$\alpha$ constraints are also considered, color-mapped by SNR. 
The blue dotted line, blue dotdash line, and blue dashed line are PLISCs of SKA with ``observing time = 20 years, SNR = 1'', ``observing time = 20 years, SNR = 3'', and ``observing time = 20 years, SNR = 10'', respectively. 
The red dotted line, red dotdash line, and red dashed line are PLISCs of LISA with ``observing time = 1 year, SNR = 1'', ``observing time = 1 year, SNR = 3'', and ``observing time = 1 year, SNR = 10'', respectively. 
From left plot to right plot, bubble wall velocity $v_w$ changes from 0.9 to 0.5.}
\label{fig:vw_A}
\end{figure}
\begin{figure}[htb]
\centering
\hspace*{-0.8cm}
\includegraphics[width=6.5in]{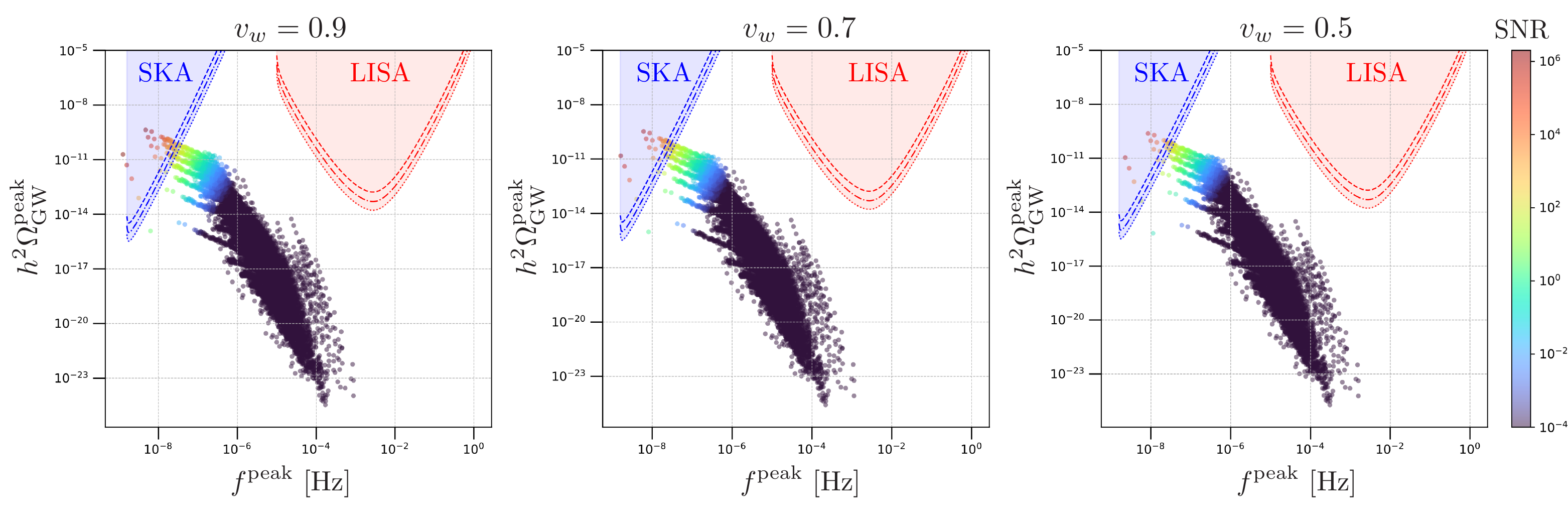}
\vspace{-30pt}
\caption{Same as Fig.~\ref{fig:vw_A}, but for charge assignment B.}
\label{fig:vw_B}
\end{figure}

In this subsection we discuss the dependence on $v_w$. 
As we introduced in Sec.~\ref{sec:nc}, we choose $v_w =0.9$ for the ``non-runaway'' FOPT for a quick and optimistic GW signal estimate. 
However, $v_w$ affects the mean bubble separation and percolation temperature, and also directly affects peak frequencies and peak amplitudes. 
Therefore, it is necessary to analyze the dependence of the conclusion made in the previous subsection on $v_w$.

\begin{figure}[htb]
\centering
\includegraphics[width=5.5in]{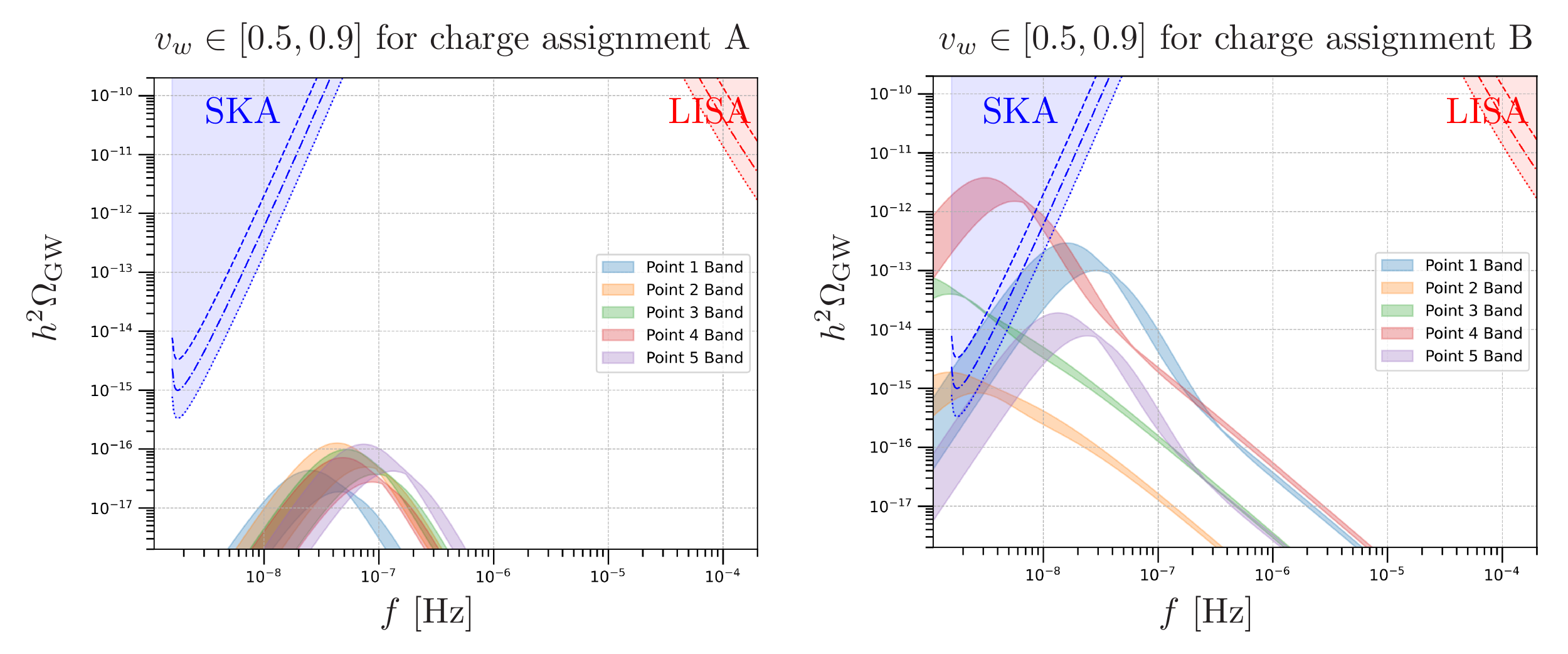}
\vspace{-20pt}
\caption{
The blue dotted line, blue dotdash line, and blue dashed line are PLISCs of SKA with ``observing time = 20 years, SNR = 1'', ``observing time = 20 years, SNR = 3'', and ``observing time = 20 years, SNR = 10'', respectively. 
The red dotted line, red dotdash line, and red dashed line are PLISCs of LISA with ``observing time = 1 year, SNR = 1'', ``observing time = 1 year, SNR = 3'', and ``observing time = 1 year, SNR = 10'', respectively. 
Left: color bands from five benchmark points in charge assignment A which are consistent with the small-scale data, $\Delta N_\text{eff}$ and Lyman-$\alpha$. These five benchmark points are selected as close as possible to PLISCs of SKA. The width of the bands comes from the varying $v_w$. 
Right: color bands from five benchmark points in charge assignment B which are consistent with the small-scale data, $\Delta N_\text{eff}$ and Lyman-$\alpha$. These five benchmark points are selected as detectable and representative. The width of the bands comes from the varying $v_w$. } 
\label{fig:color_band}
\end{figure}

To address this problem, in Fig.~\ref{fig:vw_A} and Fig.~\ref{fig:vw_B} we present the SNR distributions for two assignments with $v_w = 0.9, 0.7, 0.5$. 
We observe that for both assignments, the changes in peak frequencies and peak amplitudes with respect to $v_w$ are much smaller than an order of magnitude. 
Furthermore, the dependence of SNR on $v_w$ is also mild.

To make the dependence on $v_w$ more clear, in Fig.~\ref{fig:color_band} we present the full spectra of 5 benchmark points for charge assignment A and B.
All these benchmark points belong to the ``non-runaway'' case, and thus in Fig.~\ref{fig:color_band} we vary $v_w$ from 0.5 to 0.9 so that the full spectra form bands. 
Information of all the benchmark points are given in Tab.~\ref{tab:bps}.
Fig.~\ref{fig:color_band} (Left) shows that, for assignment A, those undetectable points are still undetectable even if we change the value of $v_w$. 
While Fig.~\ref{fig:color_band} (Right) shows that, for assignment B, even some boundary points will become undetectable when we change $v_w$, but there remains a set of points that are always detectable. 
Thus the conclusion made in Sec.~\ref{sec:results1} is robust against the value of $v_w$. 

\begin{table}[htbp]
\centering
\begin{tabular}{c|ccccc}
\hline
   & $m_\chi$ [GeV] & $m_s$ [MeV] & $m_{A'}$ [MeV] & $\alpha'$ & $\xi_\text{reh}$ \\
\hline
Charge assignment A point 1 & 433 & 0.174 & 0.7 & 0.326 & 0.2 \\
Charge assignment A point 2 & 247 & 0.160  & 0.9 & 0.140 & 0.2 \\
Charge assignment A point 3 & 265 & 0.165  & 0.9 & 0.153 & 0.2 \\
Charge assignment A point 4 & 265 & 0.144  & 0.8 & 0.146 & 0.2 \\
Charge assignment A point 5 & 215 & 0.150  & 0.9 & 0.112 & 0.3 \\
Charge assignment B point 1 & 572 & 0.465 & 1.9 & 0.0151 & 0.7 \\
Charge assignment B point 2 & 327 & 0.457 & 2.2 & 0.00624 & 0.3 \\
Charge assignment B point 3 & 327 & 0.457 & 2.2 & 0.00624 & 0.5 \\
Charge assignment B point 4 & 657 & 0.450 & 1.7 & 0.0189 & 0.7 \\
Charge assignment B point 5 & 705 & 0.472 & 1.7 & 0.0216 & 0.4 \\
\hline
\end{tabular}
\caption{Information of all the benchmark points in Fig.~\ref{fig:color_band}.}
\label{tab:bps}
\end{table}

\subsection{Favored parameter space\label{sec:results3}}

\begin{figure}[htb]
\centering
\hspace*{-0.8cm}
\includegraphics[width=6.5in]{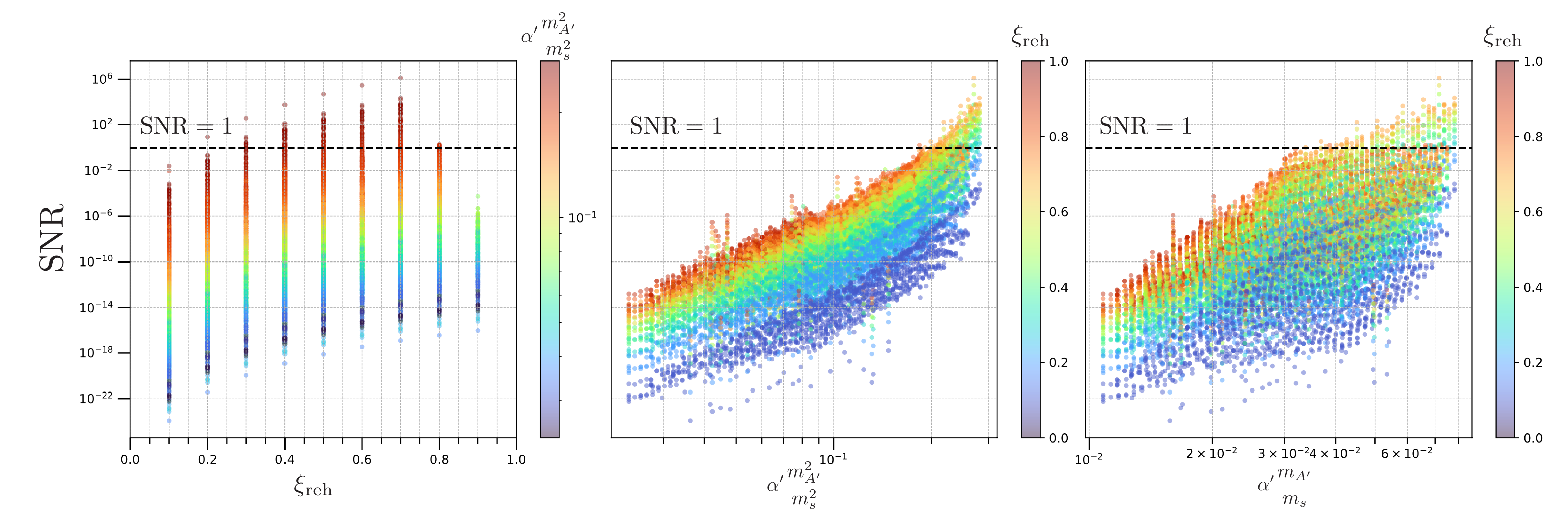}
\vspace{-30pt}
\caption{In this figure we consider charge assignment B. All the points are consistent with 
small-scale data, $\Delta N_\text{eff}$, and Lyman-$\alpha$. 
Left: SNR as functions of $\xi_\text{reh}$ color-mapped by $\alpha' \frac{m^2_{A'}}{m^2_s}$. 
Middle: SNR as functions of $\alpha' \frac{m^2_{A'}}{m^2_s}$ color-mapped by $\xi_\text{reh}$. 
Right: SNR as functions of $\alpha' \frac{m_{A'}}{m_s}$ color-mapped by $\xi_\text{reh}$. 
}
\label{fig:para}
\end{figure}

In this section we discuss in which region of the parameter space one can obtain detectable GWs while remaining consistent with 
small-scale data, $\Delta N_\text{eff}$, and Lyman-$\alpha$. 
First, the charge assignment should be similar to assignment B, which has a large Higgs charge $Q_S$ and small DR charge $Q_\psi$. This requirement helps make the model safe from the Lyman-$\alpha$ limit, as we show in Sec.~\ref{sec:results1}. 

To further dissect the favored parameter space, in Fig.~\ref{fig:para} we present the SNR of all the points in charge assignment B which are consistent with small-scale data, $\Delta N_\text{eff}$, and Lyman-$\alpha$. 
The $x$-axes are chosen to be $\xi_\text{reh}$, $\alpha' \frac{m^2_{A'}}{m^2_s}$, and $\alpha' \frac{m_{A'}}{m_s}$, respectively. We fix bubble wall velocity $v_w = 0.9$ because the discussion in the previous subsection already shows a mild dependence on $v_w$. 
Distributions in Fig.~\ref{fig:para} indicate the following favored parameter space: 
\begin{eqnarray}
 0.2 \lesssim \xi_\text{reh} \lesssim 0.8 \ , \  0.2  \lesssim \alpha' \frac{m^2_{A'}}{m^2_s} \ , \  0.03  \lesssim \alpha' \frac{m_{A'}}{m_s}   
\end{eqnarray}
The upper bound of $\xi_\text{reh}$ can be understood from Fig.~\ref{fig:scanB}, while its lower bound simply says that the dark sector cannot be too cold; otherwise, the energy density in the dark sector will be too small to generate detectable GWs. 
The lower bounds of $\alpha' \frac{m^2_{A'}}{m^2_s}$ and $\alpha' \frac{m_{A'}}{m_s}$ reflect a general property of FOPT in the Higgs model -- a strong phase transition only happens when gauge coupling is stronger than Higgs self-coupling.

It is useful to understand the surviving parameter region from a simple semi-analytic point of view. The relevant requirements can be summarized as an allowed corridor for the late-time temperature ratio $\xi_\infty$. First, the dark-radiation contribution to the effective number of neutrino species gives
\begin{eqnarray}
\Delta N_{\rm eff} = \frac{8}{7} \left(\frac{4}{11}\right)^{-4/3} \xi_\infty^4\, .
\end{eqnarray}
Using the current bound $\Delta N_{\rm eff}<0.29$, one obtains
\begin{eqnarray} \label{eq:xi_N_bound}
\xi_\infty  \lesssim  0.51\, .
\end{eqnarray}
Second, the ETHOS parameter induced by DM--DR scattering is
\begin{eqnarray}
a_{\rm dark}=3.07\times 10^7\,\xi_\infty^2Q_\chi^2 Q_\psi^2 g'^4\left(\frac{\rm GeV}{m_\chi}\right)\left(\frac{\rm MeV}{m_{A'}}\right)^4{\rm Mpc}^{-1}\, .
\end{eqnarray}
Together with the Lyman-$\alpha$ limit
$a_{\rm dark}<30\,\xi_\infty^{-4}~{\rm Mpc}^{-1}$, this gives another upper bound,
\begin{eqnarray} \label{eq:xi_L_bound}
\xi_\infty<\xi_{Ly}\equiv\left[\frac{30}{3.07\times 10^7\,Q_\chi^2 Q_\psi^2 g'^4\left(\frac{\rm GeV}{m_\chi}\right)\left(\frac{\rm MeV}{m_{A'}}\right)^4}\right]^{1/6}\, .
\end{eqnarray}
On the other hand, the GW signal prefers a sufficiently large dark-sector temperature. Schematically, the peak amplitude can be written as
\begin{eqnarray}    \label{eq:omega_hat_def}
h^2\Omega_{\rm GW}^{\rm peak}=\xi_\infty^4\, \times h^2\widehat{\Omega}_{\rm GW}^{\rm peak},
\end{eqnarray}
where $\widehat{\Omega}_{\rm GW}^{\rm peak}$ contains the dependence on the phase-transition strength, $\beta/H_\ast$, and $v_w$, but the explicit $\xi_\infty^4$ factor has been extracted. The phase-transition strength is mainly controlled by the gauge-induced barrier, and therefore by the combination $\alpha' m_{A'}^2/m_s^2$, as also observed in the numerical scan. Requiring the signal to be larger than the detector threshold at the relevant frequency (denoted by $h^2\Omega_{\rm det}^{\rm th}$), gives the approximate lower bound
\begin{eqnarray} \label{eq:xi_GW_bound}
\xi_\infty > \xi_{\rm GW} \equiv \left( \frac{h^2\Omega_{\rm det}^{\rm th}}  {h^2\widehat{\Omega}_{\rm GW}^{\rm peak}} \right)^{1/4}\, .
\end{eqnarray} 
Thus the simultaneous requirements can be expressed as
\begin{eqnarray} \label{eq:xi_corridor}
\xi_{\rm GW}<\xi_\infty< \min\left(0.51,\ \xi_{Ly}\right)\, .
\end{eqnarray}
Equivalently, the largest GW amplitude allowed by $\Delta N_{\rm eff}$ and Lyman-$\alpha$ scales as
\begin{eqnarray}  \label{eq:omega_GW_max_scaling}
h^2\Omega_{\rm GW,max}^{\rm peak} \simeq 
\min\left[ 0.51^4,\,
\left( \frac{30}{ 3.07\times 10^7\, Q_\chi^2 Q_\psi^2 g'^4  \left(\frac{\rm GeV}{m_\chi}\right) \left(\frac{\rm MeV}{m_{A'}}\right)^4} \right)^{2/3} \right]\times h^2 \widehat{\Omega}_{\rm GW}^{\rm peak}\, .
\end{eqnarray}
This expression makes transparent why the two charge assignments behave differently. 
In charge assignment A, $Q_\psi=1$, so the DM--DR scattering rate is relatively large. 
The Lyman-$\alpha$ condition then forces $\xi_\infty$ to be small, and thus the GW signal is strongly reduced because $h^2\Omega_{\rm GW}^{\rm peak}\propto \xi_\infty^4$. 
As a result, the parameter region that is sufficiently loud in GWs is removed once the Lyman-$\alpha$ and $\Delta N_{\rm eff}$ constraints are imposed simultaneously. 
In charge assignment B, however, $Q_\psi=1/2$ suppresses $a_{\rm dark}$ by a factor of four at fixed $g'$, $m_\chi$, $m_{A'}$ and $\xi_\infty$, thereby relaxing the Lyman-$\alpha$ upper bound on $\xi_\infty$. 
At the same time, the larger dark-Higgs charge $Q_S=3$ allows a stronger Abelian-Higgs phase transition for suitable values of $\alpha' m_{A'}^2/m_s^2$ (see the difference between Fig.~\ref{fig:scanA} (left) and Fig.~\ref{fig:scanB} (left)). 
Therefore, charge assignment B has a much larger parameter space in which it can maintain a sufficiently large $h^2\Omega_{\rm GW}^{\rm peak}$ while keeping $a_{\rm dark}\xi_\infty^4$ below the Lyman-$\alpha$ limit. 

\section{Conclusion\label{sec:conclu}}

In this work we studied the detection of gravitational waves generated by a secluded dark sector. 
The portal between the dark sector and the visible sector is turned off, and thus the entropy of each sector is separately conserved. 
The dark sector is charged under a dark $U(1)'$ that is spontaneously broken. 
The dark matter self-interaction induced by this dark $U(1)'$ can be consistent with the small-scale structure observations. 
Furthermore, the symmetry breaking process of this dark $U(1)'$, which can be a first-order phase transition, might generate detectable gravitational wave signals.
However, this secluded dark sector is constrained by the $\Delta N_\text{eff}$ measurement and the Lyman-$\alpha$ observations. 

We consider two charge assignments, A and B, in this work. 
$\Delta N_\text{eff}$ imposes uniform bounds on both charge assignments, but the Lyman-$\alpha$ bound on charge assignment B is much weaker than the Lyman-$\alpha$ bound on charge assignment A. 
Through a parameter space scan, we find that charge assignment A is highly constrained by the combination of these two constraints and thus is unable to generate any gravitational waves accessible at SKA or LISA, but charge assignment B still has a small parameter space that is detectable by SKA. 

Although $\Delta N_\text{eff}$ imposes a strong and uniform restriction on the secluded self-interacting dark sector, $\Delta N_\text{eff}$ alone cannot make all these dark sector scenarios undetectable by future GW experiments. 
Dark-sector scenarios that can escape the Lyman-$\alpha$ bound are still promising targets for future experiments such as SKA.

\section*{Acknowledgements}
M. Z. appreciates helpful discussions with Yue-Lin Sming Tsai, Chi Zhang, and Nils Sch{\"o}neberg. This work was supported by the Natural Science Foundation of China (NSFC) under Grants Nos. 12105118, 11947118 and 12335005, by the Research Fund for Outstanding Talents (Grant No. 5101029470335) from Henan Normal University, and by the Peng-Huan-Wu Theoretical Physics Innovation Center (Grant No. 12047503) funded by the National Natural Science Foundation of China.

\vspace{0.5cm}

\bibliographystyle{JHEP}
\bibliography{ref}

@article{Murgia:2018now,
	archiveprefix = {arXiv},
	author = {Murgia, Riccardo and Ir{\v{s}}i{\v{c}}, Vid and Viel, Matteo},
	doi = {10.1103/PhysRevD.98.083540},
	eprint = {1806.08371},
	journal = {Phys. Rev. D},
	number = {8},
	pages = {083540},
	primaryclass = {astro-ph.CO},
	title = {{Novel constraints on noncold, nonthermal dark matter from Lyman- {\ensuremath{\alpha}} forest data}},
	volume = {98},
	year = {2018}}

@article{Sandick:2021gew,
	archiveprefix = {arXiv},
	author = {Sandick, Pearl and Es Haghi, Barmak Shams and Sinha, Kuver},
	doi = {10.1103/PhysRevD.104.083523},
	eprint = {2108.08329},
	journal = {Phys. Rev. D},
	number = {8},
	pages = {083523},
	primaryclass = {astro-ph.CO},
	title = {{Asymmetric reheating by primordial black holes}},
	volume = {104},
	year = {2021}}

@article{Adshead:2016xxj,
	archiveprefix = {arXiv},
	author = {Adshead, Peter and Cui, Yanou and Shelton, Jessie},
	doi = {10.1007/JHEP06(2016)016},
	eprint = {1604.02458},
	journal = {JHEP},
	pages = {016},
	primaryclass = {hep-ph},
	title = {{Chilly Dark Sectors and Asymmetric Reheating}},
	volume = {06},
	year = {2016}}

@article{Blas:2011rf,
	archiveprefix = {arXiv},
	author = {Blas, Diego and Lesgourgues, Julien and Tram, Thomas},
	doi = {10.1088/1475-7516/2011/07/034},
	eprint = {1104.2933},
	journal = {JCAP},
	pages = {034},
	primaryclass = {astro-ph.CO},
	reportnumber = {CERN-PH-TH-2011-082, LAPTH-010-11},
	title = {{The Cosmic Linear Anisotropy Solving System (CLASS) II: Approximation schemes}},
	volume = {07},
	year = {2011}}

@article{Becker:2020hzj,
	archiveprefix = {arXiv},
	author = {Becker, Niklas and Hooper, Deanna C. and Kahlhoefer, Felix and Lesgourgues, Julien and Sch{\"o}neberg, Nils},
	doi = {10.1088/1475-7516/2021/02/019},
	eprint = {2010.04074},
	journal = {JCAP},
	pages = {019},
	primaryclass = {astro-ph.CO},
	reportnumber = {TTK-20-32, ULB-TH/20-13},
	title = {{Cosmological constraints on multi-interacting dark matter}},
	volume = {02},
	year = {2021}}

@article{Guo:2020grp,
	archiveprefix = {arXiv},
	author = {Guo, Huai-Ke and Sinha, Kuver and Vagie, Daniel and White, Graham},
	doi = {10.1088/1475-7516/2021/01/001},
	eprint = {2007.08537},
	journal = {JCAP},
	pages = {001},
	primaryclass = {hep-ph},
	title = {{Phase Transitions in an Expanding Universe: Stochastic Gravitational Waves in Standard and Non-Standard Histories}},
	volume = {01},
	year = {2021}}

@article{Banik:2024zwj,
	archiveprefix = {arXiv},
	author = {Banik, Amitayus and Cui, Yanou and Tsai, Yu-Dai and Tsai, Yuhsin},
	eprint = {2412.16282},
	month = {12},
	primaryclass = {hep-ph},
	reportnumber = {LA-UR-24-33252},
	title = {{The Sound of Dark Sectors in Pulsar Timing Arrays}},
	year = {2024}}

@article{Li:2025nja,
	archiveprefix = {arXiv},
	author = {Li, Jinzheng and Nath, Pran},
	eprint = {2501.14986},
	month = {1},
	primaryclass = {hep-ph},
	title = {{Supercooled Phase Transitions: Why Thermal History of Hidden Sector Matters in Analysis of Pulsar Timing Array Signals}},
	year = {2025}}

@article{Kang:2025nhe,
	archiveprefix = {arXiv},
	author = {Kang, Zhaofeng and Zhu, Jiang},
	eprint = {2501.15242},
	month = {1},
	primaryclass = {hep-ph},
	title = {{Dark Chiral Phase Transition Driven by Chemical Potential and its Gravitational Wave Test}},
	year = {2025}}

@article{Costa:2025csj,
	archiveprefix = {arXiv},
	author = {Costa, Francesco and Hoefken Zink, Jaime and Lucente, Michele and Pascoli, Silvia and Rosauro-Alcaraz, Salvador},
	eprint = {2501.15649},
	month = {1},
	primaryclass = {hep-ph},
	title = {{Supercooled Dark Scalar Phase Transitions explanation of NANOGrav data}},
	year = {2025}}

@article{Goncalves:2025uwh,
	archiveprefix = {arXiv},
	author = {Gon\c{c}alves, Jo\~ao and Marfatia, Danny and Morais, Ant\'onio P. and Pasechnik, Roman},
	eprint = {2501.11619},
	month = {1},
	primaryclass = {hep-ph},
	title = {{Supercooled phase transitions in conformal dark sectors explain NANOGrav data}},
	year = {2025}}

@article{Schmitz:2020syl,
	archiveprefix = {arXiv},
	author = {Schmitz, Kai},
	doi = {10.1007/JHEP01(2021)097},
	eprint = {2002.04615},
	journal = {JHEP},
	pages = {097},
	primaryclass = {hep-ph},
	reportnumber = {CERN-TH-2020-018},
	title = {{New Sensitivity Curves for Gravitational-Wave Signals from Cosmological Phase Transitions}},
	volume = {01},
	year = {2021}}

@article{Thrane:2013oya,
	archiveprefix = {arXiv},
	author = {Thrane, Eric and Romano, Joseph D.},
	doi = {10.1103/PhysRevD.88.124032},
	eprint = {1310.5300},
	journal = {Phys. Rev. D},
	number = {12},
	pages = {124032},
	primaryclass = {astro-ph.IM},
	title = {{Sensitivity curves for searches for gravitational-wave backgrounds}},
	volume = {88},
	year = {2013}}

@article{Guth:1981uk,
	author = {Guth, Alan H. and Weinberg, Erick J.},
	doi = {10.1103/PhysRevD.23.876},
	journal = {Phys. Rev. D},
	pages = {876},
	reportnumber = {CU-TP-183},
	title = {{Cosmological Consequences of a First Order Phase Transition in the SU(5) Grand Unified Model}},
	volume = {23},
	year = {1981}}

@article{Guth:1979bh,
	author = {Guth, Alan H. and Tye, S. H. H.},
	doi = {10.1103/PhysRevLett.44.631},
	journal = {Phys. Rev. Lett.},
	note = {[Erratum: Phys.Rev.Lett. 44, 963 (1980)]},
	pages = {631},
	reportnumber = {SLAC-PUB-2448, CLNS-79-441},
	title = {{Phase Transitions and Magnetic Monopole Production in the Very Early Universe}},
	volume = {44},
	year = {1980}}

@article{Wang:2020jrd,
	archiveprefix = {arXiv},
	author = {Wang, Xiao and Huang, Fa Peng and Zhang, Xinmin},
	doi = {10.1088/1475-7516/2020/05/045},
	eprint = {2003.08892},
	journal = {JCAP},
	pages = {045},
	primaryclass = {hep-ph},
	title = {{Phase transition dynamics and gravitational wave spectra of strong first-order phase transition in supercooled universe}},
	volume = {05},
	year = {2020}}

@article{Ellis:2018mja,
	archiveprefix = {arXiv},
	author = {Ellis, John and Lewicki, Marek and No, Jos\'e Miguel},
	doi = {10.1088/1475-7516/2019/04/003},
	eprint = {1809.08242},
	journal = {JCAP},
	pages = {003},
	primaryclass = {hep-ph},
	reportnumber = {KCL-PH-TH/2018-46, CERN-TH/2018-197, IFT-UAM/CSIC-18-94, CERN-TH-2018-197},
	title = {{On the Maximal Strength of a First-Order Electroweak Phase Transition and its Gravitational Wave Signal}},
	volume = {04},
	year = {2019}}

@article{Cai:2017tmh,
	archiveprefix = {arXiv},
	author = {Cai, Rong-Gen and Sasaki, Misao and Wang, Shao-Jiang},
	doi = {10.1088/1475-7516/2017/08/004},
	eprint = {1707.03001},
	journal = {JCAP},
	pages = {004},
	primaryclass = {astro-ph.CO},
	reportnumber = {YITP-17-67},
	title = {{The gravitational waves from the first-order phase transition with a dimension-six operator}},
	volume = {08},
	year = {2017}}

@article{Kobakhidze:2017mru,
	archiveprefix = {arXiv},
	author = {Kobakhidze, Archil and Lagger, Cyril and Manning, Adrian and Yue, Jason},
	doi = {10.1140/epjc/s10052-017-5132-y},
	eprint = {1703.06552},
	journal = {Eur. Phys. J. C},
	number = {8},
	pages = {570},
	primaryclass = {hep-ph},
	title = {{Gravitational waves from a supercooled electroweak phase transition and their detection with pulsar timing arrays}},
	volume = {77},
	year = {2017}}

@article{Megevand:2016lpr,
	archiveprefix = {arXiv},
	author = {Megevand, Ariel and Ramirez, Santiago},
	doi = {10.1016/j.nuclphysb.2017.03.009},
	eprint = {1611.05853},
	journal = {Nucl. Phys. B},
	pages = {74--109},
	primaryclass = {astro-ph.CO},
	title = {{Bubble nucleation and growth in very strong cosmological phase transitions}},
	volume = {919},
	year = {2017}}

@article{Turner:1992tz,
	author = {Turner, Michael S. and Weinberg, Erick J. and Widrow, Lawrence M.},
	doi = {10.1103/PhysRevD.46.2384},
	journal = {Phys. Rev. D},
	pages = {2384--2403},
	reportnumber = {FERMILAB-PUB-91-334-A, CU-TP-558, IASSNS-HEP-92-21},
	title = {{Bubble nucleation in first order inflation and other cosmological phase transitions}},
	volume = {46},
	year = {1992}}

@article{LISA:2017pwj,
	archiveprefix = {arXiv},
	author = {Amaro-Seoane, Pau and others},
	collaboration = {LISA},
	eprint = {1702.00786},
	month = {2},
	primaryclass = {astro-ph.IM},
	title = {{Laser Interferometer Space Antenna}},
	year = {2017}}

@article{Janssen:2014dka,
	archiveprefix = {arXiv},
	author = {Janssen, Gemma and others},
	doi = {10.22323/1.215.0037},
	editor = {Bourke, Tyler L. and others},
	eprint = {1501.00127},
	journal = {PoS},
	pages = {037},
	primaryclass = {astro-ph.IM},
	title = {{Gravitational wave astronomy with the SKA}},
	volume = {AASKA14},
	year = {2015}}

@article{Bodeker:2017cim,
	archiveprefix = {arXiv},
	author = {Bodeker, Dietrich and Moore, Guy D.},
	doi = {10.1088/1475-7516/2017/05/025},
	eprint = {1703.08215},
	journal = {JCAP},
	pages = {025},
	primaryclass = {hep-ph},
	title = {{Electroweak Bubble Wall Speed Limit}},
	volume = {05},
	year = {2017}}

@article{Caprini:2019egz,
	archiveprefix = {arXiv},
	author = {Caprini, Chiara and others},
	doi = {10.1088/1475-7516/2020/03/024},
	eprint = {1910.13125},
	journal = {JCAP},
	pages = {024},
	primaryclass = {astro-ph.CO},
	reportnumber = {DESY-19-159, IPPP/19/27, HIP-2019-14/TH, MITP/19-066, IFT-UAM/CSIC-19-139},
	title = {{Detecting gravitational waves from cosmological phase transitions with LISA: an update}},
	volume = {03},
	year = {2020}}

@article{Caprini:2015zlo,
	archiveprefix = {arXiv},
	author = {Caprini, Chiara and others},
	doi = {10.1088/1475-7516/2016/04/001},
	eprint = {1512.06239},
	journal = {JCAP},
	pages = {001},
	primaryclass = {astro-ph.CO},
	reportnumber = {DESY-15-246},
	title = {{Science with the space-based interferometer eLISA. II: Gravitational waves from cosmological phase transitions}},
	volume = {04},
	year = {2016}}

@article{Wang:2020zlf,
	archiveprefix = {arXiv},
	author = {Wang, Xiao and Huang, Fa Peng and Zhang, Xinmin},
	eprint = {2011.12903},
	month = {11},
	primaryclass = {hep-ph},
	title = {{Bubble wall velocity beyond leading-log approximation in electroweak phase transition}},
	year = {2020}}

@article{Laurent:2022jrs,
	archiveprefix = {arXiv},
	author = {Laurent, Benoit and Cline, James M.},
	doi = {10.1103/PhysRevD.106.023501},
	eprint = {2204.13120},
	journal = {Phys. Rev. D},
	number = {2},
	pages = {023501},
	primaryclass = {hep-ph},
	title = {{First principles determination of bubble wall velocity}},
	volume = {106},
	year = {2022}}

@article{Dorsch:2018pat,
	archiveprefix = {arXiv},
	author = {Dorsch, Glauber C. and Huber, Stephan J. and Konstandin, Thomas},
	doi = {10.1088/1475-7516/2018/12/034},
	eprint = {1809.04907},
	journal = {JCAP},
	pages = {034},
	primaryclass = {hep-ph},
	reportnumber = {DESY 18-162, DESY-18-162},
	title = {{Bubble wall velocities in the Standard Model and beyond}},
	volume = {12},
	year = {2018}}

@article{Konstandin:2014zta,
	archiveprefix = {arXiv},
	author = {Konstandin, Thomas and Nardini, Germano and Rues, Ingo},
	doi = {10.1088/1475-7516/2014/09/028},
	eprint = {1407.3132},
	journal = {JCAP},
	pages = {028},
	primaryclass = {hep-ph},
	reportnumber = {DESY-14-127, NSF-KITP-14-089},
	title = {{From Boltzmann equations to steady wall velocities}},
	volume = {09},
	year = {2014}}

@article{Huber:2013kj,
	archiveprefix = {arXiv},
	author = {Huber, Stephan J. and Sopena, Miguel},
	eprint = {1302.1044},
	month = {2},
	primaryclass = {hep-ph},
	title = {{An efficient approach to electroweak bubble velocities}},
	year = {2013}}

@article{Megevand:2009gh,
	archiveprefix = {arXiv},
	author = {Megevand, Ariel and Sanchez, Alejandro D.},
	doi = {10.1016/j.nuclphysb.2009.09.019},
	eprint = {0908.3663},
	journal = {Nucl. Phys. B},
	pages = {151--176},
	primaryclass = {hep-ph},
	title = {{Velocity of electroweak bubble walls}},
	volume = {825},
	year = {2010}}

@article{Moore:1995si,
	archiveprefix = {arXiv},
	author = {Moore, Guy D. and Prokopec, Tomislav},
	doi = {10.1103/PhysRevD.52.7182},
	eprint = {hep-ph/9506475},
	journal = {Phys. Rev. D},
	pages = {7182--7204},
	reportnumber = {PUPT-1544, PUP-TH-1544, LANCS-TH-9517},
	title = {{How fast can the wall move? A Study of the electroweak phase transition dynamics}},
	volume = {52},
	year = {1995}}

@article{Espinosa:2010hh,
	archiveprefix = {arXiv},
	author = {Espinosa, Jose R. and Konstandin, Thomas and No, Jose M. and Servant, Geraldine},
	doi = {10.1088/1475-7516/2010/06/028},
	eprint = {1004.4187},
	journal = {JCAP},
	pages = {028},
	primaryclass = {hep-ph},
	reportnumber = {CERN-PH-TH-2010-027},
	title = {{Energy Budget of Cosmological First-order Phase Transitions}},
	volume = {06},
	year = {2010}}

@article{Kisslinger:2015hua,
	archiveprefix = {arXiv},
	author = {Kisslinger, Leonard and Kahniashvili, Tina},
	doi = {10.1103/PhysRevD.92.043006},
	eprint = {1505.03680},
	journal = {Phys. Rev. D},
	number = {4},
	pages = {043006},
	primaryclass = {astro-ph.CO},
	title = {{Polarized Gravitational Waves from Cosmological Phase Transitions}},
	volume = {92},
	year = {2015}}

@article{Caprini:2009yp,
	archiveprefix = {arXiv},
	author = {Caprini, Chiara and Durrer, Ruth and Servant, Geraldine},
	doi = {10.1088/1475-7516/2009/12/024},
	eprint = {0909.0622},
	journal = {JCAP},
	pages = {024},
	primaryclass = {astro-ph.CO},
	title = {{The stochastic gravitational wave background from turbulence and magnetic fields generated by a first-order phase transition}},
	volume = {12},
	year = {2009}}

@article{Kahniashvili:2009mf,
	archiveprefix = {arXiv},
	author = {Kahniashvili, Tina and Kisslinger, Leonard and Stevens, Trevor},
	doi = {10.1103/PhysRevD.81.023004},
	eprint = {0905.0643},
	journal = {Phys. Rev. D},
	pages = {023004},
	primaryclass = {astro-ph.CO},
	title = {{Gravitational Radiation Generated by Magnetic Fields in Cosmological Phase Transitions}},
	volume = {81},
	year = {2010}}

@article{Kahniashvili:2008pe,
	archiveprefix = {arXiv},
	author = {Kahniashvili, Tina and Campanelli, Leonardo and Gogoberidze, Grigol and Maravin, Yurii and Ratra, Bharat},
	doi = {10.1103/PhysRevD.78.123006},
	eprint = {0809.1899},
	journal = {Phys. Rev. D},
	note = {[Erratum: Phys.Rev.D 79, 109901 (2009)]},
	pages = {123006},
	primaryclass = {astro-ph},
	title = {{Gravitational Radiation from Primordial Helical Inverse Cascade MHD Turbulence}},
	volume = {78},
	year = {2008}}

@article{Kahniashvili:2008pf,
	archiveprefix = {arXiv},
	author = {Kahniashvili, Tina and Kosowsky, Arthur and Gogoberidze, Grigol and Maravin, Yurii},
	doi = {10.1103/PhysRevD.78.043003},
	eprint = {0806.0293},
	journal = {Phys. Rev. D},
	pages = {043003},
	primaryclass = {astro-ph},
	title = {{Detectability of Gravitational Waves from Phase Transitions}},
	volume = {78},
	year = {2008}}

@article{Caprini:2006jb,
	archiveprefix = {arXiv},
	author = {Caprini, Chiara and Durrer, Ruth},
	doi = {10.1103/PhysRevD.74.063521},
	eprint = {astro-ph/0603476},
	journal = {Phys. Rev. D},
	pages = {063521},
	title = {{Gravitational waves from stochastic relativistic sources: Primordial turbulence and magnetic fields}},
	volume = {74},
	year = {2006}}

@article{Hindmarsh:2015qta,
	archiveprefix = {arXiv},
	author = {Hindmarsh, Mark and Huber, Stephan J. and Rummukainen, Kari and Weir, David J.},
	doi = {10.1103/PhysRevD.92.123009},
	eprint = {1504.03291},
	journal = {Phys. Rev. D},
	number = {12},
	pages = {123009},
	primaryclass = {astro-ph.CO},
	reportnumber = {HIP-2015-13-TH},
	title = {{Numerical simulations of acoustically generated gravitational waves at a first order phase transition}},
	volume = {92},
	year = {2015}}

@article{Giblin:2014qia,
	archiveprefix = {arXiv},
	author = {Giblin, John T. and Mertens, James B.},
	doi = {10.1103/PhysRevD.90.023532},
	eprint = {1405.4005},
	journal = {Phys. Rev. D},
	number = {2},
	pages = {023532},
	primaryclass = {astro-ph.CO},
	title = {{Gravitional radiation from first-order phase transitions in the presence of a fluid}},
	volume = {90},
	year = {2014}}

@article{Giblin:2013kea,
	archiveprefix = {arXiv},
	author = {Giblin, Jr., John T. and Mertens, James B.},
	doi = {10.1007/JHEP12(2013)042},
	eprint = {1310.2948},
	journal = {JHEP},
	pages = {042},
	primaryclass = {hep-th},
	title = {{Vacuum Bubbles in the Presence of a Relativistic Fluid}},
	volume = {12},
	year = {2013}}

@article{Hindmarsh:2013xza,
	archiveprefix = {arXiv},
	author = {Hindmarsh, Mark and Huber, Stephan J. and Rummukainen, Kari and Weir, David J.},
	doi = {10.1103/PhysRevLett.112.041301},
	eprint = {1304.2433},
	journal = {Phys. Rev. Lett.},
	pages = {041301},
	primaryclass = {hep-ph},
	reportnumber = {HIP-2013-07-TH},
	title = {{Gravitational waves from the sound of a first order phase transition}},
	volume = {112},
	year = {2014}}

@article{Huber:2008hg,
	archiveprefix = {arXiv},
	author = {Huber, Stephan J. and Konstandin, Thomas},
	doi = {10.1088/1475-7516/2008/09/022},
	eprint = {0806.1828},
	journal = {JCAP},
	pages = {022},
	primaryclass = {hep-ph},
	title = {{Gravitational Wave Production by Collisions: More Bubbles}},
	volume = {09},
	year = {2008}}

@article{Caprini:2007xq,
	archiveprefix = {arXiv},
	author = {Caprini, Chiara and Durrer, Ruth and Servant, Geraldine},
	doi = {10.1103/PhysRevD.77.124015},
	eprint = {0711.2593},
	journal = {Phys. Rev. D},
	pages = {124015},
	primaryclass = {astro-ph},
	reportnumber = {CERN-PH-TH-2007-206, SACLAY-T07-142},
	title = {{Gravitational wave generation from bubble collisions in first-order phase transitions: An analytic approach}},
	volume = {77},
	year = {2008}}

@article{Kamionkowski:1993fg,
	archiveprefix = {arXiv},
	author = {Kamionkowski, Marc and Kosowsky, Arthur and Turner, Michael S.},
	doi = {10.1103/PhysRevD.49.2837},
	eprint = {astro-ph/9310044},
	journal = {Phys. Rev. D},
	pages = {2837--2851},
	reportnumber = {IASSNS-HEP-93-44, FERMILAB-PUB-93-235-A},
	title = {{Gravitational radiation from first order phase transitions}},
	volume = {49},
	year = {1994}}

@article{Kosowsky:1992vn,
	archiveprefix = {arXiv},
	author = {Kosowsky, Arthur and Turner, Michael S.},
	doi = {10.1103/PhysRevD.47.4372},
	eprint = {astro-ph/9211004},
	journal = {Phys. Rev. D},
	pages = {4372--4391},
	reportnumber = {FERMILAB-PUB-92-295-A},
	title = {{Gravitational radiation from colliding vacuum bubbles: envelope approximation to many bubble collisions}},
	volume = {47},
	year = {1993}}

@article{Kosowsky:1992rz,
	author = {Kosowsky, Arthur and Turner, Michael S. and Watkins, Richard},
	doi = {10.1103/PhysRevLett.69.2026},
	journal = {Phys. Rev. Lett.},
	pages = {2026--2029},
	reportnumber = {FERMILAB-PUB-91-333-A-REV, FERMILAB-PUB-91-333-A},
	title = {{Gravitational waves from first order cosmological phase transitions}},
	volume = {69},
	year = {1992}}

@article{Kosowsky:1991ua,
	author = {Kosowsky, Arthur and Turner, Michael S. and Watkins, Richard},
	doi = {10.1103/PhysRevD.45.4514},
	journal = {Phys. Rev. D},
	pages = {4514--4535},
	reportnumber = {FERMILAB-PUB-91-323-A},
	title = {{Gravitational radiation from colliding vacuum bubbles}},
	volume = {45},
	year = {1992}}

@article{Wainwright:2011kj,
	archiveprefix = {arXiv},
	author = {Wainwright, Carroll L.},
	doi = {10.1016/j.cpc.2012.04.004},
	eprint = {1109.4189},
	journal = {Comput. Phys. Commun.},
	pages = {2006--2013},
	primaryclass = {hep-ph},
	title = {{CosmoTransitions: Computing Cosmological Phase Transition Temperatures and Bubble Profiles with Multiple Fields}},
	volume = {183},
	year = {2012}}

@article{Apreda:2001us,
	archiveprefix = {arXiv},
	author = {Apreda, Riccardo and Maggiore, Michele and Nicolis, Alberto and Riotto, Antonio},
	doi = {10.1016/S0550-3213(02)00264-X},
	eprint = {gr-qc/0107033},
	journal = {Nucl. Phys. B},
	pages = {342--368},
	reportnumber = {UGVA-DPT-07-1096},
	title = {{Gravitational waves from electroweak phase transitions}},
	volume = {631},
	year = {2002}}

@article{Linde:1981zj,
	author = {Linde, Andrei D.},
	doi = {10.1016/0550-3213(83)90072-X},
	journal = {Nucl. Phys. B},
	note = {[Erratum: Nucl.Phys.B 223, 544 (1983)]},
	pages = {421},
	reportnumber = {LEBEDEV-81-265},
	title = {{Decay of the False Vacuum at Finite Temperature}},
	volume = {216},
	year = {1983}}

@article{Callan:1977pt,
	author = {Callan, Jr., Curtis G. and Coleman, Sidney R.},
	doi = {10.1103/PhysRevD.16.1762},
	journal = {Phys. Rev. D},
	pages = {1762--1768},
	reportnumber = {HUTP-77-A032},
	title = {{The Fate of the False Vacuum. 2. First Quantum Corrections}},
	volume = {16},
	year = {1977}}

@article{Coleman:1977py,
	author = {Coleman, Sidney R.},
	doi = {10.1103/PhysRevD.16.1248},
	journal = {Phys. Rev. D},
	note = {[Erratum: Phys.Rev.D 16, 1248 (1977)]},
	pages = {2929--2936},
	reportnumber = {HUTP-77-A004},
	title = {{The Fate of the False Vacuum. 1. Semiclassical Theory}},
	volume = {15},
	year = {1977}}

@article{Anderson:1991zb,
	author = {Anderson, Greg W. and Hall, Lawrence J.},
	doi = {10.1103/PhysRevD.45.2685},
	journal = {Phys. Rev. D},
	pages = {2685--2698},
	reportnumber = {LBL-31169, UCB-PTH-91-41},
	title = {{The Electroweak phase transition and baryogenesis}},
	volume = {45},
	year = {1992}}

@article{Delaunay:2007wb,
	archiveprefix = {arXiv},
	author = {Delaunay, Cedric and Grojean, Christophe and Wells, James D.},
	doi = {10.1088/1126-6708/2008/04/029},
	eprint = {0711.2511},
	journal = {JHEP},
	pages = {029},
	primaryclass = {hep-ph},
	reportnumber = {CERN-PH-TH-2007-219, MCTP-07-31, SACLAY-T07-141},
	title = {{Dynamics of Non-renormalizable Electroweak Symmetry Breaking}},
	volume = {04},
	year = {2008}}

@article{Arnold:1992rz,
	archiveprefix = {arXiv},
	author = {Arnold, Peter Brockway and Espinosa, Olivier},
	doi = {10.1103/PhysRevD.47.3546},
	eprint = {hep-ph/9212235},
	journal = {Phys. Rev. D},
	note = {[Erratum: Phys.Rev.D 50, 6662 (1994)]},
	pages = {3546},
	reportnumber = {UW-PT-92-18, USM-TH-60},
	title = {{The Effective potential and first order phase transitions: Beyond leading-order}},
	volume = {47},
	year = {1993}}

@inproceedings{Quiros:1999jp,
	archiveprefix = {arXiv},
	author = {Quiros, Mariano},
	booktitle = {{ICTP Summer School in High-Energy Physics and Cosmology}},
	eprint = {hep-ph/9901312},
	month = {1},
	pages = {187--259},
	reportnumber = {IEM-FT-187-99},
	title = {{Finite temperature field theory and phase transitions}},
	year = {1999}}

@article{Dolan:1973qd,
	author = {Dolan, L. and Jackiw, R.},
	doi = {10.1103/PhysRevD.9.3320},
	journal = {Phys. Rev. D},
	pages = {3320--3341},
	reportnumber = {MIT-CTP-406},
	title = {{Symmetry Behavior at Finite Temperature}},
	volume = {9},
	year = {1974}}

@article{Dimopoulos:1997cz,
	archiveprefix = {arXiv},
	author = {Dimopoulos, P. and Farakos, K. and Koutsoumbas, G.},
	doi = {10.1007/s100520050116},
	eprint = {hep-lat/9703004},
	journal = {Eur. Phys. J. C},
	pages = {711--719},
	reportnumber = {NTUA-63-97},
	title = {{Three-dimensional lattice U(1) gauge Higgs model at low m(H)}},
	volume = {1},
	year = {1998}}

@article{Karjalainen:1996wx,
	archiveprefix = {arXiv},
	author = {Karjalainen, M. and Laine, M. and Peisa, J.},
	doi = {10.1016/S0920-5632(96)00692-5},
	editor = {Bernard, C. and Golterman, M. and Ogilvie, M. and Potvin, J.},
	eprint = {hep-lat/9608006},
	journal = {Nucl. Phys. B Proc. Suppl.},
	pages = {475--480},
	reportnumber = {HU-TFT-96-32},
	title = {{The Order of the phase transition in 3-d U(1) + Higgs theory}},
	volume = {53},
	year = {1997}}

@article{Chiang:2017zbz,
	archiveprefix = {arXiv},
	author = {Chiang, Cheng-Wei and Senaha, Eibun},
	doi = {10.1016/j.physletb.2017.09.064},
	eprint = {1707.06765},
	journal = {Phys. Lett. B},
	pages = {489--493},
	primaryclass = {hep-ph},
	reportnumber = {NCTS-PH-1723},
	title = {{On gauge dependence of gravitational waves from a first-order phase transition in classical scale-invariant $U(1)'$ models}},
	volume = {774},
	year = {2017}}

@article{Wainwright:2011qy,
	archiveprefix = {arXiv},
	author = {Wainwright, Carroll and Profumo, Stefano and Ramsey-Musolf, Michael J.},
	doi = {10.1103/PhysRevD.84.023521},
	eprint = {1104.5487},
	journal = {Phys. Rev. D},
	pages = {023521},
	primaryclass = {hep-ph},
	reportnumber = {NPAC-11-04},
	title = {{Gravity Waves from a Cosmological Phase Transition: Gauge Artifacts and Daisy Resummations}},
	volume = {84},
	year = {2011}}

@article{Costa:2022lpy,
	archiveprefix = {arXiv},
	author = {Costa, Francesco and Khan, Sarif and Kim, Jinsu},
	doi = {10.1007/JHEP12(2022)165},
	eprint = {2209.13653},
	journal = {JHEP},
	pages = {165},
	primaryclass = {hep-ph},
	reportnumber = {CERN-TH-2022-155},
	title = {{A two-component vector WIMP \textemdash{} fermion FIMP dark matter model with an extended seesaw mechanism}},
	volume = {12},
	year = {2022}}

@article{Costa:2022oaa,
	archiveprefix = {arXiv},
	author = {Costa, Francesco and Khan, Sarif and Kim, Jinsu},
	doi = {10.1007/JHEP06(2022)026},
	eprint = {2202.13126},
	journal = {JHEP},
	pages = {026},
	primaryclass = {hep-ph},
	reportnumber = {CERN-TH-2022-022},
	title = {{A two-component dark matter model and its associated gravitational waves}},
	volume = {06},
	year = {2022}}

@article{Azatov:2022tii,
	archiveprefix = {arXiv},
	author = {Azatov, Aleksandr and Barni, Giulio and Chakraborty, Sabyasachi and Vanvlasselaer, Miguel and Yin, Wen},
	doi = {10.1007/JHEP10(2022)017},
	eprint = {2207.02230},
	journal = {JHEP},
	pages = {017},
	primaryclass = {hep-ph},
	reportnumber = {SISSA 12/2022/FISI TU-1157},
	title = {{Ultra-relativistic bubbles from the simplest Higgs portal and their cosmological consequences}},
	volume = {10},
	year = {2022}}

@article{Azatov:2021ifm,
	archiveprefix = {arXiv},
	author = {Azatov, Aleksandr and Vanvlasselaer, Miguel and Yin, Wen},
	doi = {10.1007/JHEP03(2021)288},
	eprint = {2101.05721},
	journal = {JHEP},
	pages = {288},
	primaryclass = {hep-ph},
	reportnumber = {SISSA 03/2021/FISI},
	title = {{Dark Matter production from relativistic bubble walls}},
	volume = {03},
	year = {2021}}

@article{Wang:2022akn,
	archiveprefix = {arXiv},
	author = {Wang, Wenyu and Xu, Wu-Long and Yang, Jin Min},
	doi = {10.1140/epjp/s13360-023-04412-4},
	eprint = {2209.11408},
	journal = {Eur. Phys. J. Plus},
	number = {9},
	pages = {781},
	primaryclass = {hep-ph},
	title = {{A hidden self-interacting dark matter sector with first-order cosmological phase transition and gravitational wave}},
	volume = {138},
	year = {2023}}

@article{Wang:2022lxn,
	archiveprefix = {arXiv},
	author = {Wang, Wenyu and Xie, Ke-Pan and Xu, Wu-Long and Yang, Jin Min},
	doi = {10.1140/epjc/s10052-022-11077-3},
	eprint = {2204.01928},
	journal = {Eur. Phys. J. C},
	number = {12},
	pages = {1120},
	primaryclass = {hep-ph},
	title = {{Cosmological phase transitions, gravitational waves and self-interacting dark matter in the singlet extension of MSSM}},
	volume = {82},
	year = {2022}}

@article{Dent:2022bcd,
	archiveprefix = {arXiv},
	author = {Dent, James B. and Dutta, Bhaskar and Ghosh, Sumit and Kumar, Jason and Runburg, Jack},
	doi = {10.1007/JHEP08(2022)300},
	eprint = {2203.11736},
	journal = {JHEP},
	pages = {300},
	primaryclass = {hep-ph},
	reportnumber = {MI-HET-774, KIAS-P22016},
	title = {{Sensitivity to dark sector scales from gravitational wave signatures}},
	volume = {08},
	year = {2022}}

@article{Ghosh:2020ipy,
	archiveprefix = {arXiv},
	author = {Ghosh, Tathagata and Guo, Huai-Ke and Han, Tao and Liu, Hongkai},
	doi = {10.1007/JHEP07(2021)045},
	eprint = {2012.09758},
	journal = {JHEP},
	pages = {045},
	primaryclass = {hep-ph},
	title = {{Electroweak phase transition with an SU(2) dark sector}},
	volume = {07},
	year = {2021}}

@article{Ratzinger:2020koh,
	archiveprefix = {arXiv},
	author = {Ratzinger, Wolfram and Schwaller, Pedro},
	doi = {10.21468/SciPostPhys.10.2.047},
	eprint = {2009.11875},
	journal = {SciPost Phys.},
	number = {2},
	pages = {047},
	primaryclass = {astro-ph.CO},
	reportnumber = {MITP/20-056},
	title = {{Whispers from the dark side: Confronting light new physics with NANOGrav data}},
	volume = {10},
	year = {2021}}

@article{Addazi:2020zcj,
	archiveprefix = {arXiv},
	author = {Addazi, Andrea and Cai, Yi-Fu and Gan, Qingyu and Marciano, Antonino and Zeng, Kaiqiang},
	doi = {10.1007/s11433-021-1724-6},
	eprint = {2009.10327},
	journal = {Sci. China Phys. Mech. Astron.},
	number = {9},
	pages = {290411},
	primaryclass = {hep-ph},
	title = {{NANOGrav results and dark first order phase transitions}},
	volume = {64},
	year = {2021}}

@article{Fairbairn:2019xog,
	archiveprefix = {arXiv},
	author = {Fairbairn, Malcolm and Hardy, Edward and Wickens, Alastair},
	doi = {10.1007/JHEP07(2019)044},
	eprint = {1901.11038},
	journal = {JHEP},
	pages = {044},
	primaryclass = {hep-ph},
	reportnumber = {KCL-PH-TH/2019-12},
	title = {{Hearing without seeing: gravitational waves from hot and cold hidden sectors}},
	volume = {07},
	year = {2019}}

@article{Breitbach:2018ddu,
	archiveprefix = {arXiv},
	author = {Breitbach, Moritz and Kopp, Joachim and Madge, Eric and Opferkuch, Toby and Schwaller, Pedro},
	doi = {10.1088/1475-7516/2019/07/007},
	eprint = {1811.11175},
	journal = {JCAP},
	pages = {007},
	primaryclass = {hep-ph},
	reportnumber = {CERN-TH-2018-255, MITP/18-115},
	title = {{Dark, Cold, and Noisy: Constraining Secluded Hidden Sectors with Gravitational Waves}},
	volume = {07},
	year = {2019}}

@article{Bai:2018dxf,
	archiveprefix = {arXiv},
	author = {Bai, Yang and Long, Andrew J. and Lu, Sida},
	doi = {10.1103/PhysRevD.99.055047},
	eprint = {1810.04360},
	journal = {Phys. Rev. D},
	number = {5},
	pages = {055047},
	primaryclass = {hep-ph},
	reportnumber = {FERMILAB-PUB-18-600-T},
	title = {{Dark Quark Nuggets}},
	volume = {99},
	year = {2019}}

@article{Hashino:2018zsi,
	archiveprefix = {arXiv},
	author = {Hashino, Katsuya and Kakizaki, Mitsuru and Kanemura, Shinya and Ko, Pyungwon and Matsui, Toshinori},
	doi = {10.1007/JHEP06(2018)088},
	eprint = {1802.02947},
	journal = {JHEP},
	pages = {088},
	primaryclass = {hep-ph},
	reportnumber = {UT-HET-123, OU-HET-957, KIAS-P18013},
	title = {{Gravitational waves from first order electroweak phase transition in models with the U(1)$_{X}$ gauge symmetry}},
	volume = {06},
	year = {2018}}

@article{Huang:2017rzf,
	archiveprefix = {arXiv},
	author = {Huang, Fa Peng and Yu, Jiang-Hao},
	doi = {10.1103/PhysRevD.98.095022},
	eprint = {1704.04201},
	journal = {Phys. Rev. D},
	number = {9},
	pages = {095022},
	primaryclass = {hep-ph},
	reportnumber = {ACFI-T17-06},
	title = {{Exploring inert dark matter blind spots with gravitational wave signatures}},
	volume = {98},
	year = {2018}}

@article{Tsumura:2017knk,
	archiveprefix = {arXiv},
	author = {Tsumura, Koji and Yamada, Masatoshi and Yamaguchi, Yuya},
	doi = {10.1088/1475-7516/2017/07/044},
	eprint = {1704.00219},
	journal = {JCAP},
	pages = {044},
	primaryclass = {hep-ph},
	reportnumber = {KUNS-2669, EPHOU-17-004},
	title = {{Gravitational wave from dark sector with dark pion}},
	volume = {07},
	year = {2017}}

@article{Addazi:2017gpt,
	archiveprefix = {arXiv},
	author = {Addazi, Andrea and Marciano, Antonino},
	doi = {10.1088/1674-1137/42/2/023107},
	eprint = {1703.03248},
	journal = {Chin. Phys. C},
	number = {2},
	pages = {023107},
	primaryclass = {hep-ph},
	title = {{Gravitational waves from dark first order phase transitions and dark photons}},
	volume = {42},
	year = {2018}}

@article{Soni:2016yes,
	archiveprefix = {arXiv},
	author = {Soni, Amarjit and Zhang, Yue},
	doi = {10.1016/j.physletb.2017.05.077},
	eprint = {1610.06931},
	journal = {Phys. Lett. B},
	pages = {379--384},
	primaryclass = {hep-ph},
	reportnumber = {NUHEP-TH-16-05},
	title = {{Gravitational Waves From SU(N) Glueball Dark Matter}},
	volume = {771},
	year = {2017}}

@article{Schwaller:2015tja,
	archiveprefix = {arXiv},
	author = {Schwaller, Pedro},
	doi = {10.1103/PhysRevLett.115.181101},
	eprint = {1504.07263},
	journal = {Phys. Rev. Lett.},
	number = {18},
	pages = {181101},
	primaryclass = {hep-ph},
	reportnumber = {CERN-PH-TH-2015-093},
	title = {{Gravitational Waves from a Dark Phase Transition}},
	volume = {115},
	year = {2015}}

@article{Jaeckel:2016jlh,
	archiveprefix = {arXiv},
	author = {Jaeckel, Joerg and Khoze, Valentin V. and Spannowsky, Michael},
	doi = {10.1103/PhysRevD.94.103519},
	eprint = {1602.03901},
	journal = {Phys. Rev. D},
	number = {10},
	pages = {103519},
	primaryclass = {hep-ph},
	reportnumber = {IPPP-16-12, DCPT-16-24},
	title = {{Hearing the signal of dark sectors with gravitational wave detectors}},
	volume = {94},
	year = {2016}}

@article{Witten:1984rs,
	author = {Witten, Edward},
	doi = {10.1103/PhysRevD.30.272},
	journal = {Phys. Rev. D},
	pages = {272--285},
	reportnumber = {PRINT-84-0400 (IAS,PRINCETON)},
	title = {{Cosmic Separation of Phases}},
	volume = {30},
	year = {1984}}

@article{deSalas:2016ztq,
	archiveprefix = {arXiv},
	author = {de Salas, Pablo F. and Pastor, Sergio},
	doi = {10.1088/1475-7516/2016/07/051},
	eprint = {1606.06986},
	journal = {JCAP},
	pages = {051},
	primaryclass = {hep-ph},
	reportnumber = {IFIC-16-10, TTK-16-23},
	title = {{Relic neutrino decoupling with flavour oscillations revisited}},
	volume = {07},
	year = {2016}}

@article{Archidiacono:2019wdp,
	archiveprefix = {arXiv},
	author = {Archidiacono, Maria and Hooper, Deanna C. and Murgia, Riccardo and Bohr, Sebastian and Lesgourgues, Julien and Viel, Matteo},
	doi = {10.1088/1475-7516/2019/10/055},
	eprint = {1907.01496},
	journal = {JCAP},
	pages = {055},
	primaryclass = {astro-ph.CO},
	title = {{Constraining Dark Matter-Dark Radiation interactions with CMB, BAO, and Lyman-$\alpha$}},
	volume = {10},
	year = {2019}}

@article{Khrapak:2003kjw,
	author = {Khrapak, S. A. and Ivlev, A. V. and Morfill, G. E. and Zhdanov, S. K.},
	doi = {10.1103/PhysRevLett.90.225002},
	journal = {Phys. Rev. Lett.},
	number = {22},
	pages = {225002},
	title = {{Scattering in the Attractive Yukawa Potential in the Limit of Strong Interaction}},
	volume = {90},
	year = {2003}}

@article{Kahlhoefer:2017umn,
	archiveprefix = {arXiv},
	author = {Kahlhoefer, Felix and Schmidt-Hoberg, Kai and Wild, Sebastian},
	doi = {10.1088/1475-7516/2017/08/003},
	eprint = {1704.02149},
	journal = {JCAP},
	pages = {003},
	primaryclass = {hep-ph},
	reportnumber = {DESY-17-052},
	title = {{Dark matter self-interactions from a general spin-0 mediator}},
	volume = {08},
	year = {2017}}

@article{Hufnagel:2018bjp,
	archiveprefix = {arXiv},
	author = {Hufnagel, Marco and Schmidt-Hoberg, Kai and Wild, Sebastian},
	doi = {10.1088/1475-7516/2018/11/032},
	eprint = {1808.09324},
	journal = {JCAP},
	pages = {032},
	primaryclass = {hep-ph},
	reportnumber = {DESY 18-133, DESY-18-133},
	title = {{BBN constraints on MeV-scale dark sectors. Part II. Electromagnetic decays}},
	volume = {11},
	year = {2018}}

@article{Ibe:2021fed,
	archiveprefix = {arXiv},
	author = {Ibe, Masahiro and Kobayashi, Shin and Nakayama, Yuhei and Shirai, Satoshi},
	doi = {10.1007/JHEP03(2022)198},
	eprint = {2112.11096},
	journal = {JHEP},
	pages = {198},
	primaryclass = {hep-ph},
	reportnumber = {IPMU21-0087},
	title = {{Cosmological constraints on dark scalar}},
	volume = {03},
	year = {2022}}

@article{Chen:2023rrl,
	archiveprefix = {arXiv},
	author = {Chen, Zien and Ye, Kairui and Zhang, Mengchao},
	doi = {10.1103/PhysRevD.107.095027},
	eprint = {2303.11820},
	journal = {Phys. Rev. D},
	number = {9},
	pages = {095027},
	primaryclass = {hep-ph},
	title = {{Asymmetric dark matter with a spontaneously broken U(1)': Self-interaction and gravitational waves}},
	volume = {107},
	year = {2023}}

@article{Tulin:2013teo,
	archiveprefix = {arXiv},
	author = {Tulin, Sean and Yu, Hai-Bo and Zurek, Kathryn M.},
	doi = {10.1103/PhysRevD.87.115007},
	eprint = {1302.3898},
	journal = {Phys. Rev. D},
	number = {11},
	pages = {115007},
	primaryclass = {hep-ph},
	title = {{Beyond Collisionless Dark Matter: Particle Physics Dynamics for Dark Matter Halo Structure}},
	volume = {87},
	year = {2013}}

@article{Loeb:2010gj,
	archiveprefix = {arXiv},
	author = {Loeb, Abraham and Weiner, Neal},
	doi = {10.1103/PhysRevLett.106.171302},
	eprint = {1011.6374},
	journal = {Phys. Rev. Lett.},
	pages = {171302},
	primaryclass = {astro-ph.CO},
	title = {{Cores in Dwarf Galaxies from Dark Matter with a Yukawa Potential}},
	volume = {106},
	year = {2011}}

@article{Buckley:2009in,
	archiveprefix = {arXiv},
	author = {Buckley, Matthew R. and Fox, Patrick J.},
	doi = {10.1103/PhysRevD.81.083522},
	eprint = {0911.3898},
	journal = {Phys. Rev. D},
	pages = {083522},
	primaryclass = {hep-ph},
	reportnumber = {FERMILAB-PUB-09-560-T},
	title = {{Dark Matter Self-Interactions and Light Force Carriers}},
	volume = {81},
	year = {2010}}

@article{Foot:2014uba,
	archiveprefix = {arXiv},
	author = {Foot, R. and Vagnozzi, S.},
	doi = {10.1103/PhysRevD.91.023512},
	eprint = {1409.7174},
	journal = {Phys. Rev. D},
	pages = {023512},
	primaryclass = {hep-ph},
	title = {{Dissipative hidden sector dark matter}},
	volume = {91},
	year = {2015}}

@article{Foot:2014osa,
	archiveprefix = {arXiv},
	author = {Foot, R. and Vagnozzi, S.},
	doi = {10.1016/j.physletb.2015.06.063},
	eprint = {1412.0762},
	journal = {Phys. Lett. B},
	pages = {61--66},
	primaryclass = {hep-ph},
	title = {{Diurnal modulation signal from dissipative hidden sector dark matter}},
	volume = {748},
	year = {2015}}

@article{Bellazzini:2013foa,
	archiveprefix = {arXiv},
	author = {Bellazzini, Brando and Cliche, Mathieu and Tanedo, Philip},
	doi = {10.1103/PhysRevD.88.083506},
	eprint = {1307.1129},
	journal = {Phys. Rev. D},
	number = {8},
	pages = {083506},
	primaryclass = {hep-ph},
	title = {{Effective theory of self-interacting dark matter}},
	volume = {88},
	year = {2013}}

@article{Feng:2009hw,
	archiveprefix = {arXiv},
	author = {Feng, Jonathan L. and Kaplinghat, Manoj and Yu, Hai-Bo},
	doi = {10.1103/PhysRevLett.104.151301},
	eprint = {0911.0422},
	journal = {Phys. Rev. Lett.},
	pages = {151301},
	primaryclass = {hep-ph},
	reportnumber = {UCI-TR-2009-12},
	title = {{Halo Shape and Relic Density Exclusions of Sommerfeld-Enhanced Dark Matter Explanations of Cosmic Ray Excesses}},
	volume = {104},
	year = {2010}}

@article{Feng:2009mn,
	archiveprefix = {arXiv},
	author = {Feng, Jonathan L. and Kaplinghat, Manoj and Tu, Huitzu and Yu, Hai-Bo},
	doi = {10.1088/1475-7516/2009/07/004},
	eprint = {0905.3039},
	journal = {JCAP},
	pages = {004},
	primaryclass = {hep-ph},
	reportnumber = {UCI-TR-2009-06},
	title = {{Hidden Charged Dark Matter}},
	volume = {07},
	year = {2009}}

@article{Wang:2016lvj,
	archiveprefix = {arXiv},
	author = {Wang, Wenyu and Zhang, Mengchao and Zhao, Jun},
	doi = {10.1142/S0217751X18410026},
	eprint = {1604.00123},
	journal = {Int. J. Mod. Phys. A},
	number = {11},
	pages = {1841002},
	primaryclass = {hep-ph},
	title = {{Higgs exotic decays in general NMSSM with self-interacting dark matter}},
	volume = {33},
	year = {2018}}

@article{Ma:2017ucp,
	archiveprefix = {arXiv},
	author = {Ma, Ernest},
	doi = {10.1016/j.physletb.2017.06.067},
	eprint = {1704.04666},
	journal = {Phys. Lett. B},
	pages = {442--445},
	primaryclass = {hep-ph},
	reportnumber = {UCRHEP-T576, UCRHEP-T576-(APR-2017)},
	title = {{Inception of Self-Interacting Dark Matter with Dark Charge Conjugation Symmetry}},
	volume = {772},
	year = {2017}}

@article{vandenAarssen:2012vpm,
	archiveprefix = {arXiv},
	author = {van den Aarssen, Laura G. and Bringmann, Torsten and Pfrommer, Christoph},
	doi = {10.1103/PhysRevLett.109.231301},
	eprint = {1205.5809},
	journal = {Phys. Rev. Lett.},
	pages = {231301},
	primaryclass = {astro-ph.CO},
	title = {{Is dark matter with long-range interactions a solution to all small-scale problems of \textbackslash{}Lambda CDM cosmology?}},
	volume = {109},
	year = {2012}}

@article{Kamada:2020buc,
	archiveprefix = {arXiv},
	author = {Kamada, Ayuki and Kim, Hee Jung and Kuwahara, Takumi},
	doi = {10.1007/JHEP12(2020)202},
	eprint = {2007.15522},
	journal = {JHEP},
	pages = {202},
	primaryclass = {hep-ph},
	title = {{Maximally self-interacting dark matter: models and predictions}},
	volume = {12},
	year = {2020}}

@article{Ko:2014bka,
	archiveprefix = {arXiv},
	author = {Ko, P. and Tang, Yong},
	doi = {10.1016/j.physletb.2014.10.035},
	eprint = {1404.0236},
	journal = {Phys. Lett. B},
	pages = {62--67},
	primaryclass = {hep-ph},
	title = {{\ensuremath{\nu}\ensuremath{\Lambda}MDM: A model for sterile neutrino and dark matter reconciles cosmological and neutrino oscillation data after BICEP2}},
	volume = {739},
	year = {2014}}

@article{Kitahara:2016zyb,
	archiveprefix = {arXiv},
	author = {Kitahara, Teppei and Yamamoto, Yasuhiro},
	doi = {10.1103/PhysRevD.95.015008},
	eprint = {1609.01605},
	journal = {Phys. Rev. D},
	number = {1},
	pages = {015008},
	primaryclass = {hep-ph},
	reportnumber = {OU-HET-909, TTP16-036},
	title = {{Protophobic Light Vector Boson as a Mediator to the Dark Sector}},
	volume = {95},
	year = {2017}}

@article{Tulin:2012wi,
	archiveprefix = {arXiv},
	author = {Tulin, Sean and Yu, Hai-Bo and Zurek, Kathryn M.},
	doi = {10.1103/PhysRevLett.110.111301},
	eprint = {1210.0900},
	journal = {Phys. Rev. Lett.},
	number = {11},
	pages = {111301},
	primaryclass = {hep-ph},
	reportnumber = {MCTP-12-27},
	title = {{Resonant Dark Forces and Small Scale Structure}},
	volume = {110},
	year = {2013}}

@article{Kainulainen:2015sva,
	archiveprefix = {arXiv},
	author = {Kainulainen, Kimmo and Tuominen, Kimmo and Vaskonen, Ville},
	doi = {10.1103/PhysRevD.93.015016},
	eprint = {1507.04931},
	journal = {Phys. Rev. D},
	note = {[Erratum: Phys.Rev.D 95, 079901 (2017)]},
	number = {1},
	pages = {015016},
	primaryclass = {hep-ph},
	title = {{Self-interacting dark matter and cosmology of a light scalar mediator}},
	volume = {93},
	year = {2016}}

@article{Kamada:2018zxi,
	archiveprefix = {arXiv},
	author = {Kamada, Ayuki and Kaneta, Kunio and Yanagi, Keisuke and Yu, Hai-Bo},
	doi = {10.1007/JHEP06(2018)117},
	eprint = {1805.00651},
	journal = {JHEP},
	pages = {117},
	primaryclass = {hep-ph},
	reportnumber = {CTPU-PTC-18-10, UMN-TH-3716/18, FTPI-MINN-18/07, UT-18-10, UMN-TH-3716-18, FTPI-MINN-18-07},
	title = {{Self-interacting dark matter and muon $g-2$ in a gauged U$(1)_{L_{\mu} - L_{\tau}}$ model}},
	volume = {06},
	year = {2018}}

@article{Duerr:2018mbd,
	archiveprefix = {arXiv},
	author = {Duerr, Michael and Schmidt-Hoberg, Kai and Wild, Sebastian},
	doi = {10.1088/1475-7516/2018/09/033},
	eprint = {1804.10385},
	journal = {JCAP},
	pages = {033},
	primaryclass = {hep-ph},
	reportnumber = {DESY 18-051, DESY-18-051},
	title = {{Self-interacting dark matter with a stable vector mediator}},
	volume = {09},
	year = {2018}}

@article{Kamada:2018kmi,
	archiveprefix = {arXiv},
	author = {Kamada, Ayuki and Yamada, Masaki and Yanagida, Tsutomu T.},
	doi = {10.1007/JHEP03(2019)021},
	eprint = {1811.02567},
	journal = {JHEP},
	pages = {021},
	primaryclass = {hep-ph},
	reportnumber = {IPMU18-0179},
	title = {{Self-interacting dark matter with a vector mediator: kinetic mixing with the $ \mathrm{U}{(1)}_{{\left(B-L\right)}_3} $ gauge boson}},
	volume = {03},
	year = {2019}}

@article{Aboubrahim:2020lnr,
	archiveprefix = {arXiv},
	author = {Aboubrahim, Amin and Feng, Wan-Zhe and Nath, Pran and Wang, Zhu-Yao},
	doi = {10.1103/PhysRevD.103.075014},
	eprint = {2008.00529},
	journal = {Phys. Rev. D},
	number = {7},
	pages = {075014},
	primaryclass = {hep-ph},
	title = {{Self-interacting hidden sector dark matter, small scale galaxy structure anomalies, and a dark force}},
	volume = {103},
	year = {2021}}

@article{Ko:2014nha,
	archiveprefix = {arXiv},
	author = {Ko, P. and Tang, Y.},
	doi = {10.1088/1475-7516/2014/05/047},
	eprint = {1402.6449},
	journal = {JCAP},
	pages = {047},
	primaryclass = {hep-ph},
	reportnumber = {KIAS-P14010},
	title = {{Self-interacting scalar dark matter with local $Z_3$ symmetry}},
	volume = {05},
	year = {2014}}

@article{Schutz:2014nka,
	archiveprefix = {arXiv},
	author = {Schutz, Katelin and Slatyer, Tracy R.},
	doi = {10.1088/1475-7516/2015/01/021},
	eprint = {1409.2867},
	journal = {JCAP},
	pages = {021},
	primaryclass = {hep-ph},
	reportnumber = {MIT-CTP-4585},
	title = {{Self-Scattering for Dark Matter with an Excited State}},
	volume = {01},
	year = {2015}}

@article{Foot:2016wvj,
	archiveprefix = {arXiv},
	author = {Foot, Robert and Vagnozzi, Sunny},
	doi = {10.1088/1475-7516/2016/07/013},
	eprint = {1602.02467},
	journal = {JCAP},
	pages = {013},
	primaryclass = {astro-ph.CO},
	title = {{Solving the small-scale structure puzzles with dissipative dark matter}},
	volume = {07},
	year = {2016}}

@article{Boddy:2014qxa,
	archiveprefix = {arXiv},
	author = {Boddy, Kimberly K. and Feng, Jonathan L. and Kaplinghat, Manoj and Shadmi, Yael and Tait, Timothy M. P.},
	doi = {10.1103/PhysRevD.90.095016},
	eprint = {1408.6532},
	journal = {Phys. Rev. D},
	number = {9},
	pages = {095016},
	primaryclass = {hep-ph},
	reportnumber = {UCI-TR-2014-06, CERN-PH-TH-2014-162, CALT-TH-2014-148},
	title = {{Strongly interacting dark matter: Self-interactions and keV lines}},
	volume = {90},
	year = {2014}}

@article{Bringmann:2013vra,
	archiveprefix = {arXiv},
	author = {Bringmann, Torsten and Hasenkamp, Jasper and Kersten, J\"orn},
	doi = {10.1088/1475-7516/2014/07/042},
	eprint = {1312.4947},
	journal = {JCAP},
	pages = {042},
	primaryclass = {hep-ph},
	title = {{Tight bonds between sterile neutrinos and dark matter}},
	volume = {07},
	year = {2014}}

@article{Kamada:2019jch,
	archiveprefix = {arXiv},
	author = {Kamada, Ayuki and Yamada, Masaki and Yanagida, Tsutomu T.},
	doi = {10.1103/PhysRevD.102.015012},
	eprint = {1908.00207},
	journal = {Phys. Rev. D},
	number = {1},
	pages = {015012},
	primaryclass = {hep-ph},
	title = {{Unification for darkly charged dark matter}},
	volume = {102},
	year = {2020}}

@article{Kamada:2019gpp,
	archiveprefix = {arXiv},
	author = {Kamada, Ayuki and Yamada, Masaki and Yanagida, Tsutomu T.},
	doi = {10.1007/JHEP07(2019)180},
	eprint = {1905.04245},
	journal = {JHEP},
	pages = {180},
	primaryclass = {hep-ph},
	reportnumber = {CTPU-PTC-19-11},
	title = {{Unification of the standard model and dark matter sectors in [SU(5)$\times$U(1)]$^4$}},
	volume = {07},
	year = {2019}}

@article{Kang:2015aqa,
	archiveprefix = {arXiv},
	author = {Kang, Zhaofeng},
	doi = {10.1016/j.physletb.2015.10.031},
	eprint = {1505.06554},
	journal = {Phys. Lett. B},
	pages = {201--204},
	primaryclass = {hep-ph},
	title = {{View FImP miracle (by scale invariance) \`a la self-interaction}},
	volume = {751},
	year = {2015}}

@article{Tulin:2017ara,
	archiveprefix = {arXiv},
	author = {Tulin, Sean and Yu, Hai-Bo},
	doi = {10.1016/j.physrep.2017.11.004},
	eprint = {1705.02358},
	journal = {Phys. Rept.},
	pages = {1--57},
	primaryclass = {hep-ph},
	title = {{Dark Matter Self-interactions and Small Scale Structure}},
	volume = {730},
	year = {2018}}

@article{Chang:2016ntp,
	archiveprefix = {arXiv},
	author = {Chang, Jae Hyeok and Essig, Rouven and McDermott, Samuel D.},
	doi = {10.1007/JHEP01(2017)107},
	eprint = {1611.03864},
	journal = {JHEP},
	pages = {107},
	primaryclass = {hep-ph},
	reportnumber = {YITP-SB-16-44},
	title = {{Revisiting Supernova 1987A Constraints on Dark Photons}},
	volume = {01},
	year = {2017}}

@article{Boylan-Kolchin:2011lmk,
	archiveprefix = {arXiv},
	author = {Boylan-Kolchin, Michael and Bullock, James S. and Kaplinghat, Manoj},
	doi = {10.1111/j.1365-2966.2012.20695.x},
	eprint = {1111.2048},
	journal = {Mon. Not. Roy. Astron. Soc.},
	pages = {1203--1218},
	primaryclass = {astro-ph.CO},
	title = {{The Milky Way's bright satellites as an apparent failure of LCDM}},
	volume = {422},
	year = {2012}}

@article{Boylan-Kolchin:2011qkt,
	archiveprefix = {arXiv},
	author = {Boylan-Kolchin, Michael and Bullock, James S. and Kaplinghat, Manoj},
	doi = {10.1111/j.1745-3933.2011.01074.x},
	eprint = {1103.0007},
	journal = {Mon. Not. Roy. Astron. Soc.},
	pages = {L40},
	primaryclass = {astro-ph.CO},
	title = {{Too big to fail? The puzzling darkness of massive Milky Way subhaloes}},
	volume = {415},
	year = {2011}}

@article{Kochanek:2000pi,
	archiveprefix = {arXiv},
	author = {Kochanek, C. S. and White, Martin J.},
	doi = {10.1086/317149},
	eprint = {astro-ph/0003483},
	journal = {Astrophys. J.},
	pages = {514},
	title = {{A Quantitative study of interacting dark matter in halos}},
	volume = {543},
	year = {2000}}

@article{Andrade:2020lqq,
	archiveprefix = {arXiv},
	author = {Andrade, Kevin E. and Fuson, Jackson and Gad-Nasr, Sophia and Kong, Demao and Minor, Quinn and Roberts, M. Grant and Kaplinghat, Manoj},
	doi = {10.1093/mnras/stab3241},
	eprint = {2012.06611},
	journal = {Mon. Not. Roy. Astron. Soc.},
	number = {1},
	pages = {54--81},
	primaryclass = {astro-ph.CO},
	title = {{A stringent upper limit on dark matter self-interaction cross-section from cluster strong lensing}},
	volume = {510},
	year = {2021}}

@article{Elbert:2016dbb,
	archiveprefix = {arXiv},
	author = {Elbert, Oliver D. and Bullock, James S. and Kaplinghat, Manoj and Garrison-Kimmel, Shea and Graus, Andrew S. and Rocha, Miguel},
	doi = {10.3847/1538-4357/aa9710},
	eprint = {1609.08626},
	journal = {Astrophys. J.},
	number = {2},
	pages = {109},
	primaryclass = {astro-ph.GA},
	title = {{A Testable Conspiracy: Simulating Baryonic Effects on Self-Interacting Dark Matter Halos}},
	volume = {853},
	year = {2018}}

@article{Fry:2015rta,
	archiveprefix = {arXiv},
	author = {Fry, A. Bastidas and Governato, F. and Pontzen, A. and Quinn, T. and Tremmel, M. and Anderson, L. and Menon, H. and Brooks, A. M. and Wadsley, J.},
	doi = {10.1093/mnras/stv1330},
	eprint = {1501.00497},
	journal = {Mon. Not. Roy. Astron. Soc.},
	number = {2},
	pages = {1468--1479},
	primaryclass = {astro-ph.CO},
	title = {{All about baryons: revisiting SIDM predictions at small halo masses}},
	volume = {452},
	year = {2015}}

@article{Yoshida:2000bx,
	archiveprefix = {arXiv},
	author = {Yoshida, Naoki and Springel, Volker and White, Simon D. M. and Tormen, Giuseppe},
	doi = {10.1086/312707},
	eprint = {astro-ph/0002362},
	journal = {Astrophys. J. Lett.},
	pages = {L103},
	title = {{Collisional dark matter and the structure of dark halos}},
	volume = {535},
	year = {2000}}

@article{Moore:2000fp,
	archiveprefix = {arXiv},
	author = {Moore, Ben and Gelato, Sergio and Jenkins, Adrian and Pearce, F. R. and Quilis, Vicent},
	doi = {10.1086/312692},
	eprint = {astro-ph/0002308},
	journal = {Astrophys. J. Lett.},
	pages = {L21--L24},
	title = {{Collisional versus collisionless dark matter}},
	volume = {535},
	year = {2000}}

@article{Zavala:2012us,
	archiveprefix = {arXiv},
	author = {Zavala, Jesus and Vogelsberger, Mark and Walker, Matthew G.},
	doi = {10.1093/mnrasl/sls053},
	eprint = {1211.6426},
	journal = {Mon. Not. Roy. Astron. Soc.},
	pages = {L20--L24},
	primaryclass = {astro-ph.CO},
	title = {{Constraining Self-Interacting Dark Matter with the Milky Way's dwarf spheroidals}},
	volume = {431},
	year = {2013}}

@article{Elbert:2014bma,
	archiveprefix = {arXiv},
	author = {Elbert, Oliver D. and Bullock, James S. and Garrison-Kimmel, Shea and Rocha, Miguel and O\~norbe, Jose and Peter, Annika H. G.},
	doi = {10.1093/mnras/stv1470},
	eprint = {1412.1477},
	journal = {Mon. Not. Roy. Astron. Soc.},
	number = {1},
	pages = {29--37},
	primaryclass = {astro-ph.GA},
	title = {{Core formation in dwarf haloes with self-interacting dark matter: no fine-tuning necessary}},
	volume = {453},
	year = {2015}}

@article{Peter:2012jh,
	archiveprefix = {arXiv},
	author = {Peter, Annika H. G. and Rocha, Miguel and Bullock, James S. and Kaplinghat, Manoj},
	doi = {10.1093/mnras/sts535},
	eprint = {1208.3026},
	journal = {Mon. Not. Roy. Astron. Soc.},
	pages = {105},
	primaryclass = {astro-ph.CO},
	reportnumber = {NSF-KITP-12-147},
	title = {{Cosmological Simulations with Self-Interacting Dark Matter II: Halo Shapes vs. Observations}},
	volume = {430},
	year = {2013}}

@article{Kaplinghat:2015aga,
	archiveprefix = {arXiv},
	author = {Kaplinghat, Manoj and Tulin, Sean and Yu, Hai-Bo},
	doi = {10.1103/PhysRevLett.116.041302},
	eprint = {1508.03339},
	journal = {Phys. Rev. Lett.},
	number = {4},
	pages = {041302},
	primaryclass = {astro-ph.CO},
	title = {{Dark Matter Halos as Particle Colliders: Unified Solution to Small-Scale Structure Puzzles from Dwarfs to Clusters}},
	volume = {116},
	year = {2016}}

@article{Vogelsberger:2014pda,
	archiveprefix = {arXiv},
	author = {Vogelsberger, Mark and Zavala, Jesus and Simpson, Christine and Jenkins, Adrian},
	doi = {10.1093/mnras/stu1713},
	eprint = {1405.5216},
	journal = {Mon. Not. Roy. Astron. Soc.},
	number = {4},
	pages = {3684--3698},
	primaryclass = {astro-ph.CO},
	title = {{Dwarf galaxies in CDM and SIDM with baryons: observational probes of the nature of dark matter}},
	volume = {444},
	year = {2014}}

@article{Dooley:2016ajo,
	archiveprefix = {arXiv},
	author = {Dooley, Gregory A. and Peter, Annika H. G. and Vogelsberger, Mark and Zavala, Jes\'us and Frebel, Anna},
	doi = {10.1093/mnras/stw1309},
	eprint = {1603.08919},
	journal = {Mon. Not. Roy. Astron. Soc.},
	number = {1},
	pages = {710--727},
	primaryclass = {astro-ph.GA},
	title = {{Enhanced Tidal Stripping of Satellites in the Galactic Halo from Dark Matter Self-Interactions}},
	volume = {461},
	year = {2016}}

@article{Dave:2000ar,
	archiveprefix = {arXiv},
	author = {Dave, Romeel and Spergel, David N. and Steinhardt, Paul J. and Wandelt, Benjamin D.},
	doi = {10.1086/318417},
	eprint = {astro-ph/0006218},
	journal = {Astrophys. J.},
	pages = {574--589},
	title = {{Halo properties in cosmological simulations of selfinteracting cold dark matter}},
	volume = {547},
	year = {2001}}

@article{Robertson:2018anx,
	archiveprefix = {arXiv},
	author = {Robertson, Andrew and Harvey, David and Massey, Richard and Eke, Vincent and McCarthy, Ian G. and Jauzac, Mathilde and Li, Baojiu and Schaye, Joop},
	doi = {10.1093/mnras/stz1815},
	eprint = {1810.05649},
	journal = {Mon. Not. Roy. Astron. Soc.},
	number = {3},
	pages = {3646--3662},
	primaryclass = {astro-ph.CO},
	title = {{Observable tests of self-interacting dark matter in galaxy clusters: cosmological simulations with SIDM and baryons}},
	volume = {488},
	year = {2019}}

@article{Spergel:1999mh,
	archiveprefix = {arXiv},
	author = {Spergel, David N. and Steinhardt, Paul J.},
	doi = {10.1103/PhysRevLett.84.3760},
	eprint = {astro-ph/9909386},
	journal = {Phys. Rev. Lett.},
	pages = {3760--3763},
	title = {{Observational evidence for selfinteracting cold dark matter}},
	volume = {84},
	year = {2000}}

@article{Colin:2002nk,
	archiveprefix = {arXiv},
	author = {Colin, Pedro and Avila-Reese, Vladimir and Valenzuela, Octavio and Firmani, Claudio},
	doi = {10.1086/344259},
	eprint = {astro-ph/0205322},
	journal = {Astrophys. J.},
	pages = {777--793},
	title = {{Structure and subhalo population of halos in a selfinteracting dark matter cosmology}},
	volume = {581},
	year = {2002}}

@article{Harvey:2015hha,
	archiveprefix = {arXiv},
	author = {Harvey, David and Massey, Richard and Kitching, Thomas and Taylor, Andy and Tittley, Eric},
	doi = {10.1126/science.1261381},
	eprint = {1503.07675},
	journal = {Science},
	pages = {1462--1465},
	primaryclass = {astro-ph.CO},
	title = {{The non-gravitational interactions of dark matter in colliding galaxy clusters}},
	volume = {347},
	year = {2015}}

@article{Burkert:2000di,
	archiveprefix = {arXiv},
	author = {Burkert, Andreas},
	doi = {10.1086/312674},
	eprint = {astro-ph/0002409},
	journal = {Astrophys. J. Lett.},
	pages = {L143--L146},
	title = {{The Structure and evolution of weakly selfinteracting cold dark matter halos}},
	volume = {534},
	year = {2000}}

@article{Sagunski:2020spe,
	archiveprefix = {arXiv},
	author = {Sagunski, Laura and Gad-Nasr, Sophia and Colquhoun, Brian and Robertson, Andrew and Tulin, Sean},
	doi = {10.1088/1475-7516/2021/01/024},
	eprint = {2006.12515},
	journal = {JCAP},
	pages = {024},
	primaryclass = {astro-ph.CO},
	reportnumber = {TTK-20-16},
	title = {{Velocity-dependent Self-interacting Dark Matter from Groups and Clusters of Galaxies}},
	volume = {01},
	year = {2021}}

@article{Yoshida:2000uw,
	archiveprefix = {arXiv},
	author = {Yoshida, Naoki and Springel, Volker and White, Simon D. M. and Tormen, Giuseppe},
	doi = {10.1086/317306},
	eprint = {astro-ph/0006134},
	journal = {Astrophys. J. Lett.},
	pages = {L87--L90},
	title = {{Weakly self-interacting dark matter and the structure of dark halos}},
	volume = {544},
	year = {2000}}

@article{Nadler:2023nrd,
	archiveprefix = {arXiv},
	author = {Nadler, Ethan O. and Yang, Daneng and Yu, Hai-Bo},
	doi = {10.3847/2041-8213/ad0e09},
	eprint = {2306.01830},
	journal = {Astrophys. J. Lett.},
	number = {2},
	pages = {L39},
	primaryclass = {astro-ph.GA},
	title = {{A Self-interacting Dark Matter Solution to the Extreme Diversity of Low-mass Halo Properties}},
	volume = {958},
	year = {2023}}

@article{Minor:2020hic,
	archiveprefix = {arXiv},
	author = {Minor, Quinn E. and Gad-Nasr, Sophia and Kaplinghat, Manoj and Vegetti, Simona},
	doi = {10.1093/mnras/stab2247},
	eprint = {2011.10627},
	journal = {Mon. Not. Roy. Astron. Soc.},
	number = {2},
	pages = {1662--1683},
	primaryclass = {astro-ph.GA},
	title = {{An unexpected high concentration for the dark substructure in the gravitational lens SDSSJ0946+1006}},
	volume = {507},
	year = {2021}}

@article{Correa:2020qam,
	archiveprefix = {arXiv},
	author = {Correa, Camila A.},
	doi = {10.1093/mnras/stab506},
	eprint = {2007.02958},
	journal = {Mon. Not. Roy. Astron. Soc.},
	number = {1},
	pages = {920--937},
	primaryclass = {astro-ph.GA},
	title = {{Constraining velocity-dependent self-interacting dark matter with the Milky Way\textquoteright{}s dwarf spheroidal galaxies}},
	volume = {503},
	year = {2021}}

@article{Kaplinghat:2019dhn,
	archiveprefix = {arXiv},
	author = {Kaplinghat, Manoj and Ren, Tao and Yu, Hai-Bo},
	doi = {10.1088/1475-7516/2020/06/027},
	eprint = {1911.00544},
	journal = {JCAP},
	pages = {027},
	primaryclass = {astro-ph.GA},
	title = {{Dark Matter Cores and Cusps in Spiral Galaxies and their Explanations}},
	volume = {06},
	year = {2020}}

@article{Kamada:2016euw,
	archiveprefix = {arXiv},
	author = {Kamada, Ayuki and Kaplinghat, Manoj and Pace, Andrew B. and Yu, Hai-Bo},
	doi = {10.1103/PhysRevLett.119.111102},
	eprint = {1611.02716},
	journal = {Phys. Rev. Lett.},
	number = {11},
	pages = {111102},
	primaryclass = {astro-ph.GA},
	title = {{How the Self-Interacting Dark Matter Model Explains the Diverse Galactic Rotation Curves}},
	volume = {119},
	year = {2017}}

@article{Ghosh:2023ilw,
	archiveprefix = {arXiv},
	author = {Ghosh, Dilip Kumar and Ghosh, Purusottam and Jeesun, Sk and Srivastava, Rahul},
	eprint = {2312.16304},
	month = {12},
	primaryclass = {hep-ph},
	title = {{Hubble Tension and Cosmological Imprints of $U(1)_X$ Gauge Symmetry: $U(1)_{B_3-3 L_i}$ as a case study}},
	year = {2023}}

@article{Coleman:1973jx,
	author = {Coleman, Sidney R. and Weinberg, Erick J.},
	doi = {10.1103/PhysRevD.7.1888},
	journal = {Phys. Rev. D},
	pages = {1888--1910},
	title = {{Radiative Corrections as the Origin of Spontaneous Symmetry Breaking}},
	volume = {7},
	year = {1973}}

@article{Ren:2018jpt,
	archiveprefix = {arXiv},
	author = {Ren, Tao and Kwa, Anna and Kaplinghat, Manoj and Yu, Hai-Bo},
	doi = {10.1103/PhysRevX.9.031020},
	eprint = {1808.05695},
	journal = {Phys. Rev. X},
	number = {3},
	pages = {031020},
	primaryclass = {astro-ph.GA},
	title = {{Reconciling the Diversity and Uniformity of Galactic Rotation Curves with Self-Interacting Dark Matter}},
	volume = {9},
	year = {2019}}

@article{Colquhoun:2020adl,
	archiveprefix = {arXiv},
	author = {Colquhoun, Brian and Heeba, Saniya and Kahlhoefer, Felix and Sagunski, Laura and Tulin, Sean},
	doi = {10.1103/PhysRevD.103.035006},
	eprint = {2011.04679},
	journal = {Phys. Rev. D},
	number = {3},
	pages = {035006},
	primaryclass = {hep-ph},
	reportnumber = {TTK-20-39},
	title = {{Semiclassical regime for dark matter self-interactions}},
	volume = {103},
	year = {2021}}

@article{Yang:2022mxl,
	archiveprefix = {arXiv},
	author = {Yang, Daneng and Nadler, Ethan O. and Yu, Hai-Bo},
	doi = {10.3847/1538-4357/acc73e},
	eprint = {2211.13768},
	journal = {Astrophys. J.},
	number = {2},
	pages = {67},
	primaryclass = {astro-ph.GA},
	title = {{Strong Dark Matter Self-interactions Diversify Halo Populations within and surrounding the Milky Way}},
	volume = {949},
	year = {2023}}

@article{Husdal:2016haj,
	archiveprefix = {arXiv},
	author = {Husdal, Lars},
	doi = {10.3390/galaxies4040078},
	eprint = {1609.04979},
	journal = {Galaxies},
	number = {4},
	pages = {78},
	primaryclass = {astro-ph.CO},
	title = {{On Effective Degrees of Freedom in the Early Universe}},
	volume = {4},
	year = {2016}}

@article{Bringmann:2023opz,
	archiveprefix = {arXiv},
	author = {Bringmann, Torsten and Depta, Paul Frederik and Konstandin, Thomas and Schmidt-Hoberg, Kai and Tasillo, Carlo},
	doi = {10.1088/1475-7516/2023/11/053},
	eprint = {2306.09411},
	journal = {JCAP},
	pages = {053},
	primaryclass = {astro-ph.CO},
	reportnumber = {DESY-23-077},
	title = {{Does NANOGrav observe a dark sector phase transition?}},
	volume = {11},
	year = {2023}}

@article{Dutta:2022knf,
	archiveprefix = {arXiv},
	author = {Dutta, Manoranjan and Narendra, Nimmala and Sahu, Narendra and Shil, Sujay},
	doi = {10.1103/PhysRevD.106.095017},
	eprint = {2202.04704},
	journal = {Phys. Rev. D},
	number = {9},
	pages = {095017},
	primaryclass = {hep-ph},
	title = {{Asymmetric self-interacting dark matter via Dirac leptogenesis}},
	volume = {106},
	year = {2022}}

@article{Falkowski:2011xh,
	archiveprefix = {arXiv},
	author = {Falkowski, Adam and Ruderman, Joshua T. and Volansky, Tomer},
	doi = {10.1007/JHEP05(2011)106},
	eprint = {1101.4936},
	journal = {JHEP},
	pages = {106},
	primaryclass = {hep-ph},
	reportnumber = {LPT-ORSAY-11-09},
	title = {{Asymmetric Dark Matter from Leptogenesis}},
	volume = {05},
	year = {2011}}

@article{An:2009vq,
	archiveprefix = {arXiv},
	author = {An, Haipeng and Chen, Shao-Long and Mohapatra, Rabindra N. and Zhang, Yue},
	doi = {10.1007/JHEP03(2010)124},
	eprint = {0911.4463},
	journal = {JHEP},
	pages = {124},
	primaryclass = {hep-ph},
	reportnumber = {UMD-40762-471, UMD-PP-09-062, IC-2009-090},
	title = {{Leptogenesis as a Common Origin for Matter and Dark Matter}},
	volume = {03},
	year = {2010}}

@article{Baldes:2017gzw,
	archiveprefix = {arXiv},
	author = {Baldes, Iason and Petraki, Kalliopi},
	doi = {10.1088/1475-7516/2017/09/028},
	eprint = {1703.00478},
	journal = {JCAP},
	pages = {028},
	primaryclass = {hep-ph},
	reportnumber = {DESY-17-034, NIKHEF-2017-009},
	title = {{Asymmetric thermal-relic dark matter: Sommerfeld-enhanced freeze-out, annihilation signals and unitarity bounds}},
	volume = {09},
	year = {2017}}

@article{Graesser:2011wi,
	archiveprefix = {arXiv},
	author = {Graesser, Michael L. and Shoemaker, Ian M. and Vecchi, Luca},
	doi = {10.1007/JHEP10(2011)110},
	eprint = {1103.2771},
	journal = {JHEP},
	pages = {110},
	primaryclass = {hep-ph},
	reportnumber = {LA-UR-11-00565},
	title = {{Asymmetric WIMP dark matter}},
	volume = {10},
	year = {2011}}

@article{Tulin:2012re,
	archiveprefix = {arXiv},
	author = {Tulin, Sean and Yu, Hai-Bo and Zurek, Kathryn M.},
	doi = {10.1088/1475-7516/2012/05/013},
	eprint = {1202.0283},
	journal = {JCAP},
	pages = {013},
	primaryclass = {hep-ph},
	reportnumber = {MCTP-12-03},
	title = {{Oscillating Asymmetric Dark Matter}},
	volume = {05},
	year = {2012}}

@article{Buckley:2011ye,
	archiveprefix = {arXiv},
	author = {Buckley, Matthew R. and Profumo, Stefano},
	doi = {10.1103/PhysRevLett.108.011301},
	eprint = {1109.2164},
	journal = {Phys. Rev. Lett.},
	pages = {011301},
	primaryclass = {hep-ph},
	reportnumber = {FERMILAB-PUB-11-437-A},
	title = {{Regenerating a Symmetry in Asymmetric Dark Matter}},
	volume = {108},
	year = {2012}}

@article{Li:2024wqj,
	archiveprefix = {arXiv},
	author = {Li, Song and Yang, Jin Min and Zhang, Mengchao and Zhu, Rui},
	eprint = {2405.18226},
	month = {5},
	primaryclass = {hep-ph},
	title = {{Theoretical bounds on dark Higgs mass in a self-interacting dark matter model with $U(1)'$}},
	year = {2024}}

@article{Han:2023olf,
	archiveprefix = {arXiv},
	author = {Han, Chengcheng and Xie, Ke-Pan and Yang, Jin Min and Zhang, Mengchao},
	doi = {10.1103/PhysRevD.109.115025},
	eprint = {2306.16966},
	journal = {Phys. Rev. D},
	number = {11},
	pages = {115025},
	primaryclass = {hep-ph},
	title = {{Self-interacting dark matter implied by nano-Hertz gravitational waves}},
	volume = {109},
	year = {2024}}

@article{Buckley:2014hja,
	archiveprefix = {arXiv},
	author = {Buckley, Matthew R. and Zavala, Jes\'us and Cyr-Racine, Francis-Yan and Sigurdson, Kris and Vogelsberger, Mark},
	doi = {10.1103/PhysRevD.90.043524},
	eprint = {1405.2075},
	journal = {Phys. Rev. D},
	number = {4},
	pages = {043524},
	primaryclass = {astro-ph.CO},
	title = {{Scattering, Damping, and Acoustic Oscillations: Simulating the Structure of Dark Matter Halos with Relativistic Force Carriers}},
	volume = {90},
	year = {2014}}

@article{Cyr-Racine:2012tfp,
	archiveprefix = {arXiv},
	author = {Cyr-Racine, Francis-Yan and Sigurdson, Kris},
	doi = {10.1103/PhysRevD.87.103515},
	eprint = {1209.5752},
	journal = {Phys. Rev. D},
	number = {10},
	pages = {103515},
	primaryclass = {astro-ph.CO},
	title = {{Cosmology of atomic dark matter}},
	volume = {87},
	year = {2013}}

@article{Cyr-Racine:2013fsa,
	archiveprefix = {arXiv},
	author = {Cyr-Racine, Francis-Yan and de Putter, Roland and Raccanelli, Alvise and Sigurdson, Kris},
	doi = {10.1103/PhysRevD.89.063517},
	eprint = {1310.3278},
	journal = {Phys. Rev. D},
	number = {6},
	pages = {063517},
	primaryclass = {astro-ph.CO},
	title = {{Constraints on Large-Scale Dark Acoustic Oscillations from Cosmology}},
	volume = {89},
	year = {2014}}

@article{Bai:2021ibt,
	archiveprefix = {arXiv},
	author = {Bai, Yang and Korwar, Mrunal},
	doi = {10.1103/PhysRevD.105.095015},
	eprint = {2109.14765},
	journal = {Phys. Rev. D},
	number = {9},
	pages = {095015},
	primaryclass = {hep-ph},
	title = {{Cosmological constraints on first-order phase transitions}},
	volume = {105},
	year = {2022}}

@article{Blennow:2012de,
	archiveprefix = {arXiv},
	author = {Blennow, Mattias and Fernandez-Martinez, Enrique and Mena, Olga and Redondo, Javier and Serra, Paolo},
	doi = {10.1088/1475-7516/2012/07/022},
	eprint = {1203.5803},
	journal = {JCAP},
	pages = {022},
	primaryclass = {hep-ph},
	reportnumber = {CERN-PH-TH-2012-070, MPP-2012-56},
	title = {{Asymmetric Dark Matter and Dark Radiation}},
	volume = {07},
	year = {2012}}

@article{Gori:2020xvq,
	archiveprefix = {arXiv},
	author = {Gori, Stefania and Perez, Gilad and Tobioka, Kohsaku},
	doi = {10.1007/JHEP08(2020)110},
	eprint = {2005.05170},
	journal = {JHEP},
	pages = {110},
	primaryclass = {hep-ph},
	title = {{KOTO vs. NA62 Dark Scalar Searches}},
	volume = {08},
	year = {2020}}

@article{DelNobile:2015uua,
	archiveprefix = {arXiv},
	author = {Del Nobile, Eugenio and Kaplinghat, Manoj and Yu, Hai-Bo},
	doi = {10.1088/1475-7516/2015/10/055},
	eprint = {1507.04007},
	journal = {JCAP},
	pages = {055},
	primaryclass = {hep-ph},
	title = {{Direct Detection Signatures of Self-Interacting Dark Matter with a Light Mediator}},
	volume = {10},
	year = {2015}}

@article{Depta:2020zbh,
	archiveprefix = {arXiv},
	author = {Depta, Paul Frederik and Hufnagel, Marco and Schmidt-Hoberg, Kai},
	doi = {10.1088/1475-7516/2021/04/011},
	eprint = {2011.06519},
	journal = {JCAP},
	pages = {011},
	primaryclass = {hep-ph},
	reportnumber = {DESY-20-160, DESY 20-160, ULB-TH/20-15},
	title = {{Updated BBN constraints on electromagnetic decays of MeV-scale particles}},
	volume = {04},
	year = {2021}}

@article{Vogelsberger:2012ku,
	archiveprefix = {arXiv},
	author = {Vogelsberger, Mark and Zavala, Jesus and Loeb, Abraham},
	doi = {10.1111/j.1365-2966.2012.21182.x},
	eprint = {1201.5892},
	journal = {Mon. Not. Roy. Astron. Soc.},
	pages = {3740},
	primaryclass = {astro-ph.CO},
	title = {{Subhaloes in Self-Interacting Galactic Dark Matter Haloes}},
	volume = {423},
	year = {2012}}

@article{Rocha:2012jg,
	archiveprefix = {arXiv},
	author = {Rocha, Miguel and Peter, Annika H. G. and Bullock, James S. and Kaplinghat, Manoj and Garrison-Kimmel, Shea and Onorbe, Jose and Moustakas, Leonidas A.},
	doi = {10.1093/mnras/sts514},
	eprint = {1208.3025},
	journal = {Mon. Not. Roy. Astron. Soc.},
	pages = {81--104},
	primaryclass = {astro-ph.CO},
	title = {{Cosmological Simulations with Self-Interacting Dark Matter I: Constant Density Cores and Substructure}},
	volume = {430},
	year = {2013}}

@article{Planck:2018vyg,
	archiveprefix = {arXiv},
	author = {Aghanim, N. and others},
	collaboration = {Planck},
	doi = {10.1051/0004-6361/201833910},
	eprint = {1807.06209},
	journal = {Astron. Astrophys.},
	note = {[Erratum: Astron.Astrophys. 652, C4 (2021)]},
	pages = {A6},
	primaryclass = {astro-ph.CO},
	title = {{Planck 2018 results. VI. Cosmological parameters}},
	volume = {641},
	year = {2020}}

@article{Moore:1999nt,
	archiveprefix = {arXiv},
	author = {Moore, B. and Ghigna, S. and Governato, F. and Lake, G. and Quinn, Thomas R. and Stadel, J. and Tozzi, P.},
	doi = {10.1086/312287},
	eprint = {astro-ph/9907411},
	journal = {Astrophys. J. Lett.},
	pages = {L19--L22},
	title = {{Dark matter substructure within galactic halos}},
	volume = {524},
	year = {1999}}

@article{Klypin:1999uc,
	archiveprefix = {arXiv},
	author = {Klypin, Anatoly A. and Kravtsov, Andrey V. and Valenzuela, Octavio and Prada, Francisco},
	doi = {10.1086/307643},
	eprint = {astro-ph/9901240},
	journal = {Astrophys. J.},
	pages = {82--92},
	title = {{Where are the missing Galactic satellites?}},
	volume = {522},
	year = {1999}}

@article{Oman:2015xda,
	archiveprefix = {arXiv},
	author = {Oman, Kyle A. and others},
	doi = {10.1093/mnras/stv1504},
	eprint = {1504.01437},
	journal = {Mon. Not. Roy. Astron. Soc.},
	number = {4},
	pages = {3650--3665},
	primaryclass = {astro-ph.GA},
	title = {{The unexpected diversity of dwarf galaxy rotation curves}},
	volume = {452},
	year = {2015}}

@article{Moore:1999gc,
	archiveprefix = {arXiv},
	author = {Moore, Ben and Quinn, Thomas R. and Governato, Fabio and Stadel, Joachim and Lake, George},
	doi = {10.1046/j.1365-8711.1999.03039.x},
	eprint = {astro-ph/9903164},
	journal = {Mon. Not. Roy. Astron. Soc.},
	pages = {1147--1152},
	title = {{Cold collapse and the core catastrophe}},
	volume = {310},
	year = {1999}}

@article{Moore:1994yx,
	author = {Moore, B.},
	doi = {10.1038/370629a0},
	journal = {Nature},
	pages = {629},
	title = {{Evidence against dissipationless dark matter from observations of galaxy haloes}},
	volume = {370},
	year = {1994}}

@article{Flores:1994gz,
	archiveprefix = {arXiv},
	author = {Flores, Ricardo A. and Primack, Joel R.},
	doi = {10.1086/187350},
	eprint = {astro-ph/9402004},
	journal = {Astrophys. J. Lett.},
	pages = {L1--4},
	reportnumber = {SCIPP-93-01-REV, SCIPP-93-01},
	title = {{Observational and theoretical constraints on singular dark matter halos}},
	volume = {427},
	year = {1994}}

@article{Springel:2006vs,
	archiveprefix = {arXiv},
	author = {Springel, Volker and Frenk, Carlos S. and White, Simon D. M.},
	doi = {10.1038/nature04805},
	eprint = {astro-ph/0604561},
	journal = {Nature},
	pages = {1137},
	title = {{The large-scale structure of the Universe}},
	volume = {440},
	year = {2006}}

@article{Blumenthal:1984bp,
	author = {Blumenthal, George R. and Faber, S. M. and Primack, Joel R. and Rees, Martin J.},
	doi = {10.1038/311517a0},
	editor = {Srednicki, M. A.},
	journal = {Nature},
	pages = {517--525},
	reportnumber = {SLAC-PUB-3307},
	title = {{Formation of Galaxies and Large Scale Structure with Cold Dark Matter}},
	volume = {311},
	year = {1984}}

@article{Clowe:2006eq,
	archiveprefix = {arXiv},
	author = {Clowe, Douglas and Bradac, Marusa and Gonzalez, Anthony H. and Markevitch, Maxim and Randall, Scott W. and Jones, Christine and Zaritsky, Dennis},
	doi = {10.1086/508162},
	eprint = {astro-ph/0608407},
	journal = {Astrophys. J. Lett.},
	pages = {L109--L113},
	reportnumber = {SLAC-PUB-12078},
	title = {{A direct empirical proof of the existence of dark matter}},
	volume = {648},
	year = {2006}}

@article{Cyr-Racine:2015ihg,
	archiveprefix = {arXiv},
	author = {Cyr-Racine, Francis-Yan and Sigurdson, Kris and Zavala, Jesus and Bringmann, Torsten and Vogelsberger, Mark and Pfrommer, Christoph},
	doi = {10.1103/PhysRevD.93.123527},
	eprint = {1512.05344},
	journal = {Phys. Rev. D},
	number = {12},
	pages = {123527},
	primaryclass = {astro-ph.CO},
	title = {{ETHOS\textemdash{}an effective theory of structure formation: From dark particle physics to the matter distribution of the Universe}},
	volume = {93},
	year = {2016}}

@article{Athron:2024xrh,
	archiveprefix = {arXiv},
	author = {Athron, Peter and Balazs, Csaba and Fowlie, Andrew and Morris, Lachlan and Searle, William and Xiao, Yang and Zhang, Yang},
	eprint = {2412.04881},
	month = {12},
	primaryclass = {astro-ph.CO},
	title = {{PhaseTracer2: from the effective potential to gravitational waves}},
	year = {2024}}

@article{Ghosh:2024cxi,
	archiveprefix = {arXiv},
	author = {Ghosh, Dilip Kumar and Ghosh, Purusottam and Jeesun, Sk and Srivastava, Rahul},
	doi = {10.1103/PhysRevD.110.075032},
	eprint = {2404.10077},
	journal = {Phys. Rev. D},
	number = {7},
	pages = {075032},
	primaryclass = {hep-ph},
	title = {{Neff at CMB challenges U(1)X light gauge boson scenarios}},
	volume = {110},
	year = {2024}}

@article{Blinnikov:1982eh,
	author = {Blinnikov, S. I. and Khlopov, M. Yu.},
	journal = {Sov. J. Nucl. Phys.},
	pages = {472},
	reportnumber = {ITEP-11-1982},
	title = {{ON POSSIBLE EFFECTS OF 'MIRROR' PARTICLES}},
	volume = {36},
	year = {1982}}

@article{Beylin:2020bsz,
	archiveprefix = {arXiv},
	author = {Beylin, Vitaly and Khlopov, Maxim and Kuksa, Vladimir and Volchanskiy, Nikolay},
	doi = {10.3390/universe6110196},
	eprint = {2010.13678},
	journal = {Universe},
	number = {11},
	pages = {196},
	primaryclass = {hep-ph},
	title = {{New physics of strong interaction and Dark Universe}},
	volume = {6},
	year = {2020}}

@article{Croon:2018erz,
	archiveprefix = {arXiv},
	author = {Croon, Djuna and Sanz, Ver\'onica and White, Graham},
	doi = {10.1007/JHEP08(2018)203},
	eprint = {1806.02332},
	journal = {JHEP},
	pages = {203},
	primaryclass = {hep-ph},
	title = {{Model Discrimination in Gravitational Wave spectra from Dark Phase Transitions}},
	volume = {08},
	year = {2018}}

@article{Kamada:2023iol,
	archiveprefix = {arXiv},
	author = {Kamada, Ayuki and Kuwahara, Takumi and Patel, Ami},
	doi = {10.1007/JHEP11(2023)105},
	eprint = {2303.17961},
	journal = {JHEP},
	pages = {105},
	primaryclass = {hep-ph},
	title = {{Quantum theory of dark matter scattering}},
	volume = {11},
	year = {2023}}

@article{Athron:2023xlk,
	archiveprefix = {arXiv},
	author = {Athron, Peter and Bal\'azs, Csaba and Fowlie, Andrew and Morris, Lachlan and Wu, Lei},
	doi = {10.1016/j.ppnp.2023.104094},
	eprint = {2305.02357},
	journal = {Prog. Part. Nucl. Phys.},
	pages = {104094},
	primaryclass = {hep-ph},
	title = {{Cosmological phase transitions: From perturbative particle physics to gravitational waves}},
	volume = {135},
	year = {2024}}

@article{Caputo:2025aac,
	archiveprefix = {arXiv},
	author = {Caputo, Andrea and Janka, Hans-Thomas and Raffelt, Georg and Yun, Seokhoon},
	doi = {10.1103/PhysRevLett.134.151002},
	eprint = {2502.01731},
	journal = {Phys. Rev. Lett.},
	number = {15},
	pages = {151002},
	primaryclass = {hep-ph},
	title = {{Cooling the Shock: New Supernova Constraints on Dark Photons}},
	volume = {134},
	year = {2025}}

\end{document}